\documentclass[11pt,twoside,a4paper]{article}

\usepackage{hyperref}
\usepackage[nojss]{jss}
\usepackage[utf8]{inputenc}

\usepackage[numbered,framed,autolinebreaks,useliterate]{matlab-prettifier}

\usepackage{graphicx}
\usepackage{subfig}
\usepackage{amssymb}
\usepackage{commath}
\usepackage{dsfont}
\usepackage{bm}
\usepackage{booktabs}
\usepackage{float}

\lstdefinestyle{mystyle}{
numbers=left,
numberstyle=\small,
numbersep=7pt,
language=Matlab,
style=Matlab-editor,
basicstyle=\ttfamily\tiny,
xleftmargin=-12pt,
aboveskip=-6pt,
belowskip=-6pt,
frame=none
}

\newcommand{\R}{\ensuremath{\mathbb{R}}}           % real numbers
\newcommand{\N}{\ensuremath{\mathbb{N}}}           % natural numbers
\renewcommand{\vec}[1]{\bm{#1}}

\newcommand{\indfunc}[1]{\ensuremath{\mathds{1}_{#1} }}
\renewcommand{\SS}{\Omega}

\newcommand{\SF}{\Sigma}

\newcommand{\PM}{\mathcal{P}}

\newcommand{\randvar}[1]{#1}

\newcommand{\randvec}[1]{\vec{\randvar{#1}}}

\newcommand{\pdf}[3]{#1(#2, #3)}

\newcommand{\cdf}[1]{P(#1)}

\newcommand{\expval}[2]{\mathbb{E}_{#2} \left\lbrace #1 \right\rbrace}

\newcommand{\var}[2]{\mathbb{V}_{#2} \left\lbrace #1 \right\rbrace}

\newcommand{\KL}[2]{\mathcal{D}_{\text{KL}} \left(#1 \, || \,  #2 \right)}

\newcommand{\estim}[1]{\widehat{#1}}

\title{\pkg{CEopt}: A MATLAB Package for Non-convex Optimization with the Cross-Entropy Method}

\Plaintitle{\pkg{CEopt}: A MATLAB Package for Non-convex Optimization with the Cross-Entropy Method}

\Shorttitle{\pkg{CEopt}: A MATLAB Pack. for Non-convex Opt. with the Cross-Entropy Method}

\author{Americo Cunha~Jr\\ Rio de Janeiro State University\\
    \And 
    Marcos Vinicius Issa\\ Rio de Janeiro State University\\ 
    \AND
    Julio Cesar Basilio\\ Rio de Janeiro State University\\
    \And 
    Jos\'{e} Geraldo Telles Ribeiro\\ Rio de Janeiro State University\\
    }

\Plainauthor{A. Cunha~Jr et al.}
\Abstract{This paper introduces \pkg{CEopt} (\url{https://ceopt.org}), a MATLAB tool leveraging the Cross-Entropy method for non-convex optimization. Due to the relative simplicity of the algorithm, it provides a kind of transparent ``gray-box'' optimization solver, with intuitive control parameters. Unique in its approach, \pkg{CEopt} effectively handles both equality and inequality constraints using an augmented Lagrangian method, offering robustness and scalability for moderately sized complex problems. Through select case studies, the package's applicability and effectiveness in various optimization scenarios are showcased, marking \pkg{CEopt} as a practical addition to optimization research and application toolsets.
}
\Address{  
  Americo Cunha~Jr\\
  Department of Applied Mathematics\\
  Rio de Janeiro State University -- UERJ\\
  Rua S\~{a}o Francisco Xavier, 524, Rio de Janeiro, RJ - Brasil - 20550-900\\
  E-mail: \email{americo.cunha@uerj.br}\\
}

\begin{document}

% --------------------------------------------------------------
\section{Introduction}
\label{sec:Intro}

Optimization underpins numerous advances across many disciplines like machine learning, control systems, and scientific computing, aiming universally to maximize outcomes or minimize costs under given constraints \citep{bonnans2009}. Challenges in this domain fall into two primary categories: convex and non-convex. While convex optimization benefits from a more straightforward approach to finding global solutions, non-convex optimization introduces significant challenges. The last category is characterized by a landscape peppered with local optima and abrupt discontinuities, complicating the search for global optima and often necessitating more advanced, nuanced optimization strategies to effectively navigate these complexities \citep{nocedal2006}.

\subsection{Convexity and beyond in optimization}

Convex optimization, characterized by the convexity of the objective function and constraints, holds a privileged position in the optimization landscape. When an optimization problem can be elegantly cast into a convex formulation, the path to a solution is often smooth and well-paved \citep{boyd2004,bertsekas2009}. In this sense, the linear programming, quadratic programming, and semi-definite programming paradigms have revolutionized countless applications like: control theory \citep{Roux2022}; image reconstruction \citep{Vijina2020}; statistical learning \citep{Hastie2016}; etc.

However, the reality of optimization problems is not always so conveniently convex. Many real-world challenges defy the assumptions of convexity \citep{bertsekas2016}. Consider, for instance, the vast domains of data science and machine learning \citep{kroese2019,Brunton2022}; inverse problems \citep{Kaipio2004,tarantola2005,Quaranta2020p1709}; structural optimization and mechanical design \citep{Goncalves2015p165,Cunhajr2019p1057}; simulation-based optimization \citep{Cunhajr2015p849,CunhaJr2021p137}, uncertainty quantification \citep{Tenorio2017,Soize2017} and beyond. These domains teem with complex, multidimensional landscapes where non-convexity reigns supreme. The formulation of these problems often strays from convexity, leading to scenarios where the journey toward the optimal solution becomes treacherous.

Non-convex optimization introduces a myriad of complexities. The landscape becomes speckled with numerous local optima, leaving traditional gradient-based methods vulnerable to converging at suboptimal solutions. Moreover, non-differentiability, discontinuities, and other irregularities further obstruct the optimization path. In the face of such formidable challenges, the choice of an appropriate solver becomes of paramount importance \citep{bonnans2009}. 

To offer readers an overview of optimization solvers available for practical and general purpose use, we will now explore various software tools commonly referenced in literature and utilized across diverse computational platforms.

\subsection{Solvers for optimization problems}

Decades of research and development have cultivated a vibrant ecosystem of solvers for convex optimization \citep{bertsekas2015}, flourishing within the open-source community and beyond, offering seamless integration into widely used numerical packages and platforms:

\begin{itemize}
	\item Python's SciPy \citep{2020SciPy-NMeth} is a fundamental library for scientific computing, providing a rich collection of optimization tools.
	\item Julia's JuMP \citep{Lubin2023} offers a powerful modeling language for optimization problems.
	\item AMPL \citep{Fourer1996}, a comprehensive and highly flexible modeling system for solving large-scale optimization problems.
	\item MATLAB's Optimization Toolbox \citep{MatlabOptToolbox} provides a vast suite of algorithms for linear, nonlinear, and binary optimization.
	\item CVX \citep{Grant2008,cvx2014}, a MATLAB and Python tool for disciplined convex programming, streamlining the process of solving convex optimization problems.
	\item R's ROI \citep{Theussl2020p1,ROI2023} package integrates a wide range of optimization techniques, making them accessible to the R environment.
	\item PETSc for Fortran \citep{PETSC2023} brings robust numerical tools for parallel scientific computations, including optimization routines.
	\item GNU Scientific Library (GSL) \citep{GSL2018} caters to C/C++ programmers with a collection of mathematical routines, including optimization functions.
	\item NLopt \citep{NLopt2007} provides a library for nonlinear optimization, supporting a multitude of programming languages.
\end{itemize}

For those tackling convex optimization challenges, this assortment of tools often represents the end of their search. These solvers enable efficient and reliable optimization processes, circumventing significant obstacles.

Conversely, non-convex optimization, with its intrinsic complexity, demands a broader spectrum of solutions \citep{bazaraa2006}. This domain benefits from metaheuristic and evolutionary algorithms capable of overcoming the hurdles posed by multiple local optima and intricate solution spaces.

Software implementations for non-convex optimization include metaheuristic methods such as Genetic Algorithms (GA), Simulated Annealing (SA), and Particle Swarm Optimization (PSO). These methods, which can navigate through multimodal landscapes to seek global optima, are embodied in various platforms catering to different computational needs:

\begin{itemize}
	\item MATLAB's Global Optimization Toolbox \citep{MatlabGlobalOptToolbox} provides an integrated environment for applying GA, SA, and PSO, making it a powerful suite for users facing non-convex challenges.
	\item Python's rich ecosystem, in addition to SciPy \citep{2020SciPy-NMeth}, also includes libraries like DEAP for evolutionary algorithms \citep{DEAP_JMLR2012} or PySwarms for particle swarm optimization \citep{pyswarmsJOSS2018}. %These libraries allow for flexible and powerful optimization workflows, capitalizing on Python's extensive capabilities for scientific computing.
	\item For R users, packages such as GA \citep{Scrucca2013p1} and PSO \citep{PSO-R-Manual2022} offer tools to implement genetic and particle swarm optimization techniques, respectively, within the R programming environment, enabling statisticians and data scientists to incorporate sophisticated optimization routines in their analyses.
\end{itemize}

Despite the utility of these metaheuristic methods in exploring vast and complex solution spaces, they often operate as ``black-box'' solvers. This means that while they are capable of finding solutions to difficult optimization problems, the internal workings and decision processes of these algorithms often is opaque, making it challenging for users to interpret the optimization process or to adjust strategies with precision.

This backdrop of existing non-convex optimization software sets the stage for the introduction of the Cross-Entropy (CE) method and its MATLAB implementation dubbed \pkg{CEopt}. The subsequent sections will explore the theoretical underpinnings of the CE method, the architecture and features of \pkg{CEopt}, and its practical application across various domains, highlighting its potential to fill a significant gap in the non-convex optimization landscape.

\subsection{The Cross-Entropy method}

\begin{figure}[h]
	\centering
	\includegraphics[scale=0.5]{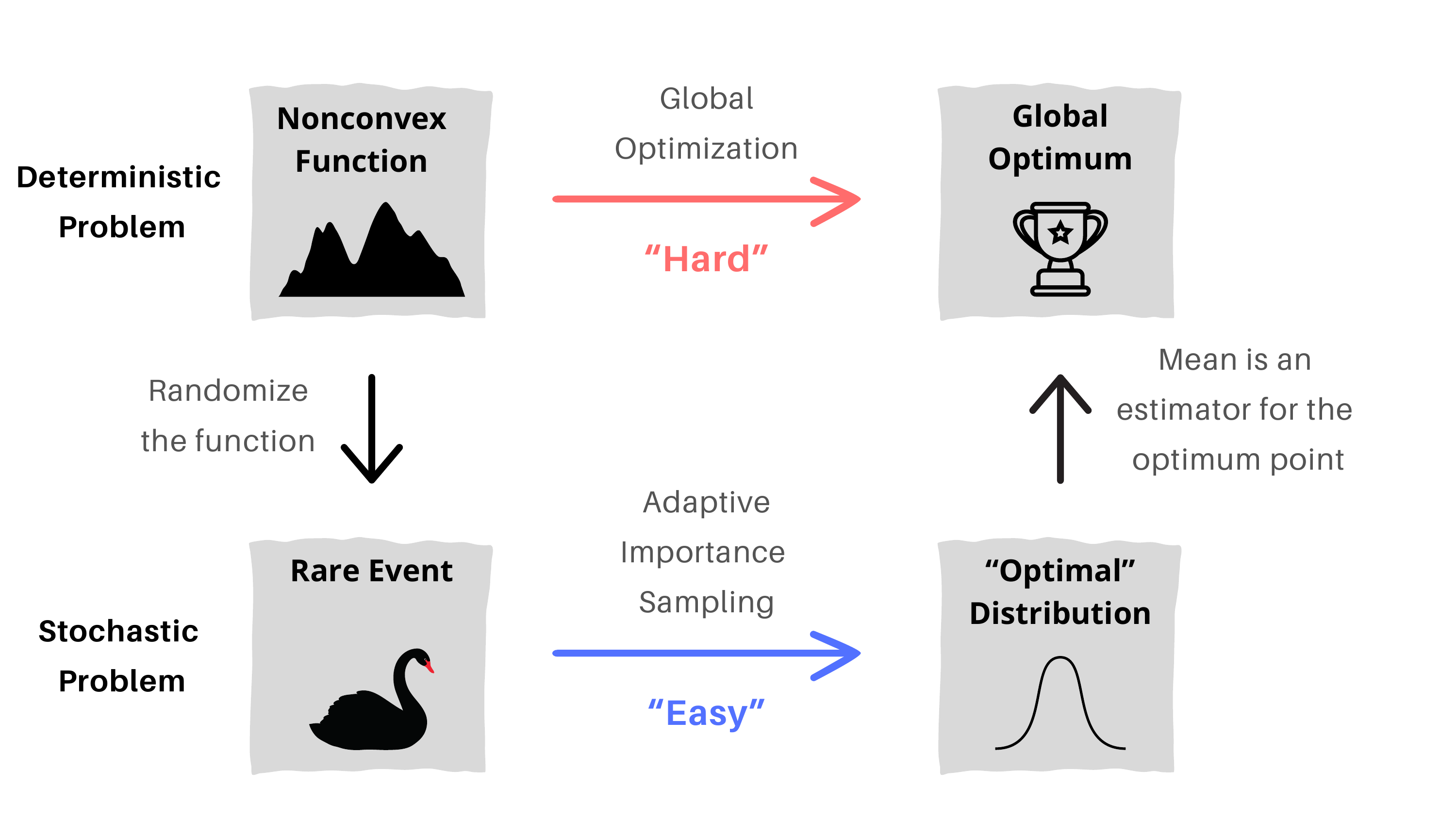}
	\caption{Cross-Entropy Method Explained: This method turns a complex optimization problem into a manageable task of estimating a rare event. By using adaptive importance sampling, it iteratively focuses on making the global optimum more likely to find. The mean of the ``optimal'' distribution estimates the best solution, showcasing how CE simplifies tough optimization challenges into more approachable ones with a probabilistic strategy.}
	\label{fig:CEprocess}
\end{figure}

The CE method embodies a systematic and lucid strategy for optimization, grounded in the robust principles of probability theory \citep{Rubinstein2004}. This method transforms deterministic optimization challenges into stochastic problems centered on the estimation of rare event probabilities, as represented in Figure~\ref{fig:CEprocess}. Through the technique of adaptive importance sampling, CE meticulously refines its pursuit of the global optimum across iterations. Such an approach not only deepens our comprehension of the intricate optimization terrain but also guarantees a focused and efficient exploration process. The probabilistic underpinnings of CE provide a flexible framework capable of addressing a broad spectrum of optimization dilemmas. From continuous to combinatorial and mixed problems, CE delineates a clear and methodical pathway to uncovering solutions, demonstrating its adaptability and broad applicability \citep{Rubinstein2005p5,Kroese2006p383}.

Contrasting with the often opaque ``black-box'' optimization methods, CE boasts  simplicity and transparency. It operates with a minimal set of control parameters, each serving a clear and defined purpose. These include the number of samples per iteration, the size of an elite subset of samples for learning, the maximum iteration count, convergence tolerances, and smoothing parameters to enhance the optimization process. The CE method's working mechanism, which is reviewed in this paper, is both elegant and understandable.

Imagine optimization as a journey through a vast landscape, with peaks and valleys representing objective function values. In this context, CE operates as a skilled navigator, charting a course to reach the optimal summit. CE begins by sampling the objective function domain according to a carefully chosen probability distribution, akin to casting a net over the landscape to catch points. Among the sampled points, CE identifies the ones that yield better results (those with lower objective function values). These elite points form the basis for learning and improvement. CE then engages in a stochastic learning process, striving to understand the landscape by updating its probability distribution.

In the CE method, the probability distribution plays a dual role, with its mean representing the method's current estimate of the optimal point. As the optimization journey unfolds, this mean undergoes a gradual shift in the direction of the global optimum, akin to a climber closing in on a summit. Simultaneously, the distribution's variance undergoes reduction, symbolizing the method's growing confidence in the vicinity where the optimal solution likely resides. This decrease in variance effectively narrows the method's exploration towards the true optimal location. In the ideal limit, the probability distribution converges to a delta function precisely centered at the global optimum.

The learning process of CE hinges on the dynamic adjustment of two pivotal parameters: the mean and the variance. These crucial updates draw wisdom from the elite samples, which represent the points that have consistently exhibited superior performance during the optimization process. The mean's adaptation aligns it more closely with the direction of the global optimum, resembling a navigator fine-tuning their course. Meanwhile, the variance diminishes in response to the method's heightened confidence, particularly in the vicinity surrounding the presumed optimal solution. This intricate interplay, finely tuning both the mean and variance, stands as a cornerstone of CE's effectiveness, guaranteeing its progressive convergence towards the global optimum across successive iterations.

\subsection{Paper contribution and organization}

In this paper, we introduce \pkg{CEopt} (\url{https://ceopt.org}), a MATLAB package developed to leverage the Cross-Entropy (CE) method for continuous non-convex optimization challenges. We detail the architecture and software engineering principles that underpin \pkg{CEopt}, designed with the goal of ensuring easy integration into diverse optimization workflows.

Through a variety of case studies, from simple to more complex scenarios, we demonstrate the utility of \pkg{CEopt}. Whether for refining machine learning models, enhancing structural optimization, or venturing into new research areas, \pkg{CEopt} provides a valuable tool for researchers and practitioners navigating the intricate landscapes of non-convex optimization.

Due to the solid probabilistic foundations and relative simplicity of CE method algorithm, \pkg{CEopt} represents a step forward in optimization software by offering a kind of ``gray-box'' MATLAB solver. The control parameters of the method transparent and user-friendly, addressing a frequent issue encountered with more opaque ``black-box'' solutions. Additionally, \pkg{CEopt} features a refined framework for addressing both equality and inequality constraints through an augmented Lagrangian method. While this provides robustness against problems laden with constraints, it is important to recognize that \pkg{CEopt}, like any tool, is not a panacea for any optimization problem, specially in large dimensions. It is, however, a significant aid in the toolkit of those tackling non-convex optimization problems, offering clarity and control where such qualities are often sought but not always found.

This work contributes to the non-convex optimization field by presenting \pkg{CEopt} as a pragmatic and accessible tool, enhancing the ability of users to engage with and solve complex optimization problems. While \pkg{CEopt} is a notable addition to the optimization toolbox, we acknowledge its role as part of a broader ensemble of methods necessary to address the wide array of challenges presented by non-convex optimization scenarios.

\pagebreak
The remainder of this paper is organized as follows:

\begin{itemize}
	\item \textbf{Historical development of the CE method:} A journey through the evolution and milestones of the CE method, setting the stage for its current applications.
	\item \textbf{Theoretical foundation of the CE method:} An exploration of the probabilistic principles and mechanisms that constitute the backbone of the CE method.
	\item \textbf{MATLAB Implementation: The \pkg{CEopt} package:} A walkthrough into the code  architecture, highlighting its modular design, user-friendliness, and integration capabilities.
	\item \textbf{Numerical experiments:} A compilation of empirical studies demonstrating \pkg{CEopt}'s versatility and power across a spectrum of optimization scenarios.
	\item \textbf{Final remarks:} Concluding observations and reflections on \pkg{CEopt}’s contribution to the field of optimization, along with a glimpse into potential expansions of the code.
\end{itemize}

% --------------------------------------------------------------

% --------------------------------------------------------------
\section{Historical development of the CE method}
\label{sec:History}

\begin{figure}[h]
	\centering
	\includegraphics[scale=1]{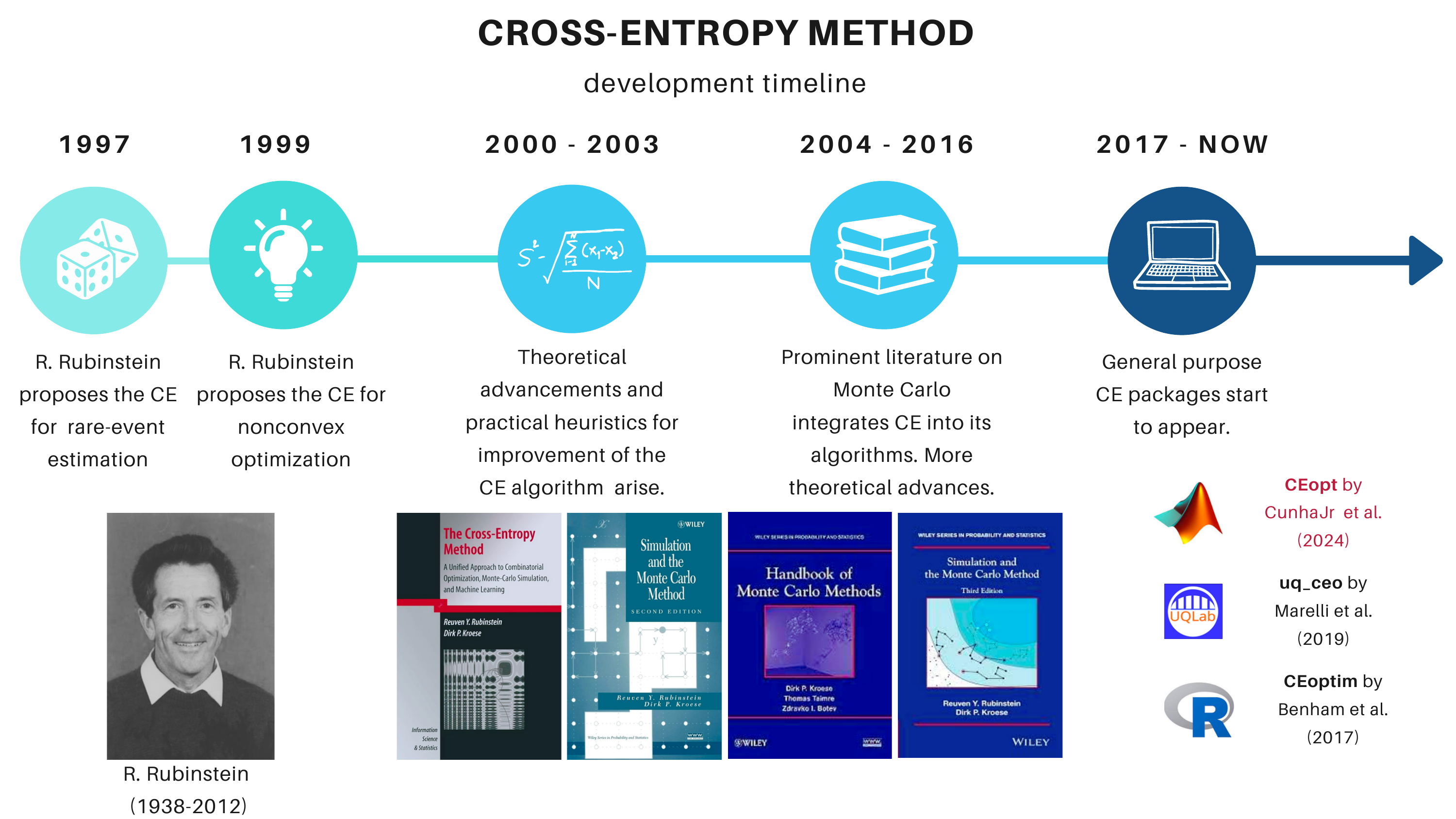}
	\caption{Cross-Entropy Method Timeline: Beginning with R. Rubinstein's initial proposal in 1997 for rare-event estimation, this timeline tracks the evolution of the method into non-convex optimization and beyond. Highlighting key theoretical advancements and the development of CE software, it illustrates the method's increasing significance in the field of optimization.}
	\label{fig:CEtimeline}
\end{figure}

The Cross-Entropy (CE) method has evolved from its humble beginnings in rare event estimation to becoming a versatile tool for non-convex optimization. This section explores the historical development of CE, tracing its journey from its pioneering work in rare event estimation by Reuven Rubinstein\footnote{Reuven Rubinstein (1938-2012) was an Israeli scientist renowned for his substantial contributions to Monte Carlo simulation, applied probability, stochastic modeling, and stochastic optimization. He authored over one hundred papers and six books during his distinguished career. Rubinstein was recognized as the founder of pioneering methods such as the score function method, stochastic counterpart method, and cross-entropy method, which have widespread applications in combinatorial optimization and simulation.  In 2010, he received the INFORMS Simulation Society's Lifetime Professional Achievement Award and, in 2011, the Lifetime Professional Award from the Operations Research Society of Israel (ORSIS) in recognition of his fundamental contributions to the fields of simulation and operations research.} to its prominent role in modern applications, theoretical advancements, and software implementations. A timeline illustrating key milestones in the CE method's evolution can be found in Figure~\ref{fig:CEtimeline}.

The roots of the CE method can be traced back to the pioneering work of Rubinstein  \citep{Rubinstein1997p89} about rare event estimation problems in the context of Monte Carlo simulations. Here, the problem is addressed using an adaptive importance sampling algorithm, allowing a significant speed-up on the estimation process. 

A key turning points in the history of CE was the realization that an adaptive importance sampling strategy could be extended to tackle optimization problems \citep{Rubinstein1999p127}. By introducing a controlled randomness in the design variables and minimizing the Kullback–Leibler divergence between a zero-variance distribution at the global optimum and the importance sampling distribution, CE could efficiently explore complex and non-convex landscapes to find high-quality optimal solutions. 

This discovery opened the door to a wide range of applications in optimization problems, in several areas of knowledge, such as clustering and classification \citep{Mannor2005,Kroese2007p137}; multi-objective optimization \citep{Bekker2011p112,Haber2017p22272}; vehicle routing problem \citep{Chepuri2005p153}; signal optimization \citep{Maher2013p76}; structural design \citep{Issa2018cnmac,Issa2023cobem,Farahmand-Tabar2023}; reliability assessment \citep{Kroese2007p37,Chan2023p102344}; energy harvesting optimization \citep{CunhaJr2021p137}; portfolio optimization \citep{Fernandez2023}; calibration of dynamical systems models for structural dynamics and finite element simulation \citep{Dantas2019icedyn,Dantas2019cobem,Raqueti2023p4799}. Additionally, during the COVID-19 pandemic, CE was employed to calibrate models used for predicting the spread of the virus \citep{CunhaJr2023p9649}, aiding in decision-making processes.

The CE method's journey was accompanied by the development of theoretical guarantees and insights, such as the convergence properties of CE for global optimization. Landmark theorems \citep{Homem-de-Mello2002,Asmussen2005p57,Margolin2005p201,Homem-de-Mello2007p381} established its legitimacy as a robust optimization and rare event estimation technique. This theoretical foundation reassured practitioners of CE's effectiveness in solving complex problems.

In 2004, the publication of the book \emph{The Cross-Entropy Method: A Unified Approach to Combinatorial Optimization, Monte-Carlo Simulation and Machine Learning} by \cite{Rubinstein2004} marked a significant milestone in the CE method's history. This comprehensive book provided a detailed account of the CE method's principles and practical applications, solidifying its position as a powerful optimization technique.

In the years following its inception, the Cross-Entropy (CE) method gained widespread recognition and accessibility through influential publications such as \citep{DeBoer2005p19, Rubinstein2007, Rubinstein2016,Kroese2013,Botev2013p35}. Its significance is underscored by its inclusion in authoritative references like the \emph{Monte Carlo Handbook} by \cite{kroese2011}, solidifying its reputation as a valuable tool in the realm of Monte Carlo simulations.

%Software Implementations
The CE method's practicality was greatly enhanced by its incorporation into software packages.  Notable implementations include \pkg{CEoptim} in R \citep{Benham2017p1}, and the function \emph{uq\_ceo} in the UQLib \citep{UQdoc_20_201} module of UQlab package \citep{Marelli2014UQLab}, available in MATLAB and Python. However, these implementations do not support complex constraint handling. \pkg{CEopt} differentiates itself by including robust features for managing both equality and inequality nonlinear constraints, making it particularly useful for engineering and control systems where such constraints are critical. Additionally, \pkg{CEopt} offers advanced stopping criteria, expanding its applicability in sophisticated optimization tasks.

%Modern Applications
Today, the CE method keep finding applications in cutting-edge fields such as robotics \citep{Xu2024p1} and power grids \citep{Zhao2024p109857}. Its ability to handle complex, non-convex optimization problems aligns well with the challenges posed by modern machine learning models, where finding optimal hyperparameters and model architectures is essential.

 % --------------------------------------------------------------

% --------------------------------------------------------------
\section{Theoretical foundation of the CE method}
\label{sec:CEtheory}

The Cross-Entropy method provides a robust framework for comprehending its operational principles and effectiveness. At its essence, CE harnesses the concept of importance sampling, a fundamental technique in probability theory and Monte Carlo methods. This concept assumes a pivotal role in CE's capacity to efficiently explore and converge towards optimal solutions. In this section, we present a comprehensive overview of the CE method's theory, elucidating how it employs an adaptive importance sampling strategy to yield high-quality solutions for non-convex optimization challenges.

\subsection{Deterministic optimization problem}

The primary problem of interest here is to find an optimal value that minimizes a certain scalar function. This problem can be compactly represented as
\begin{equation}
\vec{x}^{\star} = \arg\min_{ \displaystyle \vec{x} \in \mathcal{X} }{ F(\vec{x}) } \, ,
\label{Eq:OptDeterm}
\end{equation}
where the (cost) objective function $\vec{x} \in \mathcal{X} \mapsto F(\vec{x}) \in \R$ depends nonlinearly on the design variable vector $\vec{x} = (x_1, \cdots, x_n)$. The latter is defined within an admissible set $\mathcal{X} \subset \R^n$ that describes the constraints to be satisfied by its components. The common instances in which we deal with this problem are:
\begin{equation}
	\mathcal{X} = 
	\left\lbrace 
	\vec{x} \in \R^n 
	\,\,\vert\,\, 
	\vec{x}_{\ell} \preceq \vec{x} \preceq \vec{x}_{u}
	\right\rbrace  \, ,
	\label{Eq:AdmSet1}
\end{equation}
\begin{equation}
	\mathcal{X} = 
	\left\lbrace 
	\vec{x} \in \R^n 
	\,\,\vert\,\, 
	\vec{x}_{\ell} \preceq \vec{x} \preceq \vec{x}_{u} \, ,
	H_i(\vec{x}) = 0, \, i=1, \cdots, N_E \right\rbrace \, ,
	\label{Eq:AdmSet2}
\end{equation}
or
\begin{equation}
	\mathcal{X} = 
	\left\lbrace 
	\vec{x} \in \R^n 
	\,\,\vert\,\, 
	\vec{x}_{\ell} \preceq \vec{x} \preceq \vec{x}_{u} \, ,
	H_i(\vec{x}) = 0, \, i=1, \cdots, N_E \, ,
	G_j(\vec{x}) \leq 0, \, j=1, \cdots, N_I 
	\right\rbrace \, .
	\label{Eq:AdmSet3}
\end{equation}

In all cases, the components of $\vec{x}$ have lower and upper bounds defined by $\vec{x}_{\ell}$ and $\vec{x}_{u}$, respectively, and the generalized inequality $\preceq$ indicates that each component of the left vector is less than or equal to the respective component of the right vector. The first scenario also incorporates the simple case where $\mathcal{X} = \R^n$, by admitting the possibility of infinity bound limits. In the last two scenarios, it is also necessary to satisfy a set of (usually nonlinear) constraints, defined by the functions $\vec{x} \in \mathcal{X} \mapsto H_i(\vec{x}) \in \R (j=1, \cdots, N_E)$, in case of equality conditions, and by $\vec{x} \in \mathcal{X} \mapsto G_j(\vec{x}) \in \R (j=1, \cdots, N_I)$, for the inequality conditions.

For numerical implementation purposes, when constraints apply -- admissible sets (\ref{Eq:AdmSet2}) and (\ref{Eq:AdmSet3}) -- it is necessary to transform the problem (\ref{Eq:OptDeterm}) into an equivalent bounded unconstrained problem -- admissible set (\ref{Eq:AdmSet1}) -- seeking to minimize a modified function $\vec{x} \in \mathcal{X} \mapsto \tilde{F}(\vec{x}) \in \R$. 

In the CE literature, various strategies have been proposed for defining the modified objective function $\tilde{F}(\vec{x})$, with penalization techniques commonly favored \citep{Rubinstein2004}. Nonetheless, due to its superior stability and enhanced ability to manage a broad spectrum of constraints, we opted for the augmented Lagrangian approach \citep{Conn1991p545,Conn1997p261}, which is formulated as
\begin{equation}
\tilde{F}(\vec{x}) = F(\vec{x}) + 
\sum_{i=1}^{N_E}  \left( \lambda_i^{E} \, H_i(\vec{x})  + \frac{1}{2} \, \nu \,  H_i(\vec{x})^2 \right) -
\sum_{j=1}^{N_I} \lambda_j^{I} \, s_j \,  \ln{\left( s_j - G_j(\vec{x}) \right)} \, ,
\label{Eq:AugLag1}
\end{equation}
where $\lambda_i^{E}$ and $\lambda_j^{I}$ are the non-negative Lagrange multiplier estimates for equality and inequality constraints, respectively; $\nu > 0$ acts as a penalty factor; and $s_j$, the non-negative shifts, are given by
\begin{equation}
s_j = \frac{\lambda_j^{I}}{\nu} \, .
\label{Eq:AugLag2}
\end{equation}

This augmented Lagrangian transformation facilitates the handling of both equality and inequality constraints within an unconstrained optimization framework by substituting $F(\vec{x})$ with $\tilde{F}_k(\vec{x})$. The latter represents a sequentially indexed version of $\tilde{F}(\vec{x})$, derived from an incrementally adjusted penalty $\nu_k$. In this iterative process, a solution $\vec{x}_k$ is computed and used to update the Lagrange multipliers according to
\begin{equation}
\begin{split}
\lambda_i^{E} & \leftarrow \lambda_i^{E} + \nu_k \, H_i(\vec{x}_k) \\
\lambda_j^{I} & \leftarrow \max{\left( 0,\lambda_j^{I} + \nu_k \, G_j(\vec{x}_k) \right)} \, , \\
\end{split}
\label{Eq:AugLag3}
\end{equation}
ensuring that the penalty factor is progressively increased, i.e., $\nu_{k+1} \geq \nu_k$. This systematic augmentation of the Lagrangian underscores a commitment to refining constraint handling, thereby enhancing the CE method's applicability to complex optimization scenarios. 

\subsection{Optimization through rare event estimation}

The core concept behind the CE method is to link the non-convex optimization problem defined by (\ref{Eq:OptDeterm}) with a rare event estimation problem. By employing an optimized sampling strategy for the latter, it is possible to solve the former through stochastic calculations.

This stochastic reinterpretation of the optimization problem occurs within a probabilistic space denoted as $(\SS, \SF, \PM)$, comprising a sample space $\SS$, a $\sigma$-field $\SF$, and a probability measure $\PM$. Within this  framework, any random vector $\randvec{X} = (X_1, \cdots, X_n)$ is characterized by a distribution $\cdf{d\vec{x}}$ in $\R^n$, with a density $\vec{x} \mapsto \pdf{f}{\vec{x}}{\cdot}$ with respect to $d\vec{x}$. We denote the expected value of $\randvec{X}$ concerning the density $\pdf{f}{\vec{x}}{\cdot}$ as $\expval{\randvec{X}}{f}$, and its variance as $\var{\randvec{X}}{f}$.

To introduce randomness into the problem (\ref{Eq:OptDeterm}), we transform the vector of design variables, $\vec{x}$, into a random vector, $\randvec{X}$, governed by a probability distribution characterized by the density $\pdf{f}{\vec{x}}{\vec{u}}$. Here, $\vec{u}$ represents a vector with carries the probability distribution hyperparameters. For instance, if the mean vector $\vec{\mu} \in \R^n$ and the standard deviation vector $\vec{\sigma} \in \R^n$ are the hyperparameters of $\pdf{f}{\vec{x}}{\vec{u}}$, then $\vec{u} = (\vec{\mu},\vec{\sigma})$.

Denoting the global minimum of problem (\ref{Eq:OptDeterm}) as $\gamma^{\star} = F(\vec{x}^{\star})$, we can note that the random event $F(\vec{X}) \leq \gamma^{\star}$ is impossible since the objective function cannot yield values lower than $\gamma^{\star}$. Consequently, $\PM \left\lbrace F(\vec{X}) \leq \gamma^{\star} \right\rbrace = 0$. If we relax the reference value to a scalar $\gamma > \gamma^{\star}$, $F(\vec{X}) \leq \gamma$ typically signifies a rare event when $\gamma$ is chosen to be close to $\gamma^{\star}$, i.e.,
\begin{equation}
\PM \left\lbrace F(\vec{X}) \leq \gamma \right\rbrace \approx 0 ~~ \mbox{if} ~~ \gamma \approx \gamma^{\star} \, .
\label{Eq:ProbRareEvent}
\end{equation}

Therefore, the rare event $F(\vec{X}) \leq \gamma$ occurring for $\gamma \approx \gamma^{\star}$ is intrinsically linked to the optimization problem (\ref{Eq:OptDeterm}). It involves realizations (samples) of $\vec{X}$ that either directly solve the optimization problem or are in close proximity to the optimal solution. Consequently, by estimating this rare probability through stochastic simulation, we theoretically obtain optimal or near-optimal values (high-quality solutions) for the optimization problem.

Utilizing the indicator function
\begin{equation}
	\indfunc{\mathcal{A}}(\vec{x}) =  
	\begin{cases} 
		1, ~ \mbox{if} ~ \vec{x} \in \mathcal{A} \\ 
		0, ~ \mbox{if} ~ \vec{x} \notin \mathcal{A} \, , \\ 
		\end{cases}
\label{Eq:IndFunc}
\end{equation}
we can express the probability in (\ref{Eq:ProbRareEvent}) as
\begin{equation}
\PM \left\lbrace F(\vec{X}) \leq \gamma \right\rbrace =
\expval{\indfunc{F(\vec{X}) \leq \gamma}}{f} \, ,
\label{Eq:ProbExpValInd}
\end{equation}
and a practical and unbiased method to estimate (\ref{Eq:ProbExpValInd}) is given by the sample mean estimator
\begin{equation}
\expval{\indfunc{F(\vec{X}) \leq \gamma}}{f} \approx
\frac{1}{N_s} \sum_{k=0}^{N_s} \indfunc{F(\randvec{X}_k) \leq \gamma} \, ,
\label{Eq:SampleMean}
\end{equation}
where $\randvec{X}_k$ ($k=1, \cdots, N_s$) represents i.i.d. statistical realizations of $\randvec{X}$ drawn from $\pdf{f}{\vec{x}}{\vec{u}}$.

However, it is essential to note that using the sample mean (\ref{Eq:SampleMean}) as an estimator for a rare event like (\ref{Eq:ProbRareEvent}) can be problematic, as it can be computationally expensive and often impractical in real-world applications. For instance, consider a scenario where one million of samples (which is common in rare events) are necessary, and each evaluation of $F(\randvec{X})$ takes approximately 1 second. The total computational time needed for the estimation would be on the order of $10^6$ seconds, or roughly 11.6 days. Even with parallelization strategies to distribute the workload across multiple processors or machines, completing such a computation in a reasonable amount of time can still be challenging for very small probabilities. 

In contrast, adaptive importance sampling offers an alternative approach that can significantly reduce the computational burden by focusing computational effort on the most promising regions of the sample space. This approach is presented in the next section.

\subsection{Rare event estimation using adaptive importance sampling}

Rare event estimation using adaptive importance sampling (AIS) is a cornerstone in rare event simulation methodologies, particularly in the development of the Cross-Entropy method. The primary goal of AIS is to efficiently estimate the probability of rare events by designing a suitable importance sampling (IS) density that focuses computational effort on the regions of interest in the sample space. By allocating the majority of its mass near the rare event region, AIS optimizes the sampling process for improved accuracy and efficiency.

In this sense, consider then a generalization of the rare event probability (\ref{Eq:ProbExpValInd}) given by
\begin{equation}
\ell = \expval{S(\vec{X})}{f} = \int_{\R^n} S(\vec{x}) \, \pdf{f}{\vec{x}}{\cdot} \, d\vec{x} \, ,
\label{Eq:RareEventGen}
\end{equation}
where $S(\vec{X}) \geq 0$ denotes a certain performance function that depends on the random vector $\vec{X}$. For example, in the problem (\ref{Eq:ProbExpValInd}) we have $S(\vec{X}) = \indfunc{F(\vec{X}) \leq \gamma}$.

Aiming to reduce the computational cost of the estimation process of (\ref{Eq:RareEventGen}), we will consider a reformulation of the problem in the form
\begin{equation}
\ell = \int_{\R^n} S(\vec{x}) \, \frac{\pdf{f}{\vec{x}}{\cdot}}{\pdf{g}{\vec{x}}{\cdot}} \, \pdf{g}{\vec{x}}{\cdot} \, d\vec{x} \, ,
\label{Eq:IS1}
\end{equation}
where the IS density $\pdf{g}{\vec{x}}{\cdot}$ is such that $\pdf{g}{\vec{x}}{\cdot} = 0 \Rightarrow S(\vec{x}) \pdf{f}{\vec{x}}{\cdot} = 0$, i.e., $g$ dominates $S f$. Thus, the estimation problem is alternatively seen as
\begin{equation}
\ell = \expval{ S(\vec{x}) \, \frac{\pdf{f}{\vec{x}}{\cdot}}{\pdf{g}{\vec{x}}{\cdot}} }{g} \, ,
\label{Eq:IS2}
\end{equation}
where now the sampling is performed with respect to the IS distribution $g$, which ideally is better suited for the rare event estimation.  In this novel setting, the estimation can be done with the aid of the estimator
\begin{equation}
\hat{\ell} = \frac{1}{N_s}  \, \sum_{k=1}^{N_s} H(\vec{X}_k) \, \frac{\pdf{f}{\vec{X}_k}{\cdot}}{\pdf{g}{\vec{X}_k}{\cdot}}  \, ,
\label{Eq:IS3}
\end{equation}
where the i.i.d. samples $\randvec{X}_k$ ($k=1, \cdots, N_s$) now are obtained according to $\pdf{g}{\vec{x}}{\cdot}$.

The IS density $g$ directly affects the efficiency of $\hat{\ell}$, so a key challenge here, crucial for achieving accurate estimates of rare event probabilities, is to find the density $g$ which minimizes the variance of the estimator $\hat{\ell}$, i.e.,
\begin{equation}
\pdf{g^{\star}}{\vec{x}}{\cdot} = \arg\min_{g}{ \, \var{\hat{\ell}}{g} } \, ,
\label{Eq:IS4}
\end{equation}
which due to the i.i.d. nature of the samples is equivalent to 
\begin{equation}
\pdf{g^{\star}}{\vec{x}}{\cdot} = \arg\min_{g}{ \, \var{S(\vec{X}) \, \frac{\pdf{f}{\vec{X}}{\cdot}}{\pdf{g}{\vec{X}}{\cdot}}}{g} } \, ,
\label{Eq:IS5}
\end{equation}
a nonlinear program which has as analytic solution
\begin{equation}
\pdf{g^{\star}}{\vec{x}}{\cdot} = \frac{S(\vec{x}) \, \pdf{f}{\vec{x}}{\cdot} }{ \displaystyle \int_{\R^n} S(\vec{x}) \, \pdf{f}{\vec{x}}{\cdot} \, d\vec{x}  } \, ,
\label{Eq:IS6}
\end{equation}
since for $g = g^{\star}$ we have
\begin{equation}
\var{S(\vec{X}) \, \frac{\pdf{f}{\vec{X}}{\cdot}}{\pdf{g}{\vec{X}}{\cdot}}}{g^{\star}} = \var{ \int_{\R^n} S(\vec{x}) \, \pdf{f}{\vec{x}}{\cdot} \, d\vec{x}  }{g^{\star}} = \var{\ell}{g^{\star}} = 0 \, .
\label{Eq:IS7}
\end{equation}

However, implementing this minimization strategy is not feasible in practical problems, because the optimal IS density depends on $\ell$, which is not be known a priori. In fact, the stochastic simulation is done to get this value!

To address this challenge, an alternative approach is to restrict the search space for the IS density to a parametric family that shares similarities with the original distribution of $\vec{X}$, i.e., if $\vec{X} \sim \pdf{f}{\vec{x}}{\vec{u}}$ for a hyperparameter vector $\vec{u}$ we choose $\pdf{g}{\vec{x}}{\cdot} = \pdf{f}{\vec{x}}{\vec{v}}$ for a reference hyperparameter vector $\vec{v}$. This approach ensures that the IS density remains within a feasible and computationally tractable domain.

Thus, the IS estimator (\ref{Eq:IS3}) becomes
\begin{equation}
\hat{\ell} = \frac{1}{N_s}  \, \sum_{k=1}^{N_s} S(\vec{X}_k) \, \frac{\pdf{f}{\vec{X}_k}{\vec{u}}}{\pdf{f}{\vec{X}_k}{\vec{v}}}  \, ,
\label{Eq:IS8}
\end{equation}
with i.i.d. samples $\randvec{X}_k$ ($k=1, \cdots, N_s$) drawn from $\pdf{f}{\vec{x}}{\vec{v}}$, so that we replace the variance minimization problem in (\ref{Eq:IS5}) for the nonlinear program
\begin{equation}
\begin{split}
\vec{v}^{\star} & = \arg\min_{\vec{v}}{ \, \var{S(\vec{X}) \, \frac{\pdf{f}{\vec{X}}{\vec{u}}}{\pdf{f}{\vec{X}}{\vec{v}}}}{\vec{v}} } \\
					  & = \arg\min_{\vec{v}}{ \,  \expval{ S(\vec{X})^2 \, \frac{\pdf{f}{\vec{X}}{\vec{u}}^2}{\pdf{f}{\vec{X}}{\vec{v}}^2} }{\vec{v}}  - \expval{S(\vec{X}) \, \frac{\pdf{f}{\vec{X}}{\vec{u}}}{\pdf{f}{\vec{X}}{\vec{v}}}}{\vec{v}}^2 } \\
					  & = \arg\min_{\vec{v}}{ \,  \int_{\R^n} S(\vec{x})^2 \, \frac{\pdf{f}{\vec{x}}{\vec{u}}^2}{\pdf{f}{\vec{x}}{\vec{v}}^2} \, \pdf{f}{\vec{x}}{\vec{v}} \, d\vec{x}   - \left(\int_{\R^n} S(\vec{x}) \, \frac{\pdf{f}{\vec{x}}{\vec{u}}}{\pdf{f}{\vec{x}}{\vec{v}}} \, \pdf{f}{\vec{x}}{\vec{v}} \, d\vec{x} \right)^2 } \\
					  & = \arg\min_{\vec{v}}{ \,  \int_{\R^n} S(\vec{x})^2 \, \frac{\pdf{f}{\vec{x}}{\vec{u}}}{\pdf{f}{\vec{x}}{\vec{v}}} \, \pdf{f}{\vec{x}}{\vec{u}} \, d\vec{x}   - \left(\int_{\R^n} S(\vec{x}) \, \pdf{f}{\vec{x}}{\vec{u}} \, d\vec{x} \right)^2 } \\
					  & = \arg\min_{\vec{v}}{ \,  \int_{\R^n} S(\vec{x})^2 \, \frac{\pdf{f}{\vec{x}}{\vec{u}}}{\pdf{f}{\vec{x}}{\vec{v}}} \, \pdf{f}{\vec{x}}{\vec{u}} \, d\vec{x}} \\
                     &= \arg\min_{\vec{v}}{ \, \expval{S(\vec{X})^2 \, \frac{\pdf{f}{\vec{X}}{\vec{u}}}{\pdf{f}{\vec{X}}{\vec{v}}}}{\vec{u}} } \, ,
\end{split}
\label{Eq:IS9}
\end{equation}
which is numerically solved with aid of the stochastic counterpart program
\begin{equation}
\hat{\vec{v}}^{\star} = \arg\min_{\vec{v}}{ \, \frac{1}{N_s}  \, \sum_{k=1}^{N_s} S(\vec{X}_k)^2 \, \frac{\pdf{f}{\vec{X}_k}{\vec{u}}}{\pdf{f}{\vec{X}_k}{\vec{v}}}} \, ,
\label{Eq:IS10}
\end{equation}
with i.i.d. samples $\randvec{X}_k$ ($k=1, \cdots, N_s$) drawn from $\pdf{f}{\vec{x}}{\vec{u}}$, that is known in advance.

By confining the search space to a known parametric family, this approach streamlines the practical implementation of the estimation problem (\ref{Eq:IS1}). Consequently, it also facilitates the resolution of the original optimization problem (\ref{Eq:OptDeterm}). However, this method necessitates solving a nonlinear program numerically, introducing a level of complexity that can be mitigated by employing a strategy based on Kullback–Leibler divergence minimization, as elucidated in the subsequent section.

%Considere então uma generalização do evento raro (\ref{Eq:ProbExpValInd}) dada por
%\begin{equation}
%\ell = \expval{H(\vec{X})}{f} = \int_{\R^n} H(\vec{x}) \, \pdf{f}{\vec{x}}{\cdot} \, d\vec{x} \, ,
%\label{Eq:RareEventGen}
%\end{equation}
%onde $H(\vec{X})$ denota uma certa função de performance que depende do vetor aleatório $\vec{X}$. Por exemplo, no problema (\ref{Eq:ProbExpValInd}) temos $H(\vec{X}) = \indfunc{F(\vec{X}) \leq \gamma}$.
%
%Visando reduzir o custo computacional do processo de estimação de (\ref{Eq:RareEventGen}) vamos considerar uma reformulação do problema da forma 
%\begin{equation}
%\ell =  \int_{\R^n} H(\vec{x}) \, \frac{\pdf{f}{\vec{x}}{\cdot}}{\pdf{g}{\vec{x}}{\cdot}} \, \pdf{g}{\vec{x}}{\cdot} \, d\vec{x} \, ,
%\label{Eq:IS1}
%\end{equation}
%onde a densidade $\pdf{g}{\vec{x}}{\cdot}$ é tal que $\pdf{g}{\vec{x}}{\cdot} = 0 \Rightarrow H(\vec{x}) \pdf{f}{\vec{x}}{\cdot} = 0$, i.e, $g$ domina $H f$. Desse modo o problema de estimação passa a ser visto como
%\begin{equation}
%\ell = \expval{ H(\vec{x}) \, \frac{\pdf{f}{\vec{x}}{\cdot}}{\pdf{g}{\vec{x}}{\cdot}} \, \pdf{g}{\vec{x}}{\cdot} }{g} \, ,
%\label{Eq:IS2}
%\end{equation}
%onde agora a amostragem é realizada com relação à distribuição de amostrangem por importância $g$.

\subsection{Estimating rare events via Kullback–Leibler divergence minimization}

The Kullback-Leibler (KL) divergence, often referred to as relative entropy or the cross-entropy, serves as a measure of statistical distance between two probability distributions. Specifically, it quantifies how one distribution $g$ diverges from another distribution $f$.  Mathematically, the KL divergence between $g$ and $f$ is defined as
\begin{equation}
\KL{g}{f} = \int_{\R^n} \pdf{g}{\vec{x}}{\cdot} \, \ln{\frac{\pdf{g}{\vec{x}}{\cdot}}{\pdf{f}{\vec{x}}{\cdot}}} \, d\vec{x} \, ,
\label{Eq:KL1}
\end{equation}
which is equivalent to
\begin{equation}
\KL{g}{f} = \int_{\R^n} \pdf{g}{\vec{x}}{\cdot} \, \ln{\pdf{g}{\vec{x}}{\cdot}} \, d\vec{x} - \int_{\R^n} \pdf{g}{\vec{x}}{\cdot} \, \ln{\pdf{f}{\vec{x}}{\cdot}} \, d\vec{x}  \, .
\label{Eq:KL2}
\end{equation}

It is crucial to recognize that while KL divergence is commonly referred to as a ``distance'', it does not fully adhere to the criteria of a true distance function in the metric space sense. Notably, it lacks symmetry, meaning $\KL{g}{f} \neq \KL{f}{g}$, and it does not satisfy the triangle inequality, indicated by $\KL{g}{f} > \KL{g}{z} + \KL{z}{f}$. However, KL divergence is always non-negative, represented as $\KL{g}{f} \geq 0$, with equality occurring only when $g$ is identical to $f$, i.e., $\KL{g}{f} = 0 \Leftrightarrow g = f$. In this context, KL divergence remains a valuable tool for quantifying the difference between two probability distributions.

Here we aim to utilize the KL divergence, denoted as $\KL{g^{\star}}{f}$, to measure the divergence between the optimal IS distribution $\pdf{g^{\star}}{\vec{x}}{\vec{u}} \propto S(\vec{x}) , \pdf{f}{\vec{x}}{\vec{u}}$, defined in (\ref{Eq:IS4}), and the actual IS distribution $\pdf{f}{\vec{x}}{\vec{v}}$ underlying the stochastic program (\ref{Eq:IS10}). The objective is to adaptively adjust the hyperparameter vector $\vec{v}$ to bring $\pdf{f}{\vec{x}}{\vec{v}}$ closer and closer to $\pdf{g^{\star}}{\vec{x}}{\vec{u}}$.

This strategy entails solving the optimization problem expressed as
\begin{equation}
\begin{split}
\vec{v}^{\star} & = \arg\min_{\vec{v}}{ \, \KL{\pdf{g^{\star}}{\cdot}{\vec{u}}}{\pdf{f}{\cdot}{\vec{v}}}  } \\
                       & = \arg\min_{\vec{v}}{ \, \int_{\R^n} \pdf{g^{\star}}{\vec{x}}{\vec{u}} \, \ln{\pdf{g^{\star}}{\vec{x}}{\vec{u}}} \, d\vec{x} - \int_{\R^n} \pdf{g^{\star}}{\vec{x}}{\vec{u}} \, \ln{\pdf{f}{\vec{x}}{\vec{v}}} \, d\vec{x}  } \\
                       & = \arg\min_{\vec{v}}{ \, \int_{\R^n} S(\vec{x}) \, \pdf{f}{\vec{x}}{\vec{u}} \, \ln{\left(S(\vec{x}) \, \pdf{f}{\vec{x}}{\vec{u}} \right)} \, d\vec{x} - \int_{\R^n} S(\vec{x}) \, \pdf{f}{\vec{x}}{\vec{u}} \, \ln{\pdf{f}{\vec{x}}{\vec{v}}} \, d\vec{x}  } \\
                       & = \arg\min_{\vec{v}}{ \, - \int_{\R^n} S(\vec{x}) \, \pdf{f}{\vec{x}}{\vec{u}} \, \ln{\pdf{f}{\vec{x}}{\vec{v}}} \, d\vec{x}  } \\
                       & = \arg\max_{\vec{v}}{ \, \int_{\R^n} S(\vec{x}) \, \ln{\pdf{f}{\vec{x}}{\vec{v}}} \, \pdf{f}{\vec{x}}{\vec{u}}  \, d\vec{x}  } \\
                       & = \arg\max_{\vec{v}}{ \, \expval{S(\vec{X}) \, \ln{\pdf{f}{\vec{X}}{\vec{v}}}}{\vec{u}}  } \, .
\end{split}
\label{Eq:KL3}
\end{equation}

This can be accomplished using the stochastic program
\begin{equation}
\hat{\vec{v}}^{\star} = \arg\max_{\vec{v}}{ \, \frac{1}{N_s}  \, \sum_{k=1}^{N_s} S(\vec{X}_k) \, \, \ln{\pdf{f}{\vec{X}_k}{\vec{v}}} } \, ,
\label{Eq:KL4}
\end{equation}
where $\randvec{X}_k$ ($k=1, \cdots, N_s$) represents i.i.d. realizations obtained according to $\pdf{f}{\vec{x}}{\vec{u}}$.

For the optimization problem (\ref{Eq:OptDeterm}), we are interested in the indicator function $S(\vec{X}) = \indfunc{F(\vec{X}) \leq \gamma}$. In this case, (\ref{Eq:KL4}) simplifies to
\begin{equation}
\begin{split}
	\hat{\vec{v}}^{\star} & = 
	\arg\max_{\vec{v}}{ \, \frac{1}{N_s}  \, \sum_{k=1}^{N_s} \indfunc{F(\vec{X}_k) \leq \gamma} \, \ln{\pdf{f}{\vec{X}_k}{\vec{v}}} } \\
	& =  \arg\max_{\vec{v}}{ \,\sum_{\vec{X}_k \in \mathcal{E}} \ln{\pdf{f}{\vec{X}_k}{\vec{v}}} } \, ,
\end{split}
\label{Eq:KL5}
\end{equation}
where the elite samples set $\mathcal{E}$ comprises the samples $\vec{X}_k$ for which $F(\vec{X}_k) \leq \gamma$. It is evident that the solution to this problem is the classic maximum likelihood estimator (MLE).

This approach offers a significant advantage over the method presented in the previous section (\ref{Eq:IS10}), as it often admits an analytic solution in typical scenarios. Specifically, this is the case when the random vector $\vec{X}$ comprises independent random variables with continuous distributions in the natural exponential family or discrete distributions with finite support.

For instance, consider a scenario where each component of $\vec{X}$  is independent and follows a multivariate truncated Gaussian distribution, i.e., $\vec{X} \sim \mathcal{TN}(\vec{\mu},\vec{\sigma},\vec{x}_{\ell},\vec{x}_u)$. Here, $\vec{\mu}$ and $\vec{\sigma}$ represent the mean and standard deviation, respectively, while $\vec{x}_{\ell}$ and $\vec{x}_u$ define the lower and upper bounds. In this case, the MLE for the mean $\vec{\mu}$ is given by
\begin{equation}
	\estim{\vec{\mu}} = \displaystyle \frac{1}{N_{\mathcal{E}}} \, \sum_{\vec{X}_k \in \mathcal{E}} \vec{X}_k \, ,
\label{Eq:KL6}
\end{equation}
while the MLE for the standard deviation $\vec{\sigma}$ is computed as
\begin{equation}
\estim{\vec{\sigma}} = \displaystyle  \sqrt{diag\left(\frac{1}{N_{\mathcal{E}}} \, \sum_{\vec{X}_k \in \mathcal{E}} \left(\vec{X}_k- \estim{\vec{\mu}} \right) \left(\vec{X}_k- \estim{\vec{\mu}} \right)^T \right) } \, ,
\label{Eq:KL7}
\end{equation}
where $N_{\mathcal{E}}$ denotes the number of elements in $\mathcal{E}$.

\subsection{The classic CE algorithm}

In this section, we present an overview of the CE algorithm for optimization, emphasizing its connection to the theory of rare event estimation through KL divergence minimization. For sake of simplicity and efficiency, the multivariate truncated Gaussian distribution is used for all the random variables in $\randvec{X}$, assumed as mutually independent.

A natural approach to tackling the optimization problem (\ref{Eq:OptDeterm}) from this perspective is to solve the stochastic program (\ref{Eq:KL5}) to obtain $\hat{\vec{v}}^{\star}$. Subsequently, the distribution $\pdf{f}{\vec{x}}{\hat{\vec{v}}^{\star}}$ can be used to generate realizations that provide approximate solutions to the optimization problem. The mean of these samples, defined in (\ref{Eq:KL6}), can serve as an estimator for the global optimum $\vec{x}^{\star}$.

However, it is crucial to highlight that if the associated rare probability $F(\vec{X}_k) \leq \gamma$ is exceedingly small, the distribution $\pdf{f}{\vec{x}}{\hat{\vec{v}}^{\star}}$ may not efficiently produce realizations that approximate $\vec{x}^{\star}$. This occurs because many of the samples $\vec{X}_k$ in $\indfunc{F(\vec{X}_k) \leq \gamma}$ will yield zero, leading to a mean value $\estim{\vec{\mu}}$ far from $\vec{x}^{\star}$. One potential solution is to employ an iterative multi-level process, which instead of using just one distribution to estimate the optimum, employs a sequence of distributions $\pdf{f}{\vec{x}}{\hat{\vec{v}}_0}, \pdf{f}{\vec{x}}{\hat{\vec{v}}_1}, \cdots , \pdf{f}{\vec{x}}{\hat{\vec{v}}_t}, \cdots$ that are sequentially calculated, producing a sequence of approximations $\estim{\vec{\mu}}_0, \estim{\vec{\mu}}_1, \cdots, \estim{\vec{\mu}}_t, \cdots$ for the optimum $\vec{v}^{\star}$. Each step in this process involves a pair of estimators $(\hat{\gamma}_t, \hat{\vec{v}}_t)$ for the pair $(\gamma^{\star}, \vec{v}^{\star})$.

Ideally, these estimators converge, $(\hat{\gamma}_t, \hat{\vec{v}}_t) \, \rightarrow \, (\gamma^{\star}, \vec{v}^{\star})$, in a way that $\pdf{f}{\vec{x}}{\hat{\vec{v}}_t} \, \rightarrow \, \delta \left(\vec{x} - \vec{x}^{\star} \right)$, where $\delta \left(\vec{x} - \vec{x}^{\star} \right)$ denotes Dirac's delta distribution, with all its mass centered at $\vec{x}^{\star}$. In practice, the iterative process is terminated once a convergence criterion is met. In such scenarios, $\pdf{f}{\vec{x}}{\hat{\vec{v}}_t}$ progressively moves toward $\vec{x}^{\star}$ while also shrinking.

This estimation process is multi-level because both $\gamma_t$ and $\vec{v}_t$ are sequentially updated. To achieve this, a rarity parameter $0 < \rho < 1$ is introduced, and $\gamma_t$ is defined as the $(1-\rho)$-quantile for $F(\vec{X})$ with $\randvec{X} \sim \pdf{f}{\vec{x}}{\vec{v}_{t-1}}$. Formally, $\gamma_t$ is characterized by
\begin{equation}
\PM \left\lbrace F(\vec{X}) \leq \gamma \right\rbrace \geq \rho
~\mbox{and}~
\PM \left\lbrace F(\vec{X}) > \gamma \right\rbrace \geq 1- \rho \, ,
\label{Eq:CEAlg1}
\end{equation}
and it is estimated using the order statistic
\begin{equation}
	\estim{\gamma}_t = F_{(N_{\mathcal{E}})} \, ,
\label{Eq:CEAlg2}
\end{equation}
where $F_{(1)} \leq F_{(2)} \leq \cdots \leq F_{(N_s)}$ represents the ordered values of $F(\vec{x})$ calculated at the samples $\randvec{X}_k (k=1, \cdots, N_s)$ from $\pdf{f}{\vec{x}}{\vec{v}_{t-1}}$. In this estimator, $N_{\mathcal{E}} = \texttt{ceil}(\rho \, N_s)$ denotes the size of the elite set $\mathcal{E}_t = \lbrace \randvec{X}_k \, | \, F(\randvec{X}_k) \leq \gamma_t \rbrace$.

Subsequently, $\estim{\vec{v}}_t$ is obtained from
\begin{equation}
	\estim{\vec{v}}_t =
	\arg\max_{\vec{v}} \,\sum_{\vec{X}_k \in \mathcal{E}_t} \ln{\pdf{f}{\vec{X}_k}{\vec{v}}} \, .
	\label{Eq:CEAlg3}
\end{equation}
For a multivariate Gaussian distribution, $\estim{\vec{v}}_t = (\tilde{\vec{\mu}}_t, \tilde{\vec{\sigma}}_t)$ is computed analytically using the formulas (\ref{Eq:KL6}) and (\ref{Eq:KL7}).

Finally, these vectors are updated according to the smoothing schemes
\begin{equation}
\estim{\vec{\mu}}_t = \alpha \, \tilde{\vec{\mu}}_t + (1-\alpha) \, \estim{\vec{\mu}}_{t-1} \, ,
\end{equation}
\begin{equation}
\estim{\vec{\sigma}}_t = \beta_t \, \tilde{\vec{\sigma}}_t + (1-\beta_t) \, \estim{\vec{\sigma}}_{t-1} \, ,
\end{equation}
where $0 < \alpha \leq 1$ is the smoothing parameter and
\begin{equation}
\beta_t  =  \beta - \beta \, \left(1 - \frac{1}{t} \right)^q \, ,
\end{equation}
with $\beta \geq 0$ and $q \in \N$.

The convergence of the iterative process is controlled by the criterion
\begin{equation}
||\estim{\vec{\sigma}}_t - \estim{\vec{\sigma}}_{t-1}||_{wrms} \leq 1 \, ,
 \label{Eq:CEAlg4}
\end{equation}
where the weighted root-mean-square norm of $\vec{x} \in \R^n$ is computed as follows
\begin{equation}
||\vec{x}||_{wrms} = \sqrt{ \frac{1}{n} \, \sum_{j=1}^{n} \, \left ( w_j \, x_j \right )^{\, 2} } \, ,
 \label{Eq:CEAlg5}
\end{equation}
where $w_j$ represents the error weights defined by
\begin{equation}
w_j = \frac{1}{\texttt{atol}_j +  |x_j| \, \texttt{rtol}} \, .
\end{equation}
Here, $\texttt{atol}_j$ and $\texttt{rtol}$ represent the absolute and relative tolerances, respectively. The normalization provided by the weights in (\ref{Eq:CEAlg5}) ensures that a weighted norm of approximately 1 in (\ref{Eq:CEAlg4}) indicates convergence. This convergence criterion, commonly employed in high-quality differential equation solvers \cite{hindmarsh2005p410,Shampine1997p1}, ensures robust error control.

This algorithm is summarized as follows:
\begin{enumerate}
    \item \textbf{Initialization}: Begin by defining the necessary parameters: the number of total samples $N_{s}$, the number of elite samples $N_{\mathcal{E}}$ (which must be less than $N_{s}$), an absolute tolerance \texttt{atol}, a relative tolerance \texttt{rtol}, the maximum number of iteration levels $t_{max}$, the family of probability distributions $\pdf{f}{\cdot}{\vec{v}}$, an initial vector of hyperparameters $\estim{\vec{v}}_0$, the smoothing parameters $\alpha$, $\beta$, and $q$, and set the iteration level counter to zero, i.e., $t=0$.
    
    \item \textbf{Update Iteration Level}: Increment the iteration level counter by 1, i.e., $t = t+1$.
    
    \item \textbf{Sample Generation}: Generate $N_{s}$ independent and identically distributed (iid) samples from $\pdf{f}{\cdot}{\vec{v}_{t-1}}$, denoted by $\randvec{X}_1, \cdots, \randvec{X}_{N_s}$.
    
    \item \textbf{Objective Function Evaluation}: Evaluate the objective function $F(\vec{x})$ at each sample $\randvec{X}_1, \cdots, \randvec{X}_{N_s}$, sort the results as $F_{(1)} \leq \cdots \leq F_{(N_s)}$, and define the elite sample set $\mathcal{E}_t$ containing the $N_{\mathcal{E}}$ best points, i.e., those associated with the minimum values.
    
    \item \textbf{Update Estimators}: Update the estimators $\estim{\gamma}_{t}$ and $\estim{\vec{v}}_{t}$ using the elite sample set. Use order statistic estimator from (\ref{Eq:CEAlg2}) for $\estim{\gamma}_{t}$ and maximum likelihood  estimators from (\ref{Eq:CEAlg3}) for $\estim{\vec{v}}_{t}$. If necessary, apply a scheme for smooth updating.

\item \textbf{Iteration}: Repeat steps 2 through 5 of this algorithm until a stopping criterion is met. This can be when the maximum number of iterations $t_{\text{max}}$ is reached, or when the weighted root-mean-square norm of the difference vector $\estim{\vec{\sigma}}_t - \estim{\vec{\sigma}}_{t-1}$ falls below or equals 1, signaling convergence as per the defined criteria. For a comprehensive discussion on various stopping strategies, including those based on alternative convergence measures and practical considerations, refer to Section~\ref{code-functionality}.

\end{enumerate}

\begin{figure}[h]
	\centering
	\includegraphics[scale=1]{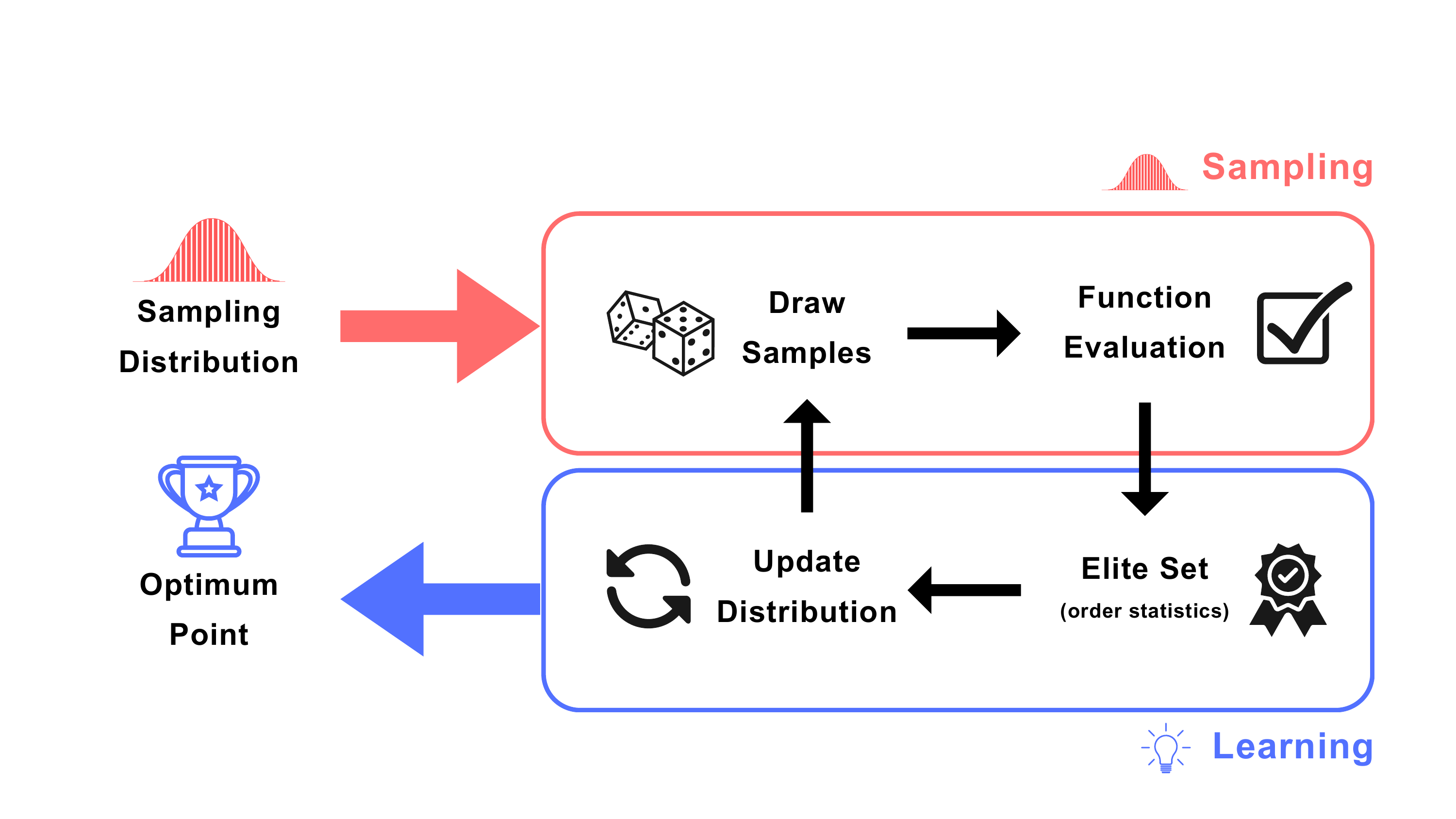}
	\caption{Cross-Entropy Method Flowchart: This diagram shows the CE method's steps, from starting with a sample distribution to finding the best solution. It involves sampling, function evaluation, refining the distribution with the best samples, and repeating these steps until the optimal solution is found.}
	\label{fig:CEalgorithm}
\end{figure}

A schematic representation of the above CE algorithm is presented in Figure~\ref{fig:CEalgorithm}.

\subsection{Remarks about CE method}

The CE algorithm possesses several noteworthy characteristics that contribute to its appeal and effectiveness in solving optimization problems. These features include:

\begin{itemize}
	\item \emph{Simplicity:} The CE algorithm is intuitively understandable and has only a few control parameters, namely $N_s$, $N_{\mathcal{E}}$, $t_{\text{max}}$, $\texttt{atol}$ and $\texttt{rtol}$. Each of these parameters has a clear interpretation, making the algorithm easy to implement and tune.

	\item \emph{Robustness:} Theoretical results establish that, under typical conditions, the CE algorithm is guaranteed to converge to the global optimum if the optimization problem has a single global extremum. This robustness ensures its reliability in a variety of optimization scenarios.

	\item \emph{Efficiency:} The CE algorithm often exhibits fast convergence rates compared to traditional global search metaheuristics such as genetic algorithms. This efficiency makes it particularly suitable for problems where computational resources are limited or where rapid optimization is desired.

	\item \emph{Generality:} Unlike some optimization methods that require specific properties of the objective function, the CE algorithm is applicable to a wide range of non-convex optimization problems. It can handle functions that are non-differentiable, discontinuous, or lack other regularity conditions.

	\item \emph{Extensibility:} The theoretical framework underlying the CE algorithm is general and can be applied to optimization problems of any finite dimension. While computational cost and the curse of dimensionality may limit its practical applicability in high-dimensional spaces, the method remains versatile and adaptable.

	\item \emph{Ease of Implementation:} Thanks to its simplicity and the availability of the MATLAB black-box package presented in this paper, implementing the CE algorithm for a particular optimization problem is straightforward. This package offers a robust and modern implementation of CE, making it easy to apply in various optimization tasks.
\end{itemize}

The mathematical foundation of the CE algorithm is well-established, with a rigorous development of its formalism over the past decades. Theoretical theorems have been formulated to rigorously establish the conditions under which the algorithm converges. While the detailed mathematical analysis is beyond the scope of this paper, interested readers can refer to \cite{Rubinstein2004, Rubinstein2016} for further exploration.

In conclusion, the CE algorithm's simplicity, robustness, efficiency, generality, extensibility, and ease of implementation make it a valuable tool for non-convex optimization problems. Its principled approach and clear control parameters make it a recommendable choice for a wide range of optimization tasks.

 % --------------------------------------------------------------

% --------------------------------------------------------------
\section{MATLAB implementation: The CEopt package}
\label{sec:CEcode}

The \pkg{CEopt} package is a MATLAB-based framework designed to solve non-convex optimization problems using the Cross-Entropy method. The development of this package is motivated by the need for an easy-to-use, efficient, modular and general purpose tool based on CE for both theoretical researchers and practitioners in the field of optimization. The \pkg{CEopt} package closely mirrors MATLAB's black box function \texttt{ga} (Genetic Algorithm), in its user-friendliness and functionality, making it a valuable tool for solving a wide range of optimization problems without requiring in-depth knowledge of the underlying algorithmic intricacies.

\subsection{Code architecture and modularity}

\begin{figure}[h]
	\centering
	\includegraphics[scale=1]{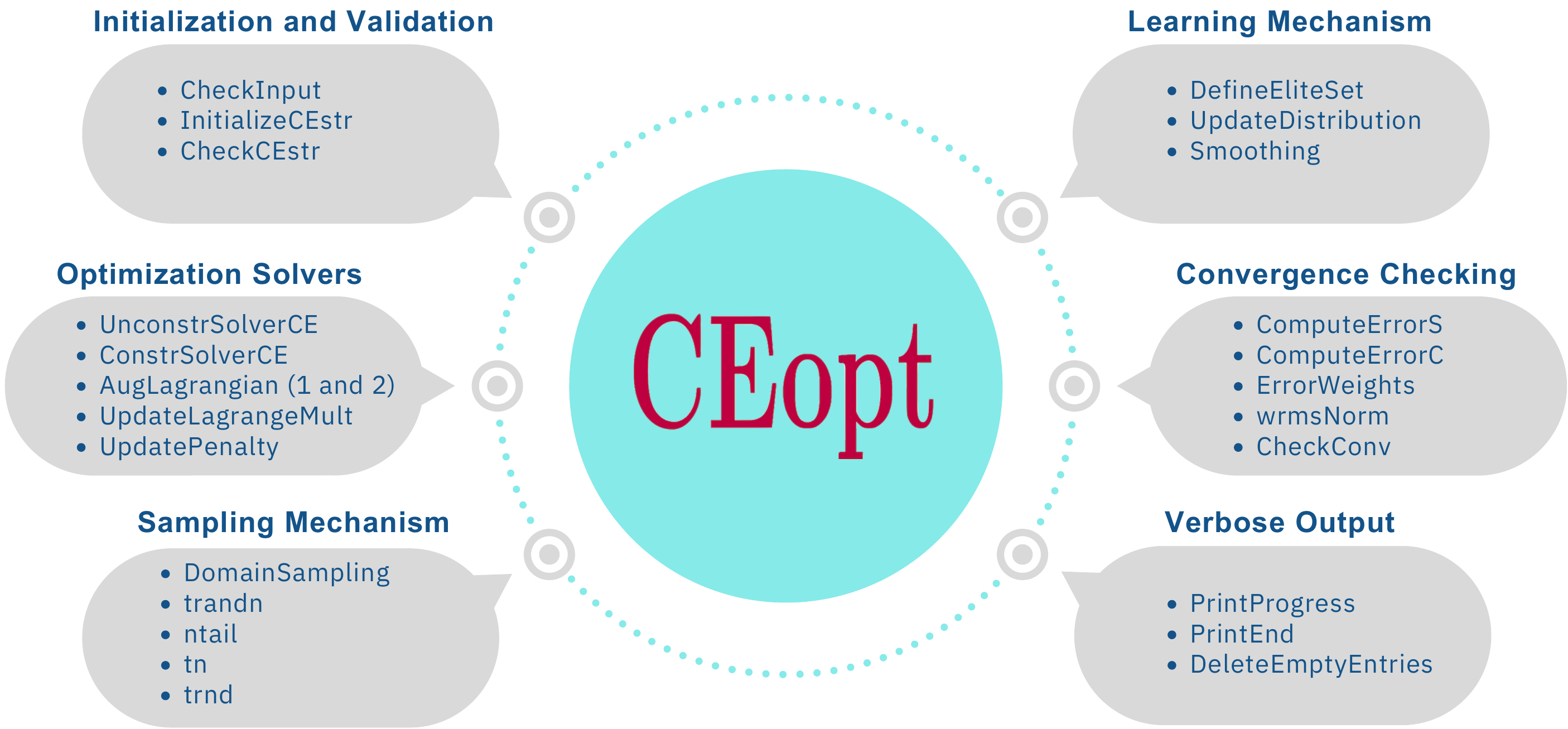}
	\caption{\pkg{CEopt} MATLAB Implementation: This diagram outlines the key components of the \pkg{CEopt} package, starting with input checks and setup, through to optimization solvers for various problem types. It highlights the sampling and learning processes used to refine solutions, along with methods for convergence verification and detailed progress reporting, illustrating the systematic approach embedded in \pkg{CEopt} for solving optimization tasks.}
	\label{fig:CEopt}
\end{figure}

The \pkg{CEopt} package is structured to maximize code modularity and readability, adopting a design that segregates the optimization process into distinct, logically organized functions, as can be seen in Figure~\ref{fig:CEopt}. This modular architecture not only enhances maintainability but also facilitates the understanding and potential customization of the code. The main function, \texttt{CEopt}, serves as the entry point for the optimization process, encapsulating the core CE method logic and iteratively refining the solution until convergence criteria are met or computational limits are reached.

Key components of the package include:
\begin{itemize}
	\item \emph{Initialization and Validation:} Sets up the initial parameters and structures and checks user inputs for consistency and correctness, ensuring robust and error-free execution.
\item \emph{Optimization Solvers:} Offers both constrained and unconstrained solvers, providing versatile options for tackling a wide range of optimization problems.
	\item \emph{Sampling Mechanism:} Generates samples from a truncated Gaussian distribution, focusing the search on promising regions of the solution space.
	\item \emph{Learning Mechanism:} Identifies and processes elite samples to update the distribution parameters, guiding the algorithm towards optimal solutions.
	\item \emph{Convergence Checking:} Monitors the optimization process, assessing convergence based on predefined criteria and computational bounds.
\item \emph{Verbose Output:} Delivers concise progress updates and diagnostic information, enhancing the optimization process transparency and ease of analysis.
\end{itemize}

\subsection{Ease of use and functionality}
\label{code-functionality}

Operating the \pkg{CEopt} package is straightforward: users need only to provide the necessary input parameters and invoke the \texttt{CEopt} function to commence the optimization process. The simplicity of use does not compromise the method's power or versatility, making it accessible to a wide audience of optimization practitioners across various domains of knowledge. The basic syntax for invoking \pkg{CEopt} with all inputs and outputs is given by:

\vspace{-2mm}
\begin{center}
\begin{minipage}{0.98\linewidth}
\begin{lstlisting}[
backgroundcolor=\color{gray!08},
style=Matlab-editor,
basicstyle=\ttfamily\footnotesize,
frame=single,
numbers=none]
[Xopt,Fopt,ExitFlag,CEstr] = CEopt(fun,xmean0,sigma0,lb,ub,nonlcon,CEstr)
\end{lstlisting}
\end{minipage}
\end{center}

Below is a detailed explanation of the inputs and outputs of the \pkg{CEopt} package:

\begin{itemize}
\item Input Parameters
\begin{itemize}
	\item \texttt{fun}: The objective function handle to be minimized or maximized;
	\item \texttt{xmean0}: Initial mean of the sampling distribution;
	\item \texttt{sigma0}: Initial standard deviation of the sampling distribution;
	\item \texttt{lb} and \texttt{ub}: Lower and upper bounds for the design variables, defining a rectangular feasible region to search the solution;
	\item \texttt{nonlcon}: Handle for nonlinear constraint functions (optional);
	\item \texttt{CEstr}: A struct containing various settings and parameters for customizing the CE~method's behavior and some features of \texttt{CEopt} implementation (optional).
\end{itemize}
\end{itemize}

\begin{itemize}
\item Output Parameters
\begin{itemize}
	\item \texttt{Xopt}: The optimal solution found by the algorithm;
	\item \texttt{Fopt}: The value of the objective function at the optimal solution;
	\item \texttt{ExitFlag}: An integer flag indicating the optimization process's termination reason;
	\item \texttt{CEstr}: Updated \texttt{CEstr} struct, also containing diagnostic information about the algorithm execution and the final state of the optimization process.
\end{itemize}
\end{itemize}

The \texttt{ExitFlag} serves as a crucial indicator of the outcome of the optimization process in the \pkg{CEopt} package. It provides concise feedback on why the optimization procedure terminated, allowing users to quickly assess the success of the optimization and understand any issues that may have arisen. Here is a detailed explanation of the possible values for \texttt{ExitFlag} and their implications:

\begin{itemize}
	\item \texttt{1 -- Maximum Number of Iterations Reached}: The optimization process stopped because it hit the maximum limit of iterations specified by the user through the \texttt{MaxIter} field in the \texttt{CEstr} structure. This may indicate that more iterations are needed to converge to a solution or that the optimization is stalled;

	\item \texttt{2 -- Solution Stalled}: The algorithm terminated because there was no significant change in the objective function value over a predefined number of iterations (\texttt{MaxStall}). This suggests that the algorithm might have converged to a global optimum solution or is trapped in a local minimum;

	\item \texttt{3 -- Maximum Number of Function Evaluations Reached}: This flag indicates that the optimization ended because the total number of objective function evaluations exceeded the user-defined maximum (\texttt{MaxFcount}). Like flag \texttt{1}, this might suggest the need for higher limits or a review of the optimization strategy;

	\item \texttt{4 -- Objective Function Tolerance Achieved}: The optimization successfully concluded because the range of objective function values across iterations fell below the specified tolerance (\texttt{TolFun}). For constrained problems, it also implies that constraint violations are small. This is typically a sign of successful convergence;

	\item \texttt{5 -- Standard Deviation Convergence}: The process terminated due to the small variation in the standard deviation of the sampling distribution (\texttt{sigma}), and, for constrained optimizations, constraint violations are also small. This indicates that the population of solutions is converging towards a point, suggesting a potential optimum solution has been found by the algorithm;

	\item \texttt{6 -- Minimum Function Value Criterion Met}: This outcome means the optimization found a solution where the objective function value is equal to or better than the target minimum function value set by the user (\texttt{MinFval}). It is a clear indication of success, especially in applications where a specific performance benchmark is targeted.
\end{itemize}

Understanding the meaning behind each \texttt{ExitFlag} value is essential for interpreting the results of the optimization process. It not only provides insights into the behavior and performance of the CE algorithm but also guides users in making decisions about potential next steps. Depending on the \texttt{ExitFlag} received, users may adjust algorithm parameters, extend computational limits, or refine their problem formulation seeking better results in next runs.

The \texttt{CEstr} structure is central to the customization and configuration of the optimization process in the \pkg{CEopt} package. It offers users a high degree of control over the behavior of the Cross-Entropy optimization algorithm, enabling the adjustment of various parameters to best fit the specific requirements of their optimization problem. Furthermore, it encapsulates a rich tapestry of historical insights into the optimization's evolutionary path, granting users the capacity to meticulously scrutinize the algorithm's trajectory, identify any emergent issues, and strategically refine parameters for optimized outcomes. Below is a detailed description of all possible fields within the \texttt{CEstr} structure:

\begin{itemize}
	\item \texttt{Verbose}: A boolean flag that, when set to \texttt{true}, enables the display of detailed information about the optimization process at each iteration on screen. This includes the current iteration number, objective function value, and other diagnostic information;

	\item \texttt{isConstrained}: A boolean flag indicating whether the optimization problem includes constraints. This influences the choice of optimization strategy within algorithm;
	
	\item \texttt{isVectorized}: A boolean flag indicating whether the objective function fun and the nonlinear constraints function nonlcon (if provided) are vectorized. Vectorized functions are capable of taking multiple input samples at once and returning their corresponding output values in a single call, which can significantly enhance computational efficiency;
	
	\item \texttt{Nvars}: Indicates the number of design variables in the optimization problem. This is set automatically based on the input parameters;

	\item \texttt{EliteFactor}: Determines the proportion of samples considered as elite. These are the best samples from the current iteration used to update the sampling distribution;

	\item \texttt{Nsamp}: The total number of samples to be drawn at each iteration. It affects the breadth of the search and can influence both the speed and the quality of convergence;

	\item \texttt{MaxIter}: Specifies the maximum number of iterations to be performed by the algorithm. This serves as a stopping criterion to prevent the algorithm from running indefinitely;

	\item \texttt{MaxStall}: Defines the maximum number of iterations without any improvement in the objective function value that the algorithm will tolerate before terminating. This helps in avoiding unnecessary computations when the iteration has converged to the global optimum or is stuck in a local minimum;

	\item \texttt{MaxFcount}: Sets an upper limit on the total number of objective function evaluations. This is another stopping criterion that ensures efficient use of computational resources;

	\item \texttt{MinFval}: Establishes a threshold for the objective function value. If the algorithm finds a solution with an objective function value below this threshold, it will terminate, assuming an acceptable solution has been found;

	\item \texttt{TolAbs}: An absolute tolerance level for the convergence criterion based on the standard deviation. It helps in determining when the changes in the solution vector between iterations are sufficiently small to consider the solution as converged;

	\item \texttt{TolRel}: A relative tolerance level for the convergence criterion based on the standard deviation, similar to \texttt{TolAbs}, but relative to the size of the solution vector. It provides a scale-invariant measure of convergence;

	\item \texttt{TolCon}: Specifies the tolerance for constraint violations. It allows for slight violations in the constraints, facilitating convergence in problems where strict adherence to constraints is challenging to achieve;

	\item \texttt{TolFun}: Indicates the tolerance for changes in the objective function value. If the change in function values between iterations is below this threshold, the process terminate;

	\item \texttt{alpha}: Smoothing parameter used for updating the mean of the sampling distribution. It controls the balance between exploration and exploitation in the search space;

	\item \texttt{beta}: Similar to \texttt{alpha}, but for the standard deviation of the sampling distribution. It adjusts the spread of the distribution, affecting the diversity of the samples generated;

	\item \texttt{q}: An exponent used in the dynamic update formula for \texttt{beta}. It allows to change the smoothing effect over time, potentially improving convergence properties;

	\item \texttt{NonlconAlgorithm}: Specifies the algorithm used for handling nonlinear constraints, such as the augmented Lagrangian method. It allows for the selection of the most appropriate method based on the nature of the constraints;

	\item \texttt{InitialPenalty}: Sets the initial penalty value for the augmented Lagrangian when handling constrained problems. It affects the initial weight given to constraint violations;

	\item \texttt{PenaltyFactor}: Determines the factor by which the penalty parameter is increased during the optimization process. This is relevant in the context of methods that adjust penalties dynamically to enforce constraint satisfaction;
	
	\item \texttt{xmean}: Stores the history of mean values of the sampling distribution over iterations. This array provides insight into how the central tendency of the sampled solutions evolves as the optimization progresses;

	\item \texttt{xmedian}: Captures the history of median values of the sampling distribution. Similar to \texttt{xmean}, but offers a different statistical perspective on the distribution's central tendency;

	\item \texttt{xbest}: Contains the best sample (solution) found until that iteration. This field is particularly useful for observing the progression towards the optimal solution across iterations;

	\item \texttt{Fmean}: Records the history of mean objective function values of the elite set at each iteration. It helps in assessing the overall performance improvement of the algorithm;

	\item \texttt{Fmedian}: Similar to \texttt{Fmean}, but tracks the median of the objective function values among the elite samples. This provides an alternative measure of central tendency, which is less sensitive to outliers;

	\item \texttt{Fbest}: Maintains the history of the best objective function value found up to each iteration. This field is crucial for understanding the optimization process's efficacy and convergence behavior;

	\item \texttt{sigma}: Keeps a record of the standard deviation of the sampling distribution over iterations. Variations in \texttt{sigma} reflect changes in the diversity of the sampled solutions and the algorithm's exploration-exploitation balance;

	\item \texttt{ErrorS}: Logs the history of the standard deviation error measure. This field is instrumental in monitoring the convergence of the sampling distribution's spread;

	\item \texttt{ErrorC}: Provides the history of the constraint violation error measure for constrained problems. It indicates how well the solutions satisfy the constraints over time;

	\item \texttt{iter}: Tracks the total number of iterations performed during the optimization process. Fundamental for managing loop execution and assessing computational effort;

	\item \texttt{stall}: Records the number of iterations without significant progress. A high stall count may indicate convergence to a suboptimal solution or necessitate parameter adjustment;

	\item \texttt{Fcount}: Counts the total number of objective function evaluations conducted. This metric is valuable for evaluating the computational cost of the optimization process;
	
	\item \texttt{ConvergenceStatus}: Reflects the outcome of the optimization process with regard to convergence. This boolean field is set to \texttt{true} if the algorithm successfully meets the convergence criteria, such as reaching a satisfactory solution within specified tolerances or achieving adequate constraint satisfaction in the case of constrained problems. Conversely, it is set to \texttt{false} if the optimization process terminates due to reaching computational limits (like the maximum number of iterations or function evaluations) without satisfying the convergence conditions.
\end{itemize}

This comprehensive list of fields within the \texttt{CEstr} structure demonstrates the flexibility and depth of customization available in the \pkg{CEopt} package. Users can fine-tune these parameters to adapt the optimization process to their specific needs, enhancing the package's utility across a broad spectrum of optimization problems. For additional information regarding code installation, licensing, and more, please visit the \pkg{CEopt} website at \url{https://ceopt.org}.

In the following section, we present several case studies that demonstrate the \pkg{CEopt} package's application across a diverse range of non-convex optimization problems. These examples are designed to showcase the versatility, effectiveness, and broad applicability of \pkg{CEopt} in addressing complex optimization challenges in various domains. Through these illustrative examples, readers will gain insights into how \pkg{CEopt} can be seamlessly integrated into practical scenarios, highlighting its value and utility in research and industry settings alike.

% --------------------------------------------------------------

% --------------------------------------------------------------
\section{Numerical experiments}
\label{sec:NumExamp}

This section of the paper is dedicated to showcasing the robust capabilities of the \pkg{CEopt} package through a series of detailed numerical experiments. Each example has been carefully selected to illustrate the package's ability to efficiently solve a variety of complex, non-convex optimization problems that are commonly encountered in both research and industrial applications. By demonstrating \pkg{CEopt} in action across different scenarios, we aim to highlight its adaptability, precision, and effectiveness. These case studies not only validate the theoretical aspects discussed in previous sections but also provide practical insights that can be directly applied in diverse fields such as machine learning, engineering design, and computational finance. The following examples will demonstrate the implementation specifics, the ease of use of \pkg{CEopt}, and its performance in achieving optimal solutions under challenging conditions.

\subsection{Example 1: A Gaussian mixture in 1D}

For pedagogical reasons, we begin with a straightforward example featuring a Gaussian mixture function. This case study is designed to demonstrate the ease of using \pkg{CEopt} as a black-box optimization command. The objective function considered here is a simple Gaussian mixture given by
\begin{align}
F(x) = - 0.8 \, \exp{\left(-(x-2)^2 \right)} - 0.5 \, \exp{\left(-(x+2)^2 \right)} + 1\, .
\end{align}

Despite its simplicity, this function has illustrative properties, allowing us to demonstrate the effectiveness of the CE method in navigating complex landscapes to locate minima, making it an ideal candidate to showcase the capabilities of \pkg{CEopt}.  The MATLAB code provided below automates the optimization process using \pkg{CEopt}, treating it as a black box that requires minimal user intervention for parameter setup.

% Example 1
\begin{center}
\texttt{MainCEoptExample1.m}\\
\begin{minipage}{0.8\linewidth}
\begin{lstlisting}[
backgroundcolor=\color{gray!08},
%numbers=left,
numberstyle=\tiny,
%numbersep=8pt,
style=Matlab-editor,
basicstyle=\ttfamily\footnotesize,
numbersep=10pt,
frame=single]
clc; clear; close all;

disp(' ------------------- ')
disp(' MainCEoptExample1.m ')
disp(' ------------------- ')

% objective function
F = @(x) -0.8*exp(-(x-2).^2) - 0.5*exp(-(x+2).^2)+1;

% bound for design variables
lb = -5;
ub =  5;

% initialize mean and std. dev. vectors
mu0    = (ub+lb)/2;
sigma0 = (ub-lb)/6;

% CE optimizer
tic
[Xopt,Fopt,ExitFlag,CEobj] = CEopt(F,mu0,sigma0,lb,ub)
toc
\end{lstlisting}
	\label{Code:Example1}
\end{minipage}
\end{center}

\begin{figure}[h]
	\centering
	\includegraphics[scale=1]{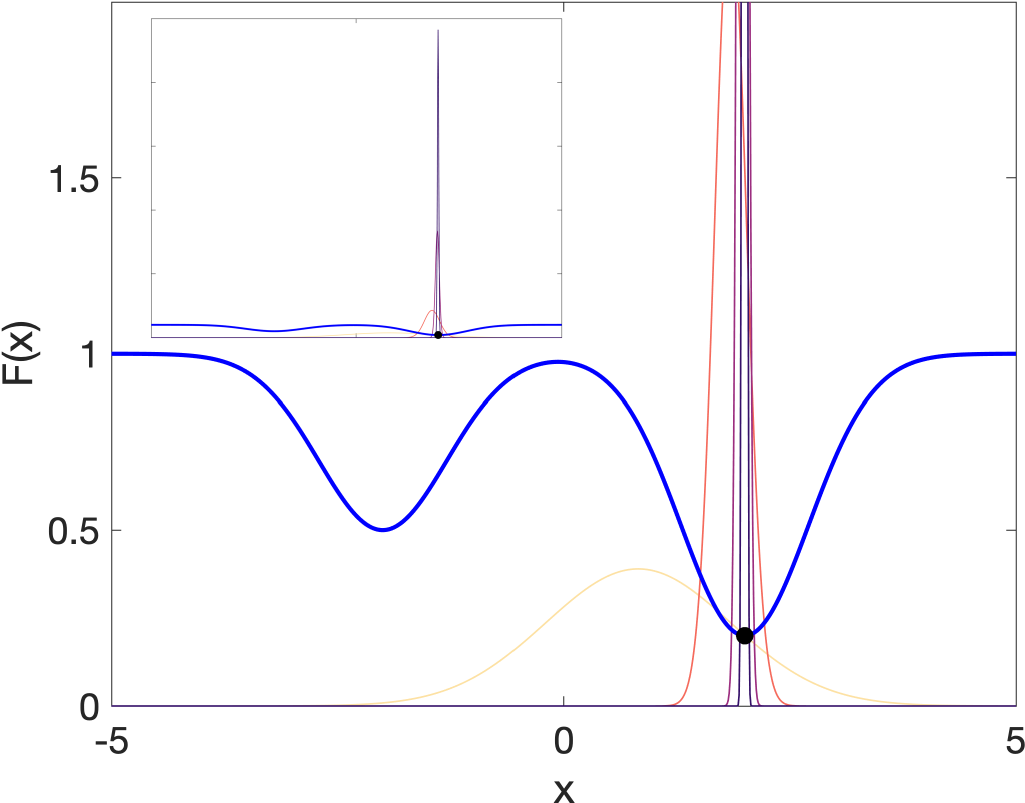}
	\caption{Adaptive Sampling in CE Optimization. The plot showcases the Cross-Entropy method optimizing a one-dimensional Gaussian mixture function (thick blue line). Gaussian distributions, from orange to dark near-black, visualize the adaptive sampling strategy. Early in the search, broader red distributions cover wider areas, while focused near-black distributions towards the end pinpoint the optimum, marked by the black dot. The inset magnifies the final stages, highlighting the precision of the narrowing search distributions.}
	\label{fig:Example1}
\end{figure}

Figure~\ref{fig:Example1} provides a detailed visualization of the optimization process. It shows not only the Gaussian mixture function but also the sequence of Gaussian distributions employed by the CE method. The adaptive nature of this sampling is clearly visible in the transition from broad, orange-colored distributions early in the search process to darker, nearly black distributions as the search converges towards the optimum. The focused distributions towards the end of the search effectively pinpoint the location of the minimum, marked by a black dot on the plot.

The inset within Figure~\ref{fig:Example1} further magnifies the final stages of this optimization process, illustrating the precision with which the CE method narrows down the search distributions. This detailed view highlights the method’s ability to finely tune the sampling strategy as it homes in on the optimum, showcasing the dynamic adaptation that is central to the CE method's approach.

\subsection{Example 2: The Peaks function in 2D}

Now we delve into the optimization of a intricate two-dimensional function characterized by its complex landscape of multiple local maxima and minima, the Peaks function:
\begin{align}
F(x_1, x_2) = & \; 3(1 - x_1)^2 \exp\left(-x_1^2 - (x_2 + 1)^2 \right) \nonumber \\
& - 10 \left(\frac{x_1}{5} - x_1^3 - x_2^5\right) \exp\left(-x_1^2 - x_2^2\right) - \frac{1}{3} \exp\left(-(x_1 + 1)^2 - x_2^2 \right) \, .
\end{align}
This function presents a challenge for many optimization algorithms due to its convoluted surface features, demanding robust strategies for navigation and convergence.

For the optimization process, the function $F(x_1,x_2)$ is implemented in MATLAB as an auxiliary function and transformed into a function handle, a standard procedure for managing complex objective functions with \pkg{CEopt}. The following MATLAB code demonstrates how users can customize the parameters of the CE method via an auxiliary structure. This customization enables users to adjust aspects such as iteration count, sample size, and elite sample proportion, adapting the optimization process to their specific requirements. Initial conditions for the mean and standard deviation are set randomly within a broad range to ensure comprehensive exploration of the potential landscape.

% Example 2
\begin{center}
\texttt{MainCEoptExample2.m}\
\begin{minipage}{0.85\linewidth}
\begin{lstlisting}[
backgroundcolor=\color{gray!08},
%numbers=left,
numberstyle=\tiny,
%numbersep=8pt,
style=Matlab-editor,
basicstyle=\ttfamily\footnotesize,
numbersep=10pt,
frame=single]
clc; clear; close all;

disp(' ------------------- ')
disp(' MainCEoptExample2.m ')
disp(' ------------------- ')

% objective function
F = @(x)PeaksFunc(x);

% bound for design variables
lb = [-3; -3];
ub = [ 3;  3];

% initialize mean and std. dev. vectors
mu0    = lb + (ub-lb).*rand(2,1);
sigma0 = 5*(ub-lb);
        
% define parameters for the CE optimizer
CEstr.isVectorized = 1;       % Vectorized function
CEstr.EliteFactor  = 0.1;     % Elite samples percentage
CEstr.Nsamp        = 50;      % Number of samples
CEstr.MaxIter      = 80;      % Maximum of iterations
CEstr.TolAbs       = 1.0e-2;  % Absolute tolerance
CEstr.TolRel       = 1.0e-2;  % Relative tolerance
CEstr.alpha        = 0.7;     % Smoothing parameter
CEstr.beta         = 0.8;     % Smoothing parameter
CEstr.q            = 10;      % Smoothing parameter

% CE optimizer
tic
[Xopt,Fopt,ExitFlag,CEstr] = CEopt(F,mu0,sigma0,lb,ub,[],CEstr)
toc

% objective function
function F = PeaksFunc(x)
    x1 = x(:,1);
    x2 = x(:,2);
     F = 3*(1-x1).^2.*exp(-x1.^2 - (x2+1).^2) ...
       - 10*(x1/5 - x1.^3 - x2.^5).*exp(-x1.^2 - x2.^2)...
       - (1/3)*exp(-(x1+1).^2 - x2.^2);
end
\end{lstlisting}
	\label{Code:Example2}
\end{minipage}
\end{center}

\begin{figure}[h]
\centering
\includegraphics[scale=1]{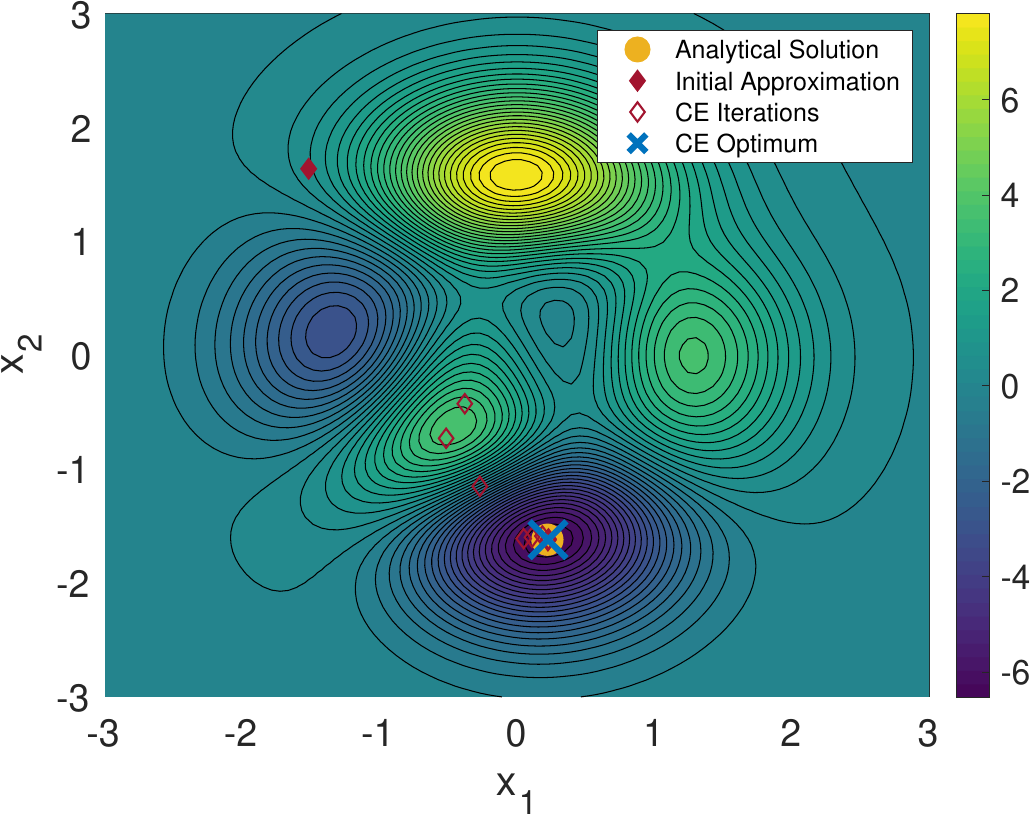}
\caption{Visualization of the Peaks Function Optimization. The color gradient from blue to yellow represents function values from low to high. Key points include the initial approximation (full red diamond), the trajectory of CEopt iterations (open red diamonds), the final CE solution (red cross), and the analytical solution (yellow circle).}
	\label{fig:Example2}
\end{figure}

The CE algorithm initiates with a broad search area, indicated by the initial approximation settings, and methodically narrows this area under the guidance of elite samples. Figure~\ref{fig:Example2} visually represents the trajectory of the CE iterations as they converge towards the global optimum. The iterations, marked by red diamonds, illustrate the positions sampled by CE method, providing a visual narrative of the algorithm's exploration and exploitation of the landscape. The ultimate convergence to the global optimum, marked by a red cross, underscores the efficacy of the CE method in navigating through and distinguishing between multiple peaks to identify the best solution.

\subsection{Example 3: A collection of 2D algebraic benchmark problems}

This section evaluates the \pkg{CEopt} solver's efficacy across a diverse set of 24 two-dimensional benchmark problems listed in Table~\ref{tab:benchmarks}. These benchmarks, known for their complex landscapes and intricate topological features, rigorously test the solver beyond the conventional limits of convex optimization methods. They encompass a wide range of mathematical models, each designed to challenge optimization algorithms in unique ways. Functions such as Ackley and Rastrigin test the solver's global search capabilities through their multi-modal landscapes, while the steep gradients and narrow valleys of the Rosenbrock and Dixon-Price functions probe its precision and convergence in more confined spaces.

The optimization trajectories and final outcomes are comprehensively depicted in Figures~\ref{fig:Example31}~and~\ref{fig:Example32}, which illustrate the \pkg{CEopt} solver's path through complex optimization landscapes. These figures provide a visual account of the solver's strategy, highlighting its dynamic adaptability and robust performance across varied challenges. \pkg{CEopt} demonstrated exceptional performance, reliably achieving or closely approximating global optima across most benchmarks. The solver showcased notable resilience against deceptive local minima, particularly in complex landscapes like the Eggholder and Cross-in-Tray, where it effectively navigated through intricate terrain to locate near-optimal solutions. While not every run achieved the global optimum due to inherent landscape challenges, the results were consistently near-optimal, underscoring the solver's capability to yield high-quality solutions.

\begin{table}[H]
\centering
\caption{Benchmark functions for unconstrained optimization testing. The functions range from simple unimodal landscapes to complex multimodal challenges, representing typical problems encountered in various scientific and engineering disciplines.}
\vspace{5mm}
\resizebox{\textwidth}{!}{%
\begin{tabular}{lcc}
\toprule
\textbf{Function Name}      & \textbf{Function Formula} $F(x_1,x_2)$ & \textbf{Domain for} $(x_1,x_2)$  \\ 
\midrule
Ackley                      & $\displaystyle -20 \exp\left(-0.2 \sqrt{0.5(x_1^2 + x_2^2)}\right) - \exp\left(0.5(\cos 2\pi x_1 + \cos 2\pi x_2)\right) + 20 + e$ & $[-32.768, 32.768] \times [-32.768, 32.768]$        \\
\vspace{4mm}
Beale                       & $\displaystyle (1.5 - x_1 + x_1 x_2)^2 + (2.25 - x_1 + x_1 x_2^2)^2 + (2.625 - x_1 + x_1 x_2^3)^2$                                  & $[-4.5, 4.5] \times [-4.5, 4.5]$              \\
\vspace{4mm}
Booth                       & $\displaystyle (x_1 + 2x_2 - 7)^2 + (2x_1 + x_2 - 5)^2$                                                                             & $[-10, 10] \times [-10, 10]$                \\
\vspace{4mm}
Bukin N.6                   & $\displaystyle 100 \sqrt{|x_2 - 0.01 x_1^2|} + 0.01 |x_1 + 10|$                                                                     & $[-15, -5] \times [-3, 3]$ \\
\vspace{4mm}
Cross-in-tray               & $\displaystyle -0.0001\left(\left|\sin(x_1)\sin(x_2)\exp\left(|100 - \frac{\sqrt{x_1^2+x_2^2}}{\pi}|\right)\right|+1\right)^{0.1}$ & $[-10, 10] \times [-10, 10]$                \\
\vspace{4mm}
Dixon-Price               & $\displaystyle (x_1 - 1)^2 + 2 \,  (2x_2^2 - x_1)^2$ & $[-10, 10] \times [-10, 10]$ \\
\vspace{4mm}
Easom                       & $\displaystyle - \cos(x_1) \cos(x_2) \exp\left(-(x_1 - \pi)^2 - (x_2 - \pi)^2\right)$                                              & $[-100, 100] \times [-100, 100]$              \\
\vspace{4mm}
Eggholder                   & $\displaystyle -(x_2 + 47) \sin(\sqrt{|x_2 + \frac{x_1}{2} + 47|}) - x_1 \sin(\sqrt{|x_1 - (x_2 + 47)|})$                          & $[-512, 512] \times [-512, 512]$              \\
\vspace{4mm}
Goldstein-Price             & $\displaystyle (1+(x_1 + x_2 + 1)^2(19 - 14x_1 + 3x_1^2 - 14x_2 + 6x_1x_2 + 3x_2^2)) \times$                                      & $[-2, 2] \times [-2, 2]$                  \\
                            & $\displaystyle (30 + (2x_1 - 3x_2)^2(18 - 32x_1 + 12x_1^2 + 48x_2 - 36x_1x_2 + 27x_2^2))$                                         &                            \\
\vspace{4mm}
Griewank                    & $\displaystyle 1 + \sum_{i=1}^2 \frac{x_i^2}{4000} - \prod_{i=1}^2 \cos\left(\frac{x_i}{\sqrt{i}}\right)$                          & $[-600, 600] \times [-600, 600]$              \\
\vspace{4mm}
Himmelblau's                & $\displaystyle (x_1^2 + x_2 - 11)^2 + (x_1 + x_2^2 - 7)^2$                                                                          & $[-5, 5] \times [-5, 5]$                  \\
\vspace{4mm}
Holder Table                & $\displaystyle -|\sin(x_1) \cos(x_2) \exp(|1 - \sqrt{x_1^2+x_2^2}/\pi|)|$                                                           & $[-10, 10] \times [-10, 10]$                \\
\vspace{4mm}
Lévi N.13                   & $\displaystyle \sin^2(3\pi x_1) + (x_1 - 1)^2 (1+\sin^2(3\pi x_2)) + (x_2 - 1)^2(1+\sin^2(2\pi x_2))$                              & $[-10, 10] \times [-10, 10]$                \\
\vspace{4mm}
Matyas                      & $\displaystyle 0.26(x_1^2 + x_2^2) - 0.48 x_1 x_2$                                                                                  & $[-10, 10] \times [-10, 10]$                \\
\vspace{4mm}
McCormick                   & $\displaystyle \sin(x_1 + x_2) + (x_1 - x_2)^2 - 1.5 x_1 + 2.5 x_2 + 1$                                                             & $[-1.5, 4] \times [-3, 3]$ \\
\vspace{4mm}
Rastrigin                   & $\displaystyle 10 \times 2 + \sum_{i=1}^2 [x_i^2 - 10 \cos(2 \pi x_i)]$                                                                      & $[-5.12, 5.12] \times [-5.12, 5.12]$            \\
\vspace{4mm}
Rosenbrock                  & $\displaystyle \sum_{i=1}^{2-1} [100(x_{i+1} - x_i^2)^2 + (1-x_i)^2]$                                                               & $[-5, 10] \times [-5, 10]$                 \\
\vspace{4mm}
Schaffer N.2                & $\displaystyle 0.5 + \frac{\sin^2(x_1^2 - x_2^2) - 0.5}{[1 + 0.001(x_1^2 + x_2^2)]^2}$                                             & $[-100, 100] \times [-100, 100]$              \\
\vspace{4mm}
Schaffer N.4                & $\displaystyle 0.5 + \frac{\cos^2(\sin|x_1^2 - x_2^2|) - 0.5}{[1 + 0.001(x_1^2 + x_2^2)]^2}$                                       & $[-100, 100] \times [-100, 100]$              \\
\vspace{4mm}
Shekel                      & $\displaystyle -\sum_{i=1}^{10} \frac{1}{c_i + \sum_{j=1}^2 (x_j - A_{ij})^2}$                                                      & $[0, 10] \times [0, 10]$                  \\
\vspace{4mm}
Sphere                      & $\displaystyle \sum_{i=1}^2 x_i^2$                                                                                                  & $[-5, 5] \times [-5, 5]$                  \\
\vspace{4mm}
Styblinski–Tang             & $\displaystyle \frac{1}{2}\sum_{i=1}^2 (x_i^4 - 16x_i^2 + 5x_i)$                                                                     & $[-5, 5] \times [-5, 5]$                  \\
\vspace{4mm}
Three-hump Camel            & $\displaystyle 2x_1^2 - 1.05x_1^4 + \frac{x_1^6}{6} + x_1x_2 + x_2^2$                                                               & $[-5, 5] \times [-5, 5]$                  \\
\vspace{4mm}
Zakharov                  & $\displaystyle \sum_{i=1}^2 x_i^2 + \left(\sum_{i=1}^2 0.5 \, i \, x_i\right)^2 + \left(\sum_{i=1}^2 0.5 \, i \, x_i\right)^4$ & $[-5, 10] \times [-5, 10]$ \\
\bottomrule
\end{tabular}}
\begin{flushleft}
\footnotesize{Note: For the Shekel function, matrix 
$A = \begin{bmatrix}4 & 1 & 8 & 6 & 3 & 2 & 5 & 8 & 6 & 7\\ 4 & 1 & 8 & 6 & 3 & 2 & 5 & 8 & 6 & 7\\ \end{bmatrix}$ and vector $\bm{c} = \frac{1}{10} \times [1, 2, 2, 4, 4, 6, 3, 7, 5, 5]$ are used.}
\end{flushleft}
\label{tab:benchmarks}
\end{table}

\begin{figure}[H]
\centering
\captionsetup[subfigure]{labelformat=empty} % Optional: removes labels from each subfigure
\subfloat[]{\includegraphics[width=0.33\textwidth]{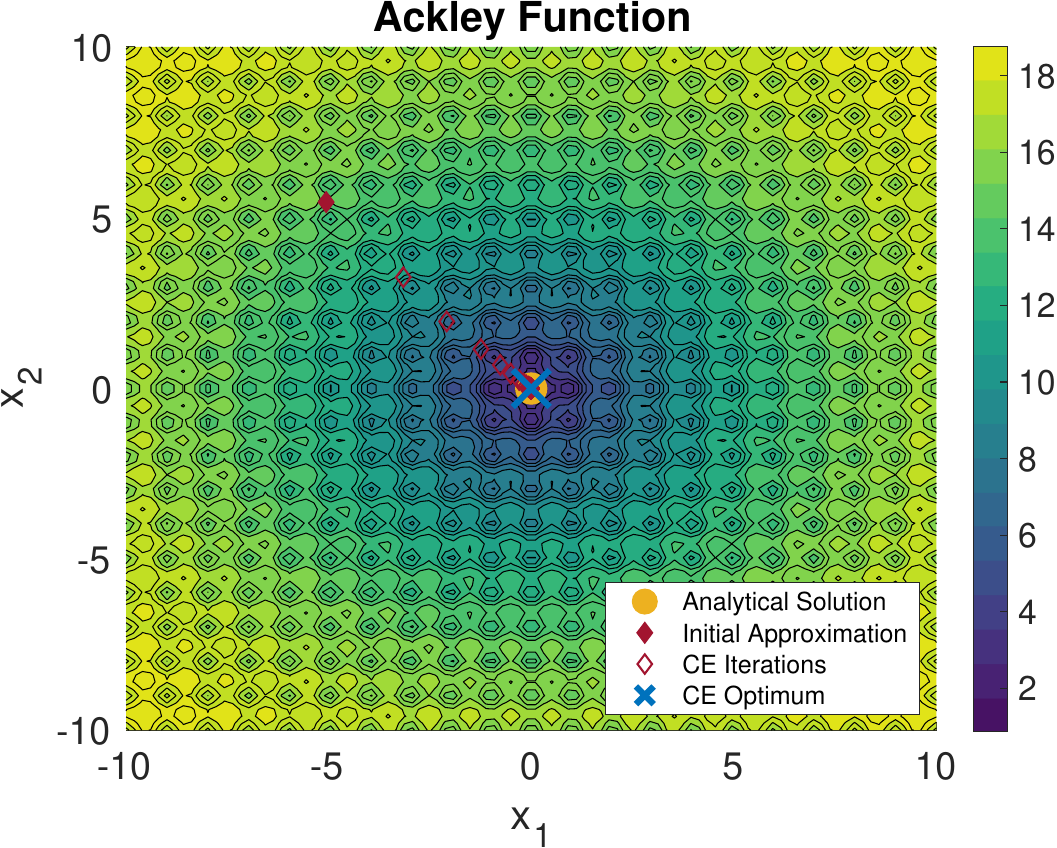}}
\subfloat[]{\includegraphics[width=0.33\textwidth]{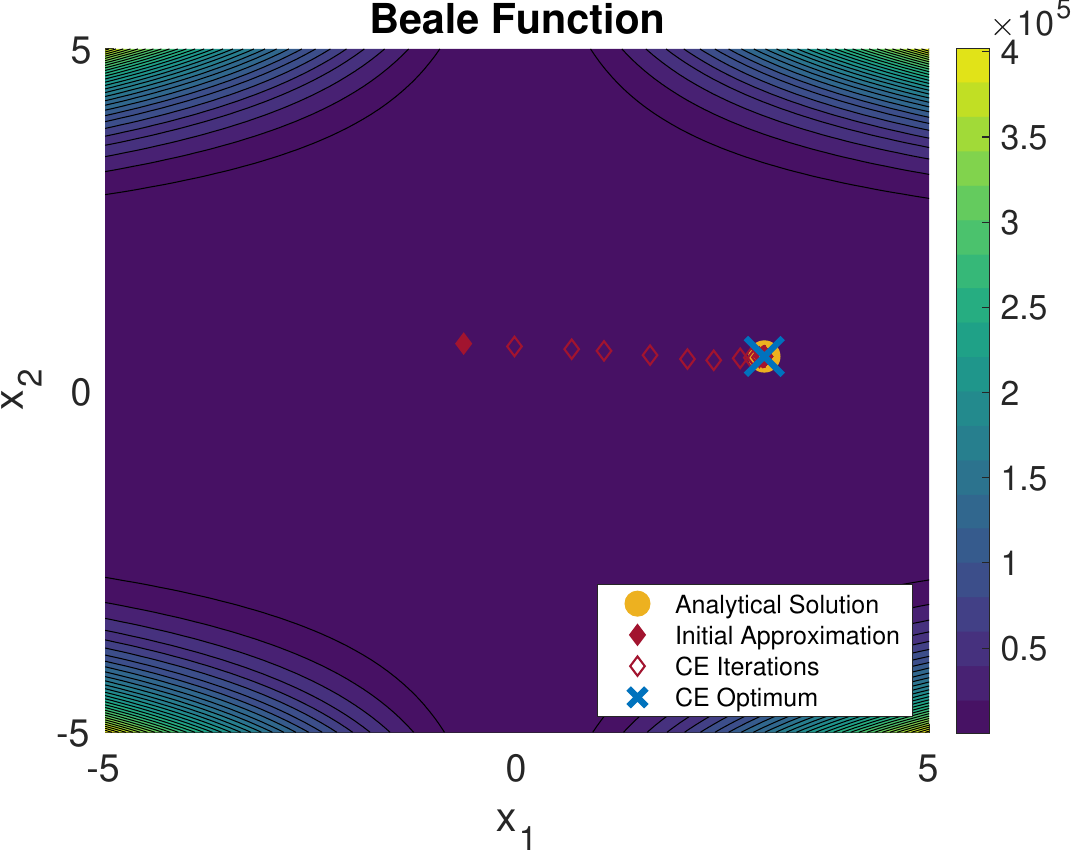}}
\subfloat[]{\includegraphics[width=0.33\textwidth]{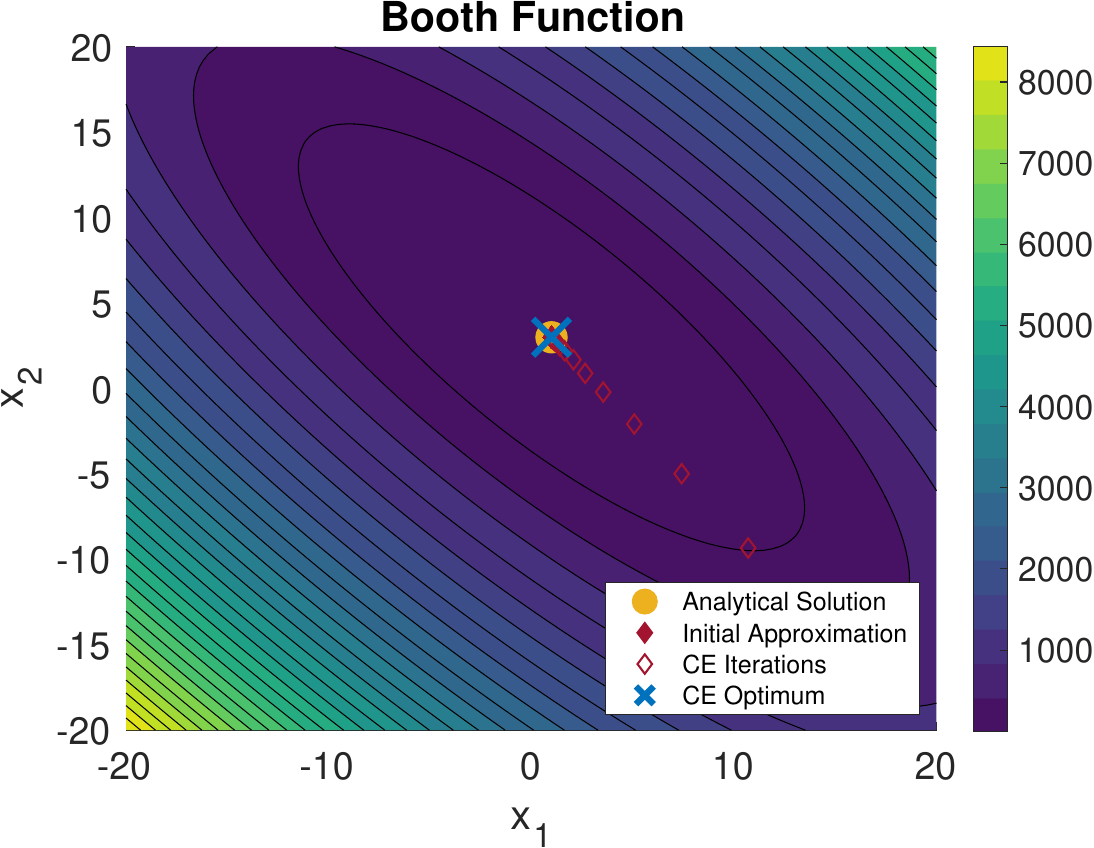}}\\
\subfloat[]{\includegraphics[width=0.33\textwidth]{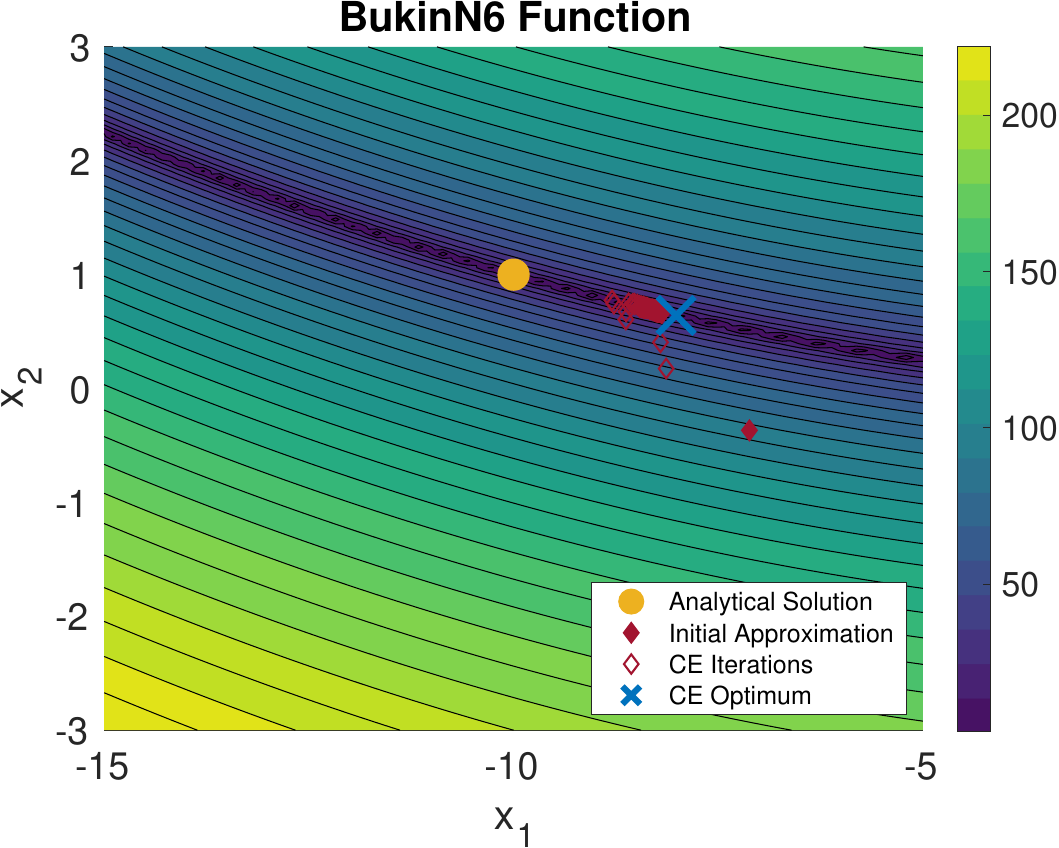}}
\subfloat[]{\includegraphics[width=0.33\textwidth]{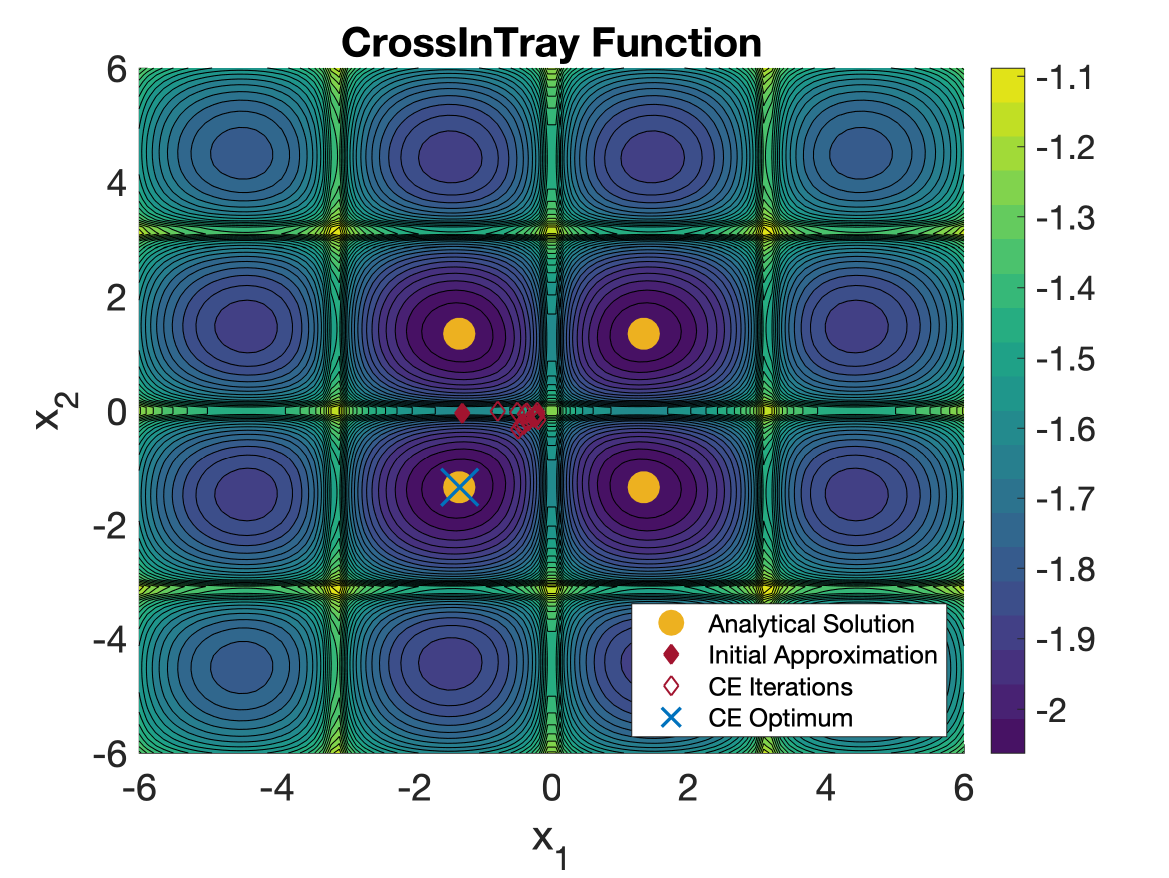}}
\subfloat[]{\includegraphics[width=0.33\textwidth]{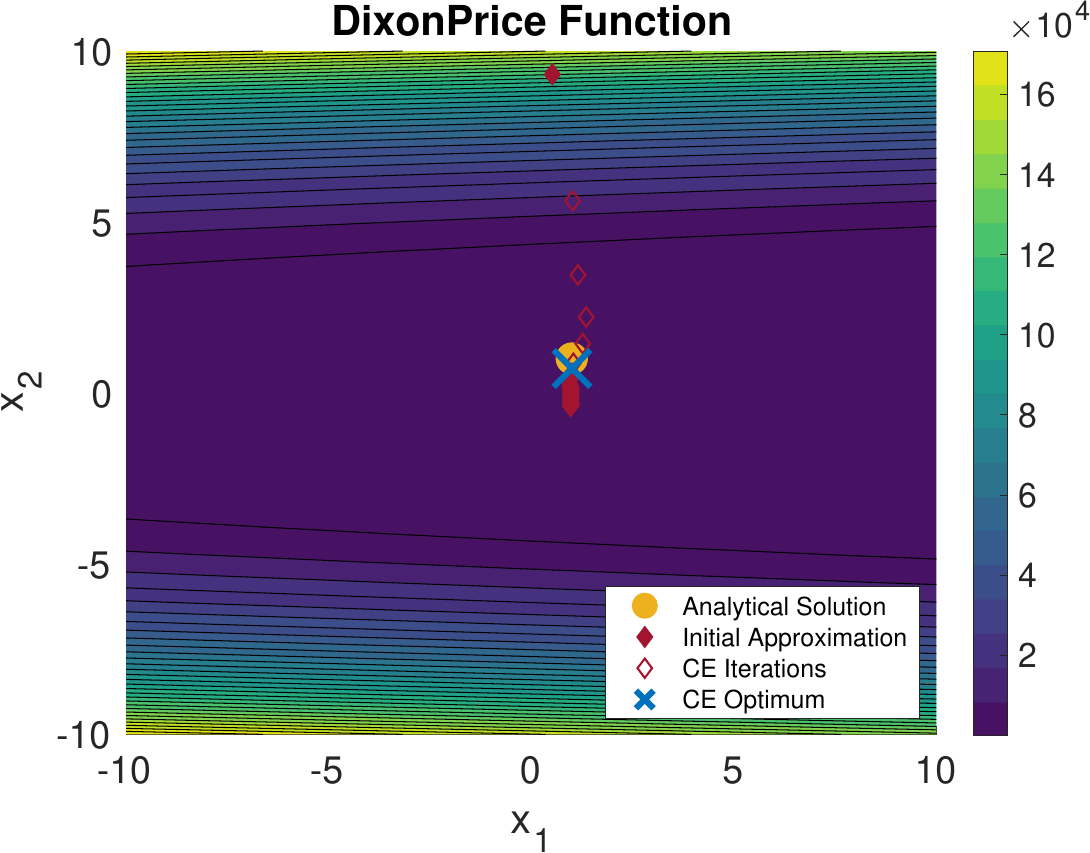}}\\
\subfloat[]{\includegraphics[width=0.33\textwidth]{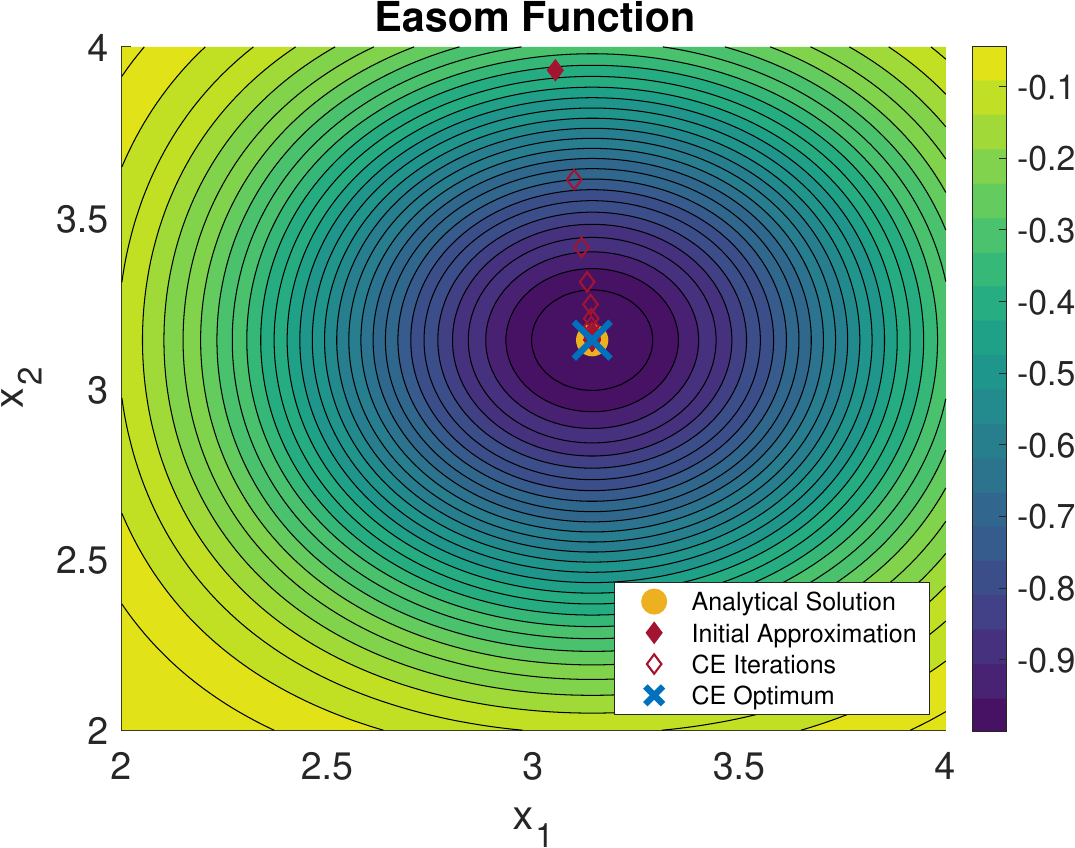}}
\subfloat[]{\includegraphics[width=0.33\textwidth]{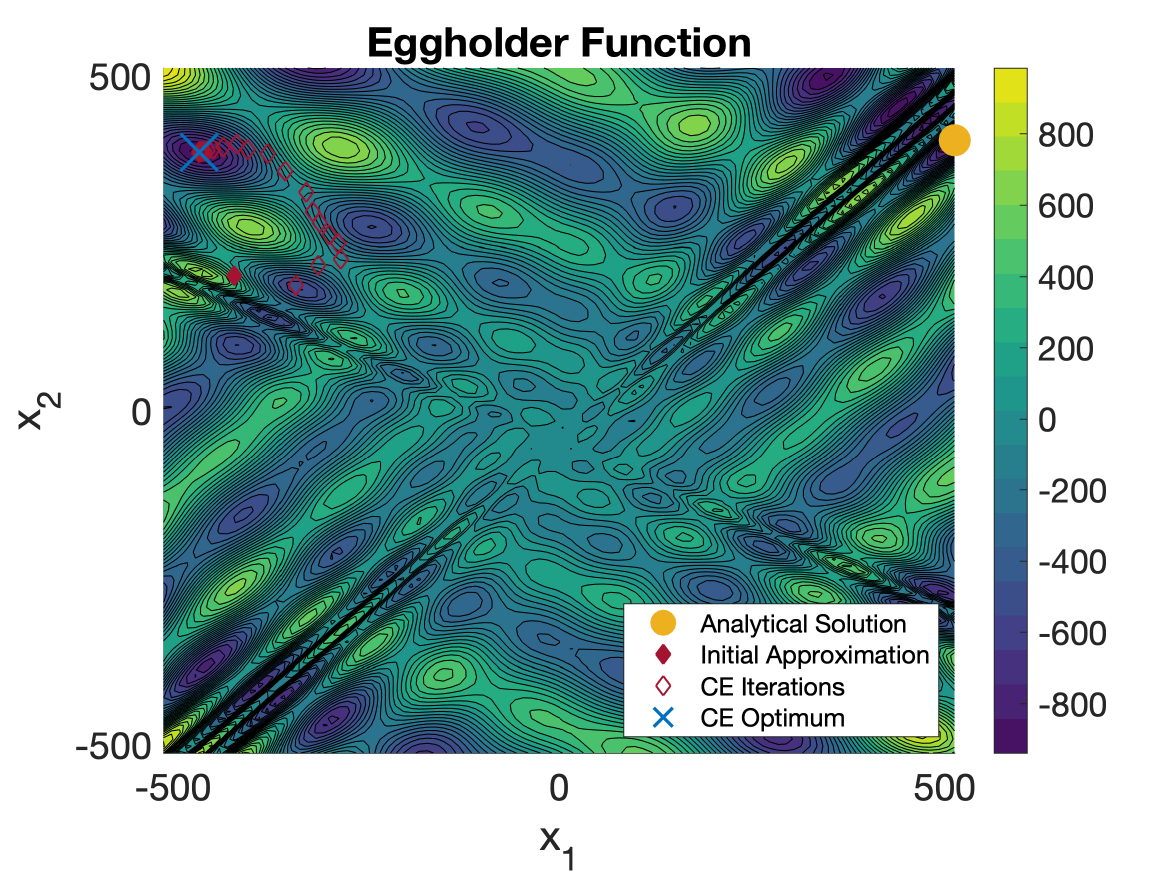}}
\subfloat[]{\includegraphics[width=0.33\textwidth]{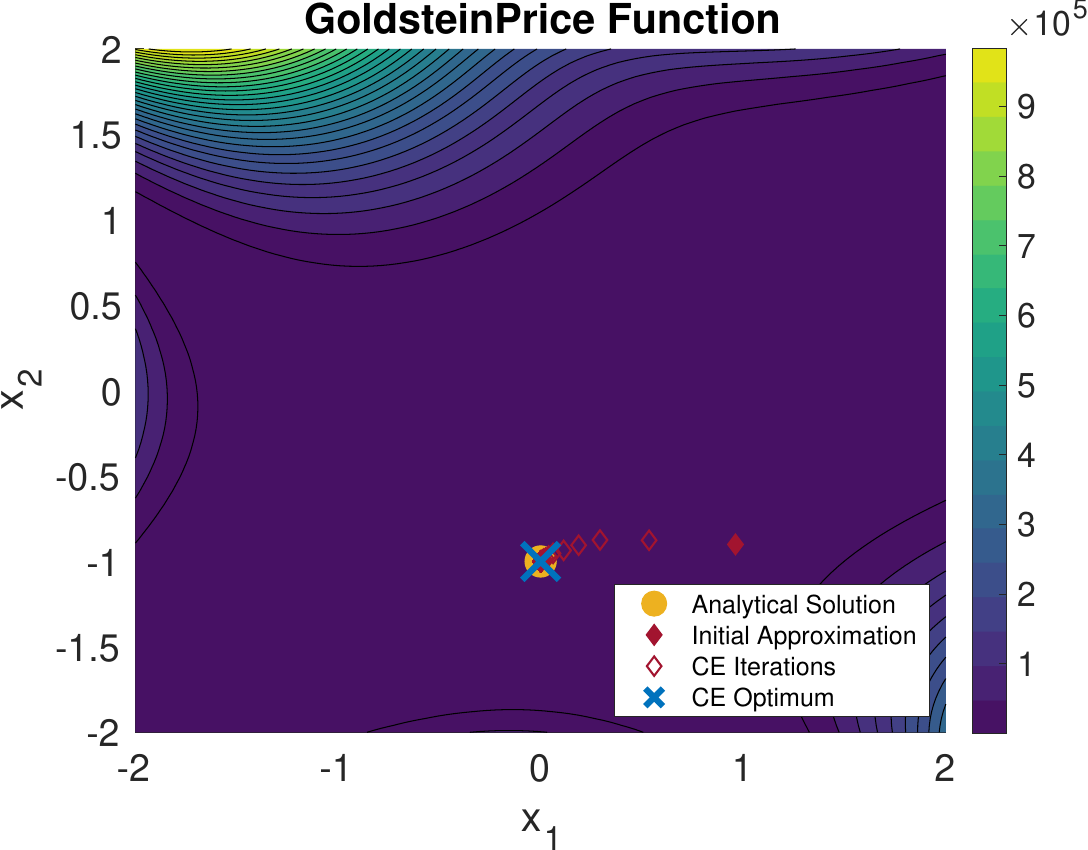}}\\
\subfloat[]{\includegraphics[width=0.33\textwidth]{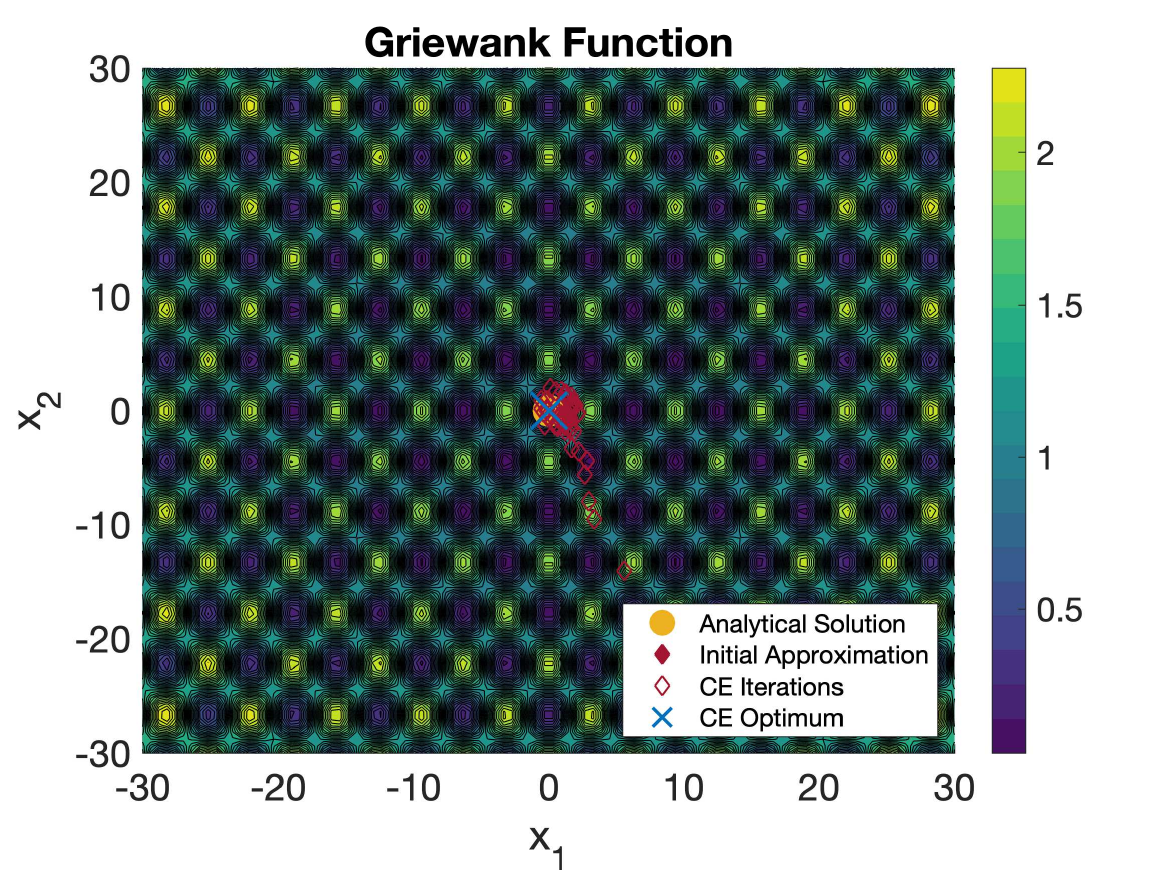}}
\subfloat[]{\includegraphics[width=0.33\textwidth]{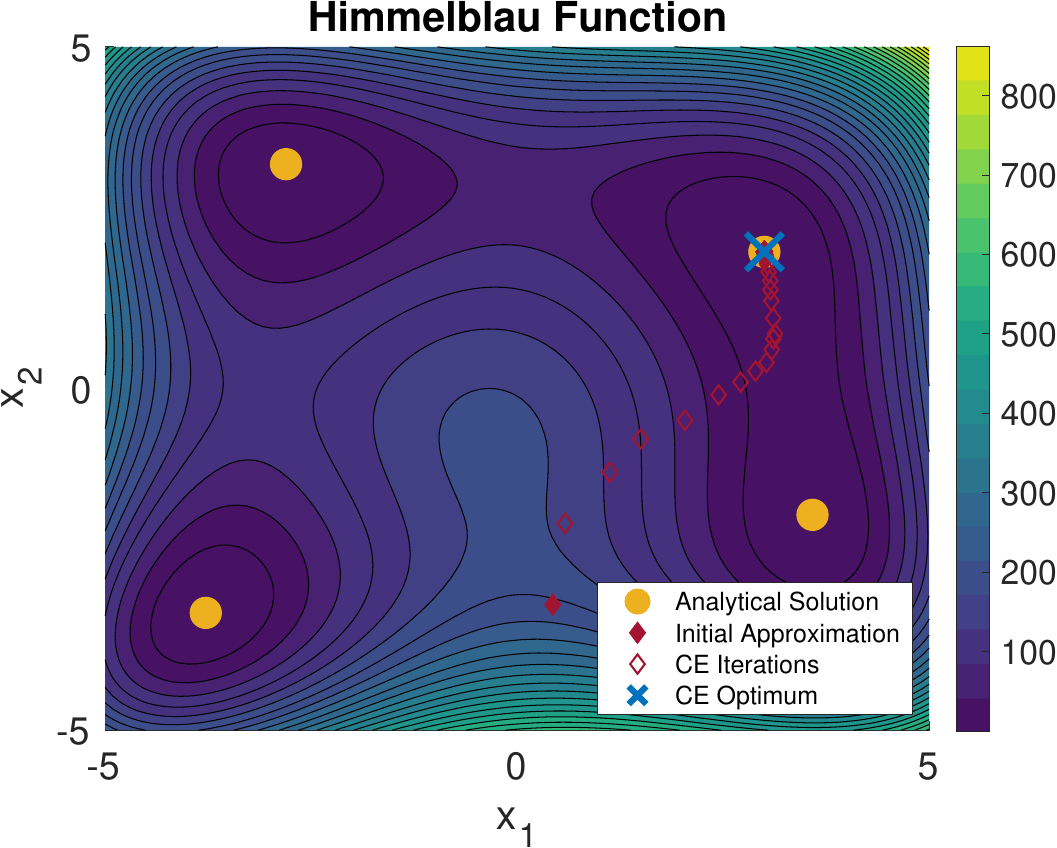}}
\subfloat[]{\includegraphics[width=0.33\textwidth]{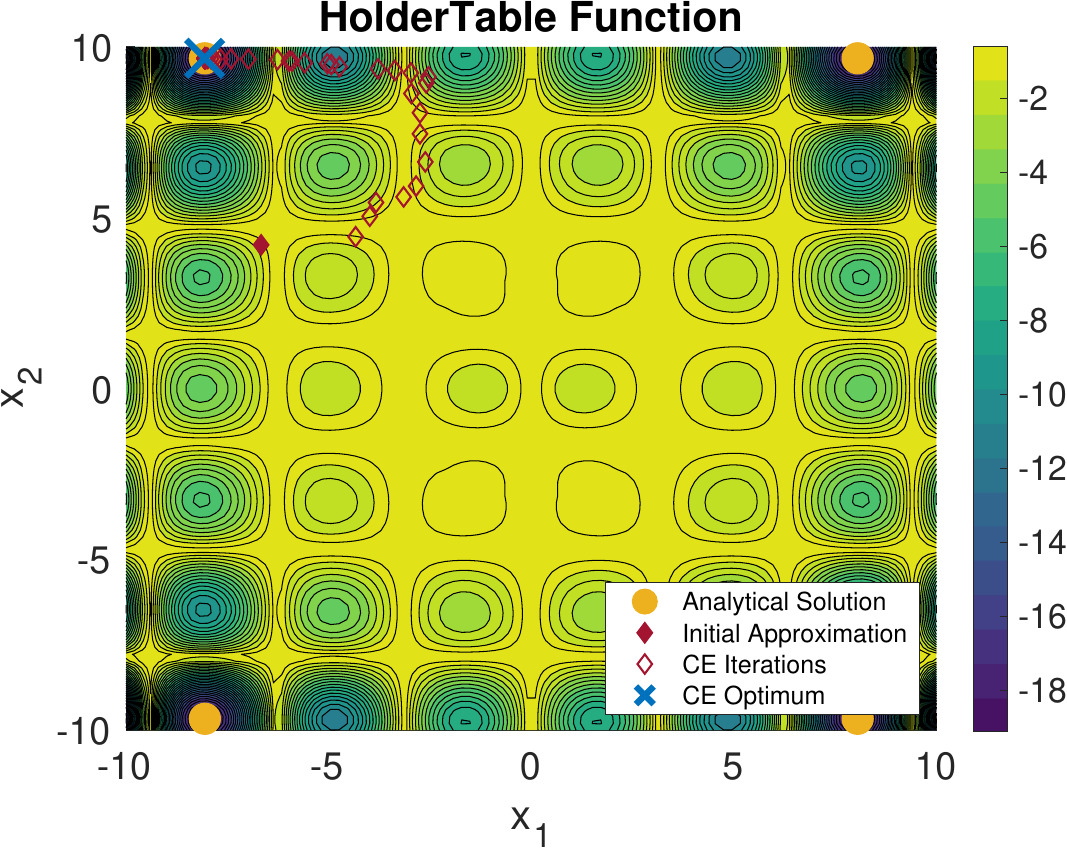}}
\caption{Optimization landscapes and paths for the first set of benchmark functions. Each subfigure shows the optimization process, with legends indicating initial points, progression paths, and final solutions, highlighting the solver’s effectiveness across diverse landscapes.}
\label{fig:Example31}
\end{figure}

\begin{figure}[H]
\centering
\captionsetup[subfigure]{labelformat=empty} % Optional: removes labels from each subfigure
\subfloat[]{\includegraphics[width=0.33\textwidth]{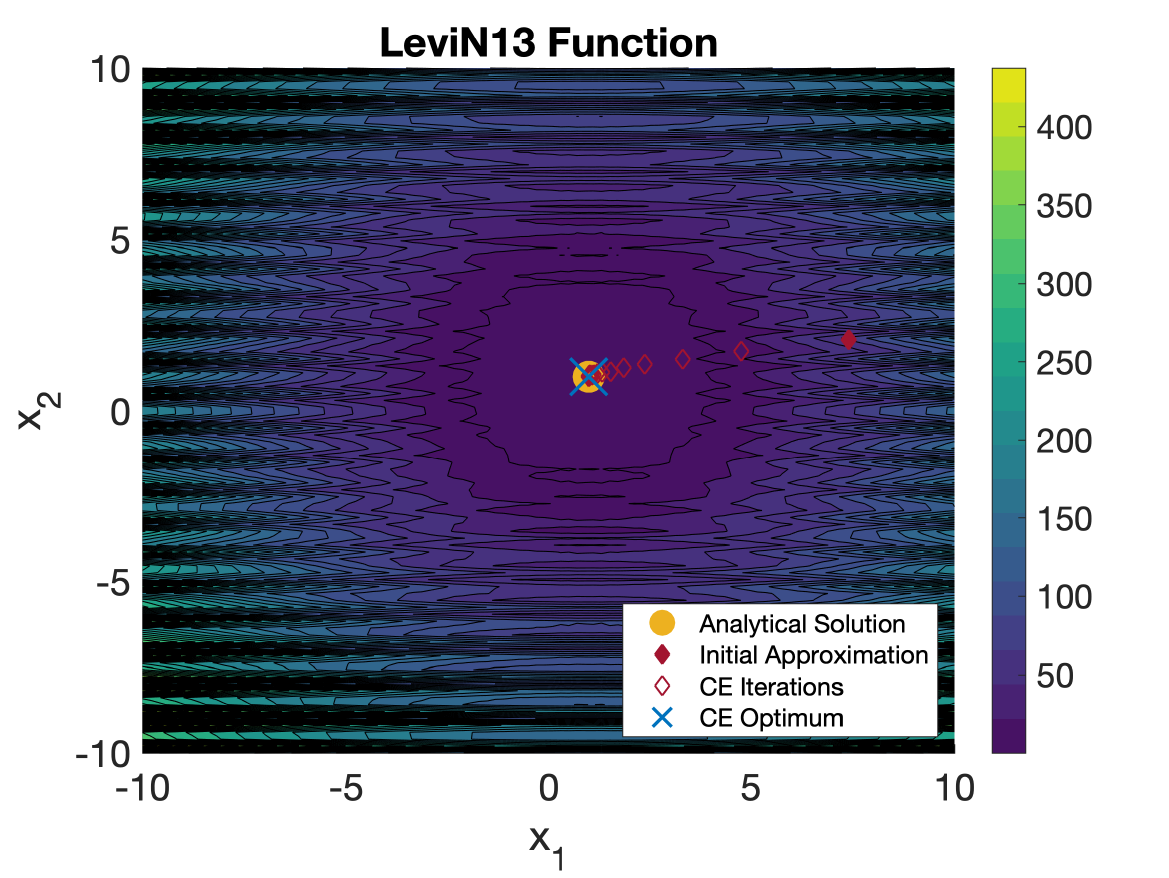}}
\subfloat[]{\includegraphics[width=0.33\textwidth]{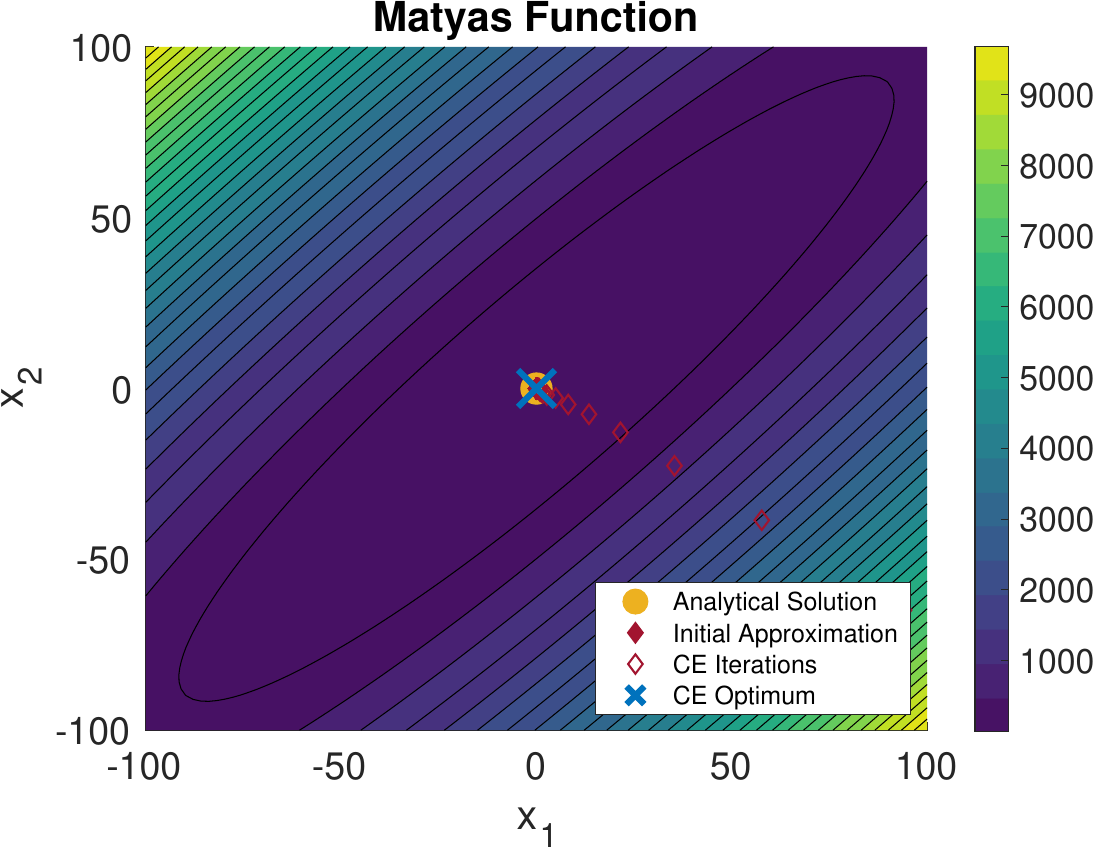}}
\subfloat[]{\includegraphics[width=0.33\textwidth]{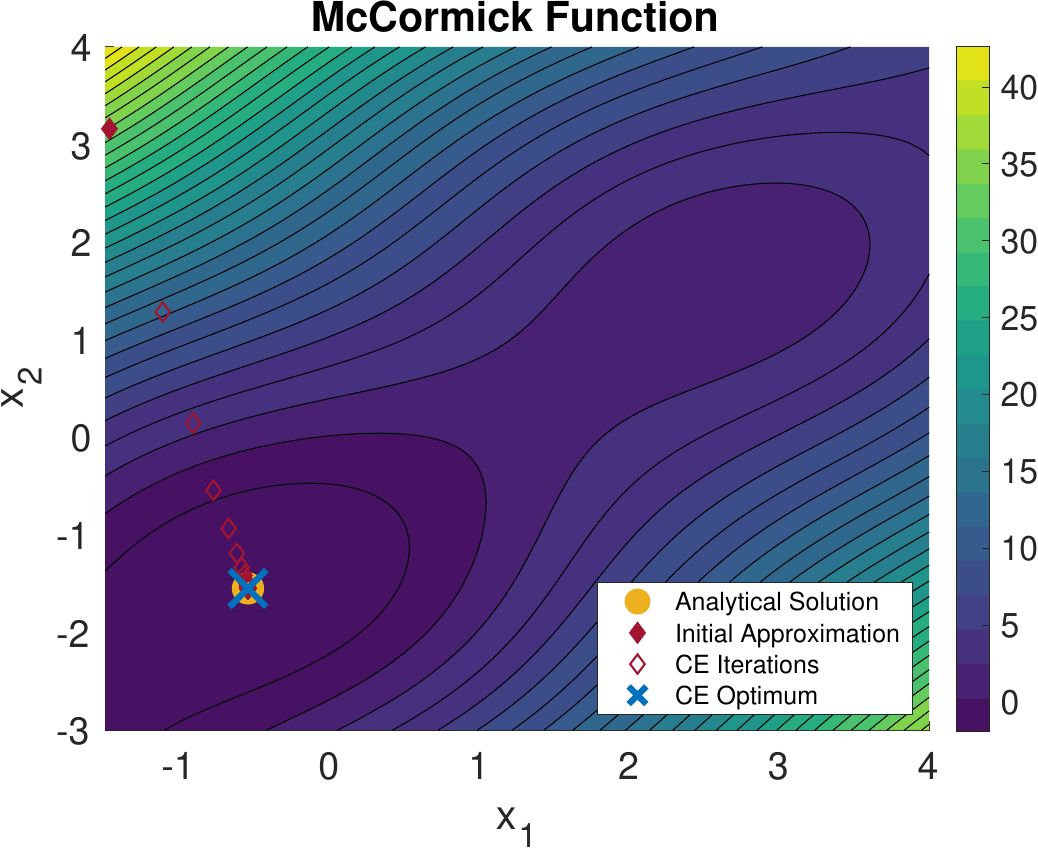}}\\
\subfloat[]{\includegraphics[width=0.33\textwidth]{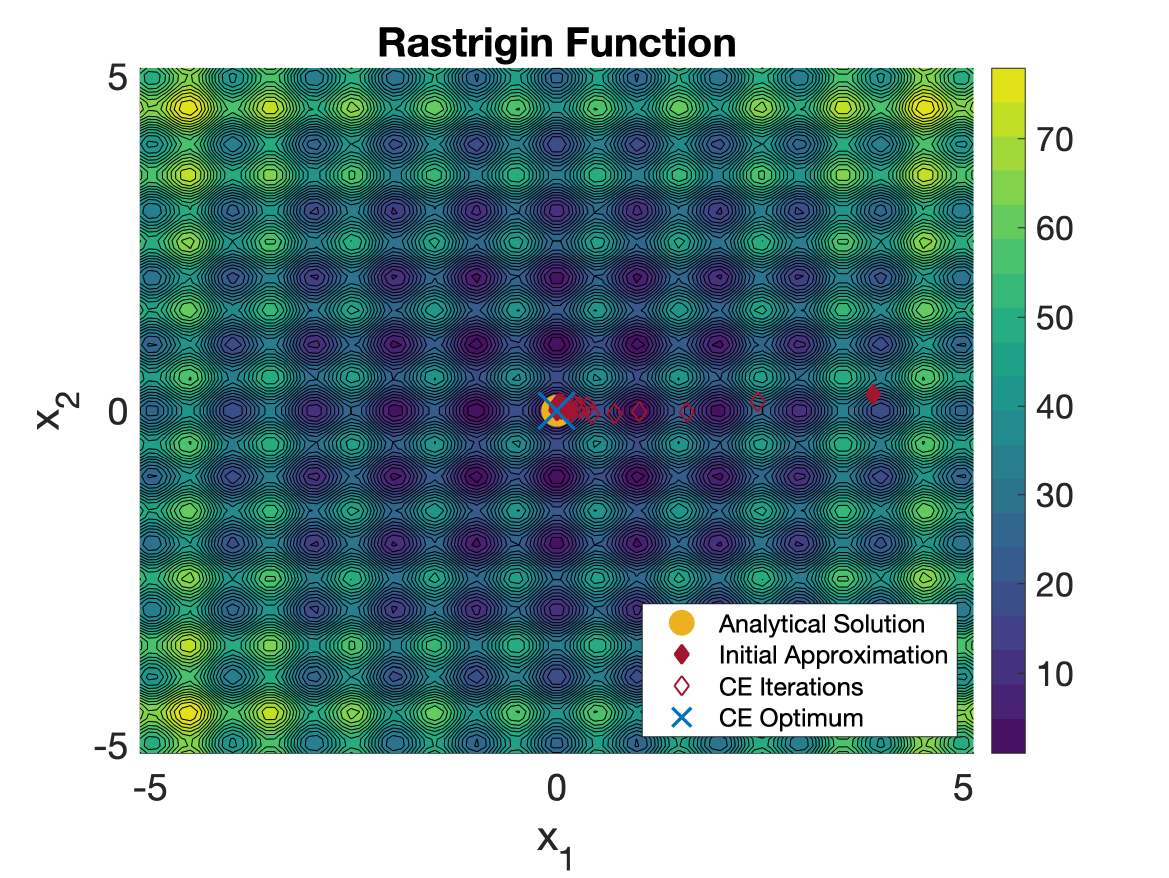}}
\subfloat[]{\includegraphics[width=0.33\textwidth]{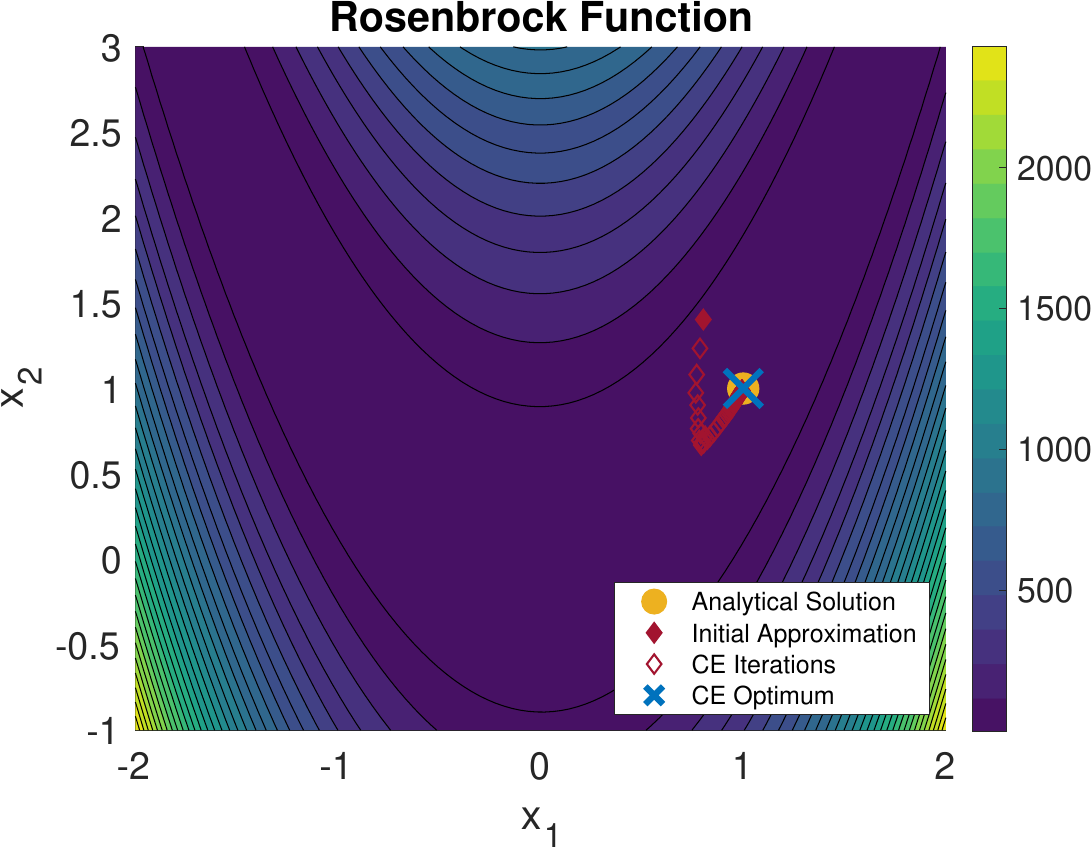}}
\subfloat[]{\includegraphics[width=0.33\textwidth]{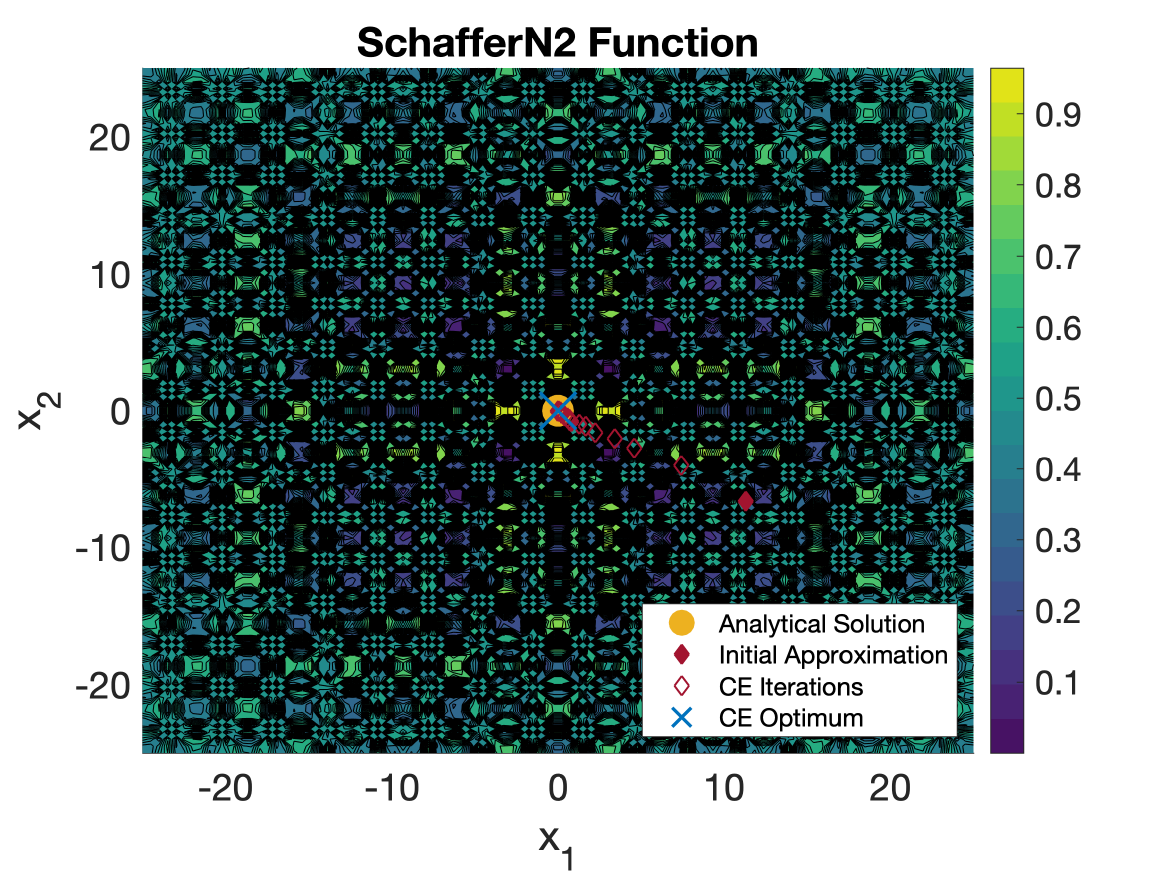}}\\
\subfloat[]{\includegraphics[width=0.33\textwidth]{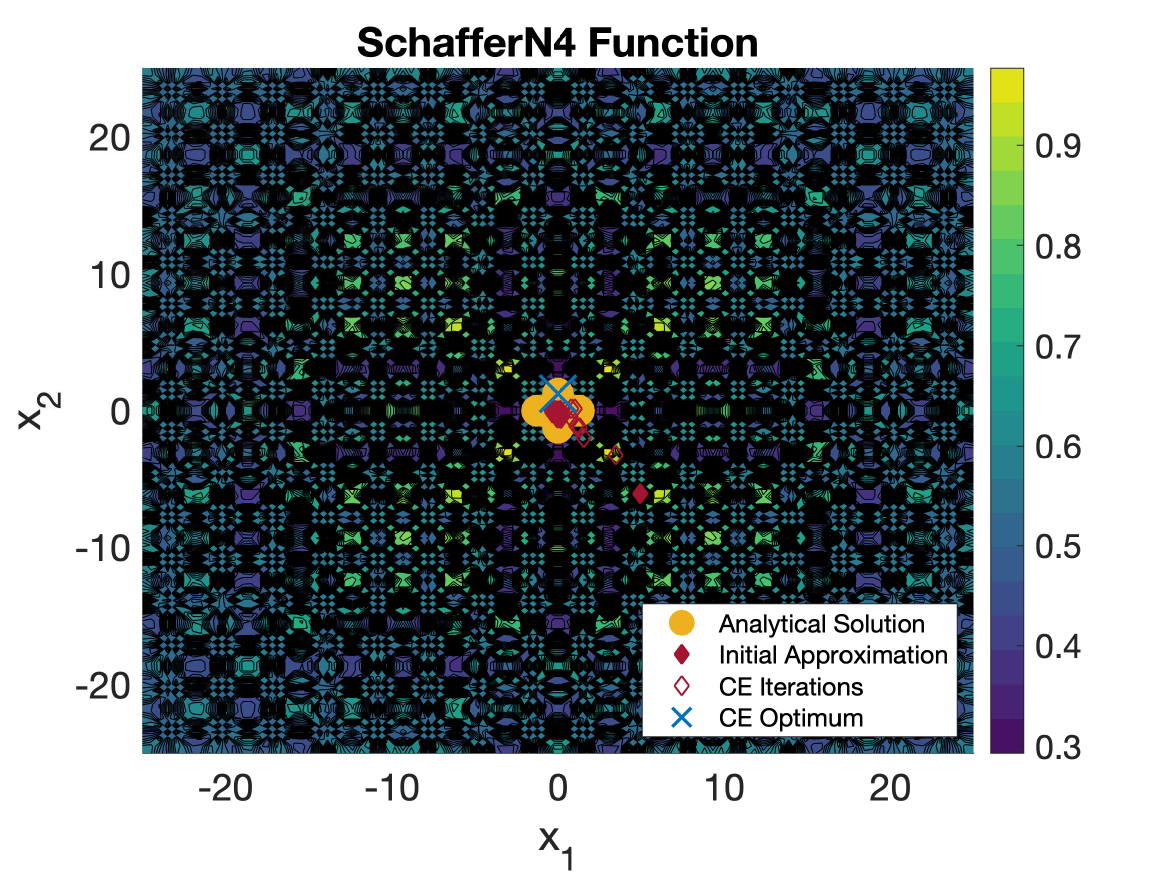}}
\subfloat[]{\includegraphics[width=0.33\textwidth]{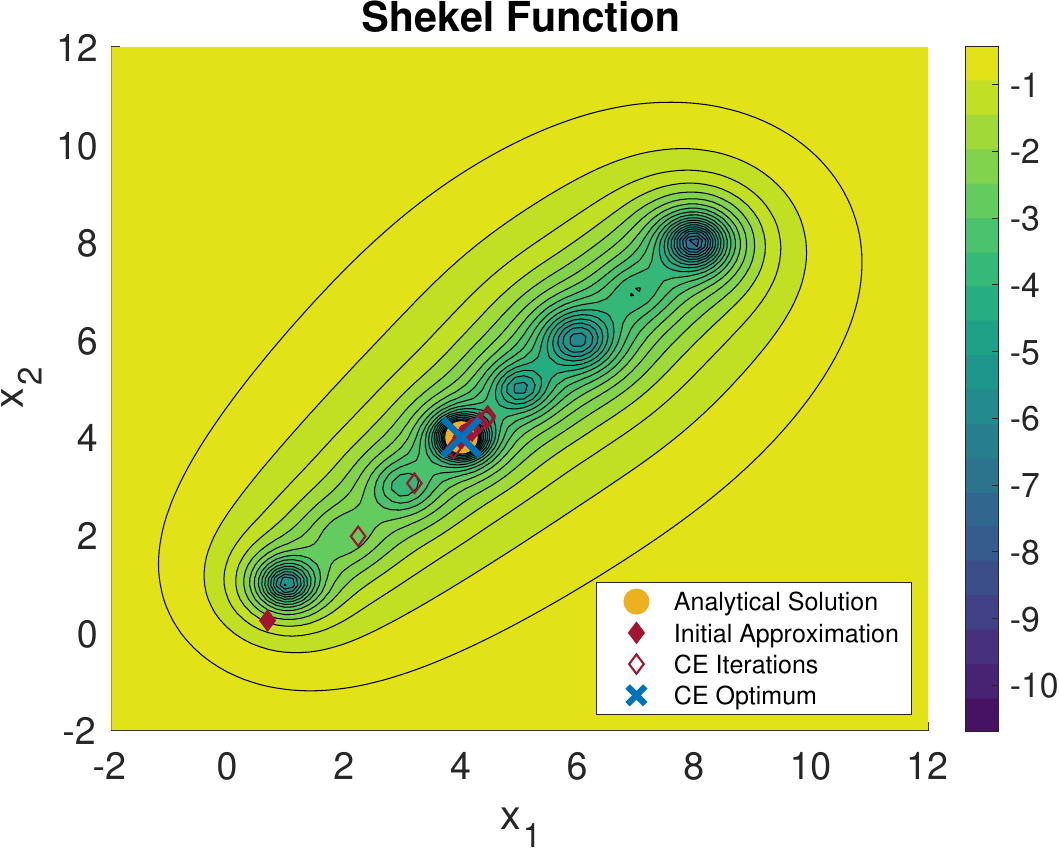}}
\subfloat[]{\includegraphics[width=0.33\textwidth]{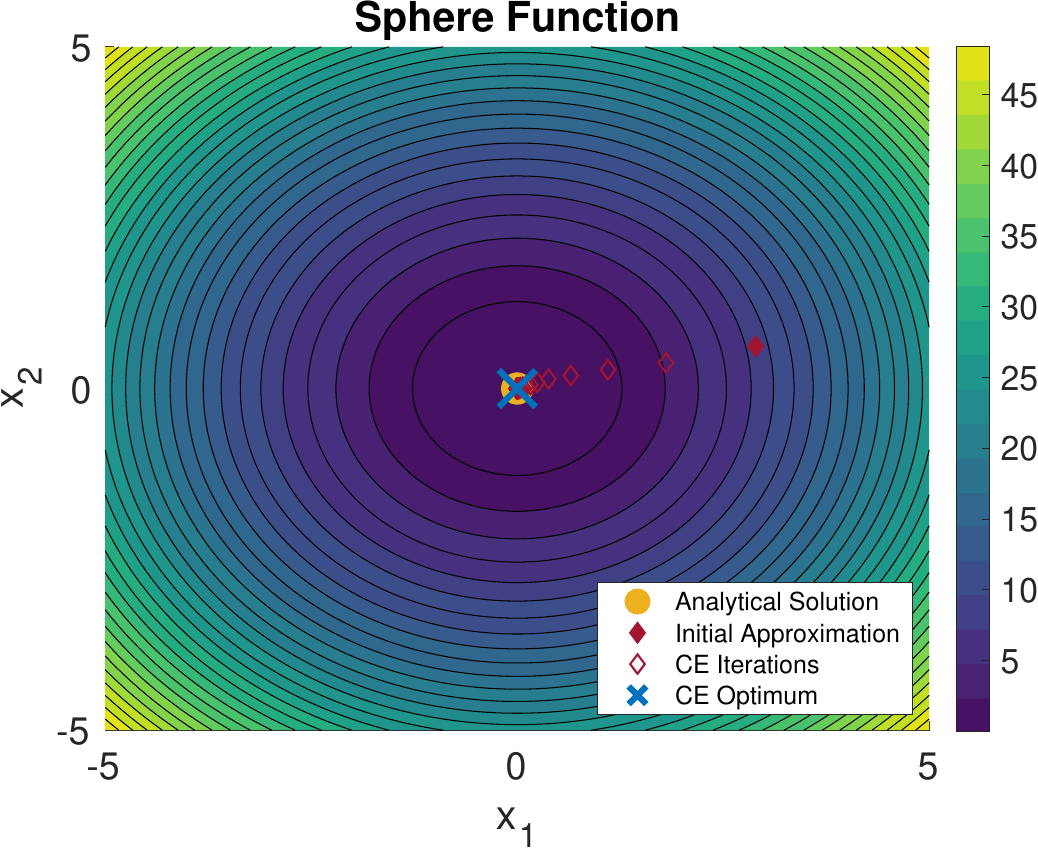}}\\
\subfloat[]{\includegraphics[width=0.33\textwidth]{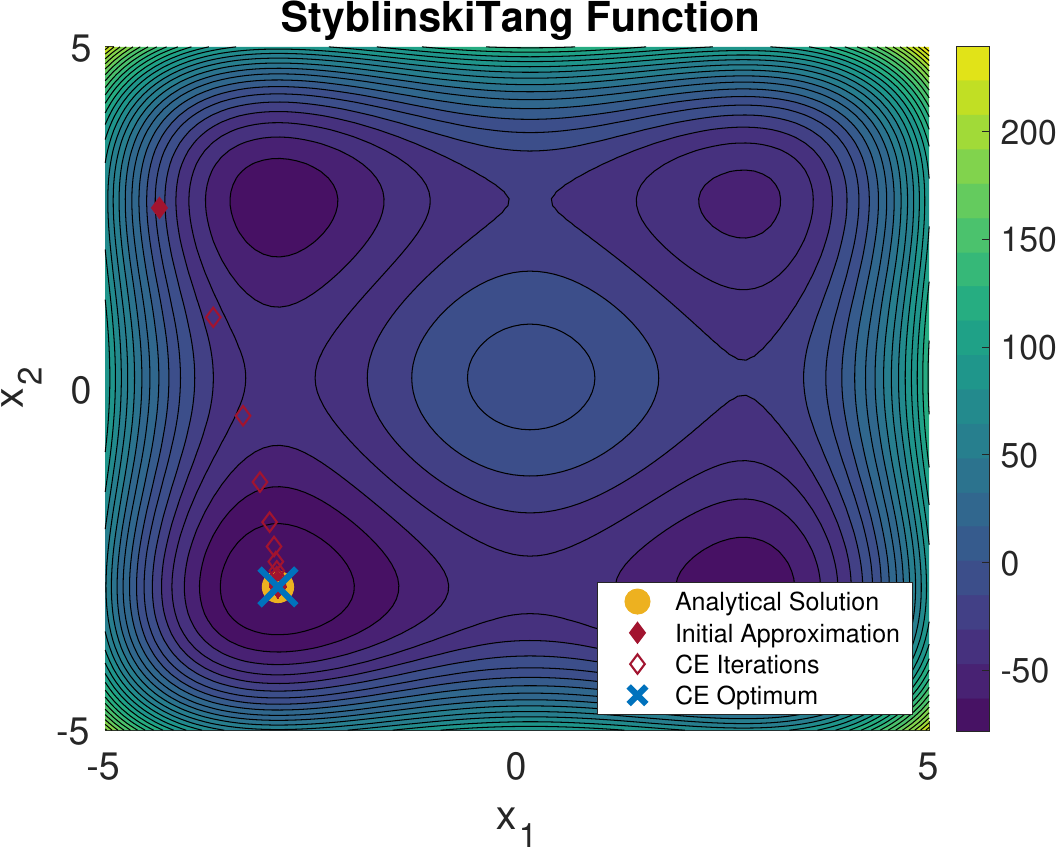}}
\subfloat[]{\includegraphics[width=0.33\textwidth]{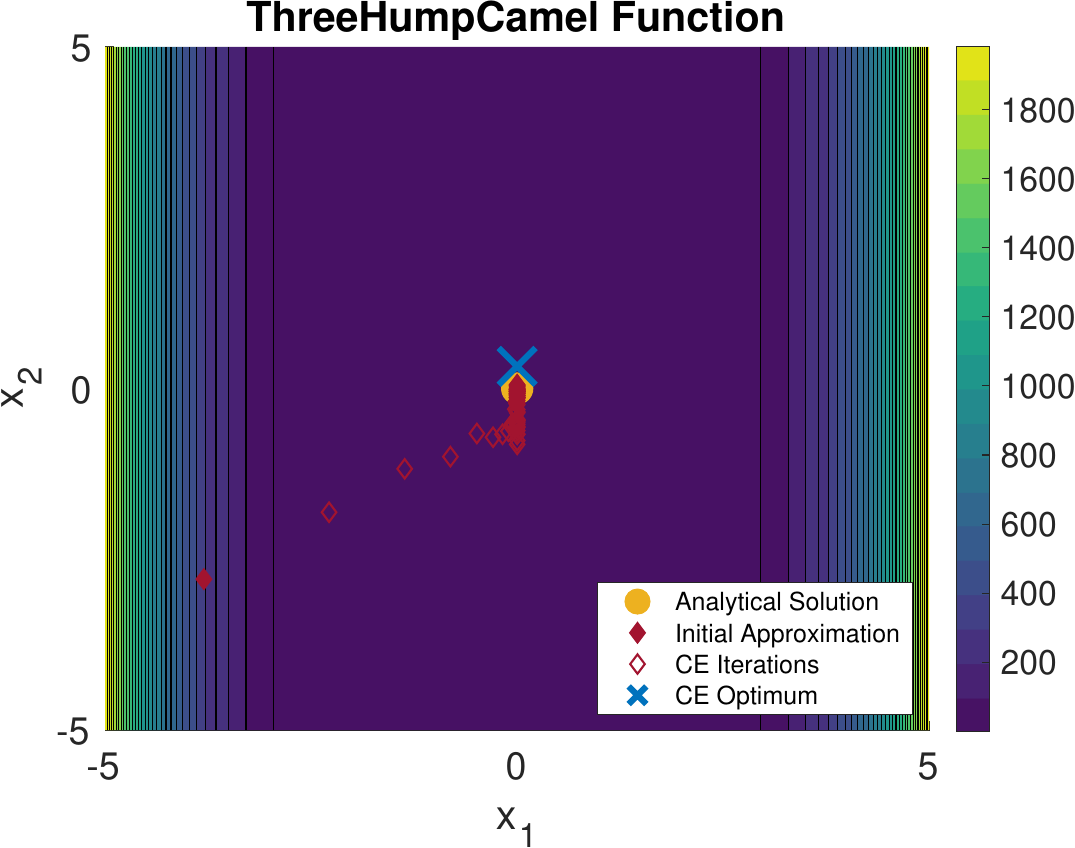}}
\subfloat[]{\includegraphics[width=0.33\textwidth]{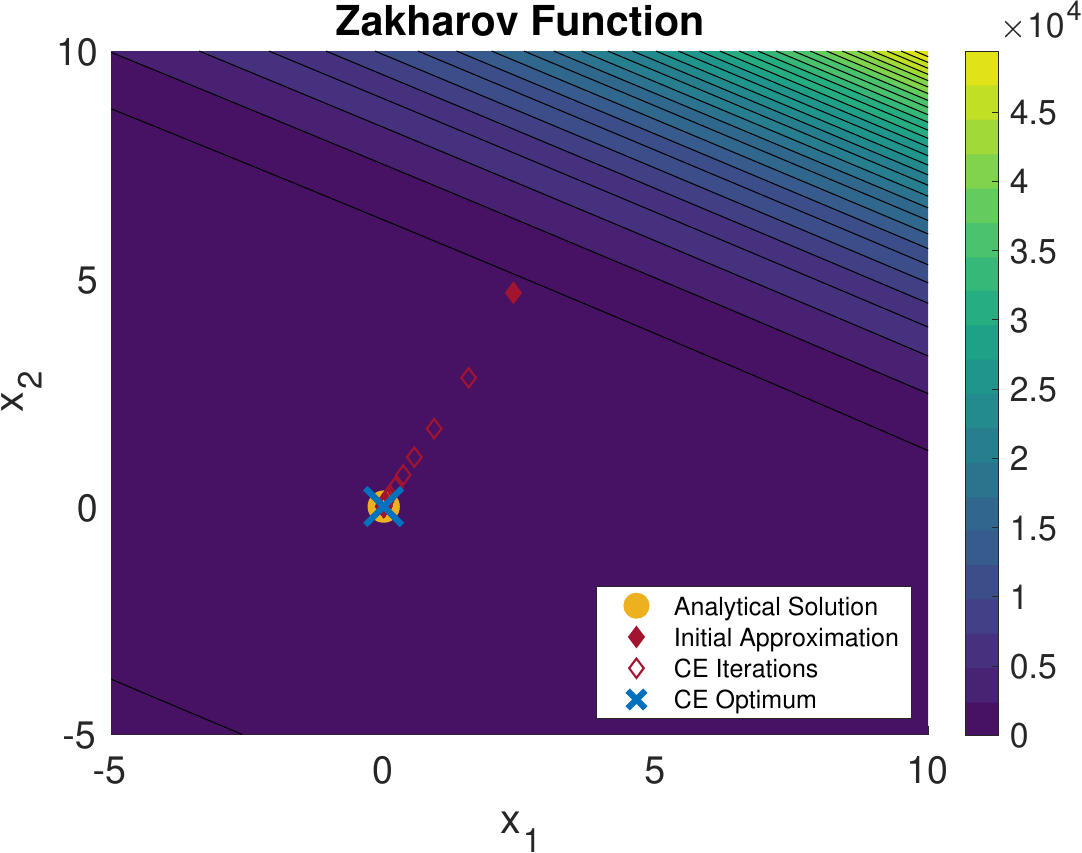}}
\caption{Continuation of optimization landscapes for the second set of benchmark functions. These visualizations demonstrate the solver's adaptability and strategic exploration, with detailed trajectories marking the course to optimal or near-optimal solutions.}
\label{fig:Example32}
\end{figure}

\begin{figure}[H]
\centering
\captionsetup[subfigure]{labelformat=empty} % Optional: removes labels from each subfigure
\subfloat[]{\includegraphics[width=0.5\textwidth]{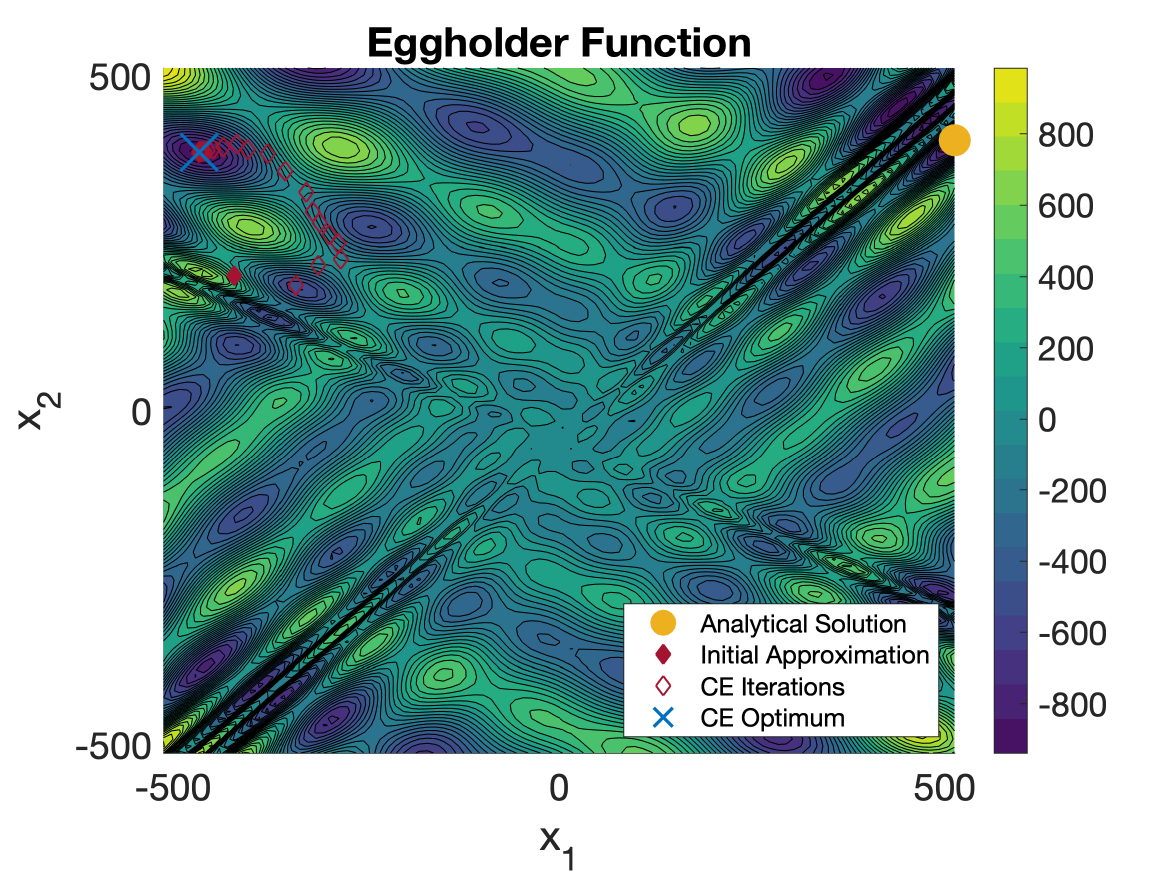}}
\subfloat[]{\includegraphics[width=0.5\textwidth]{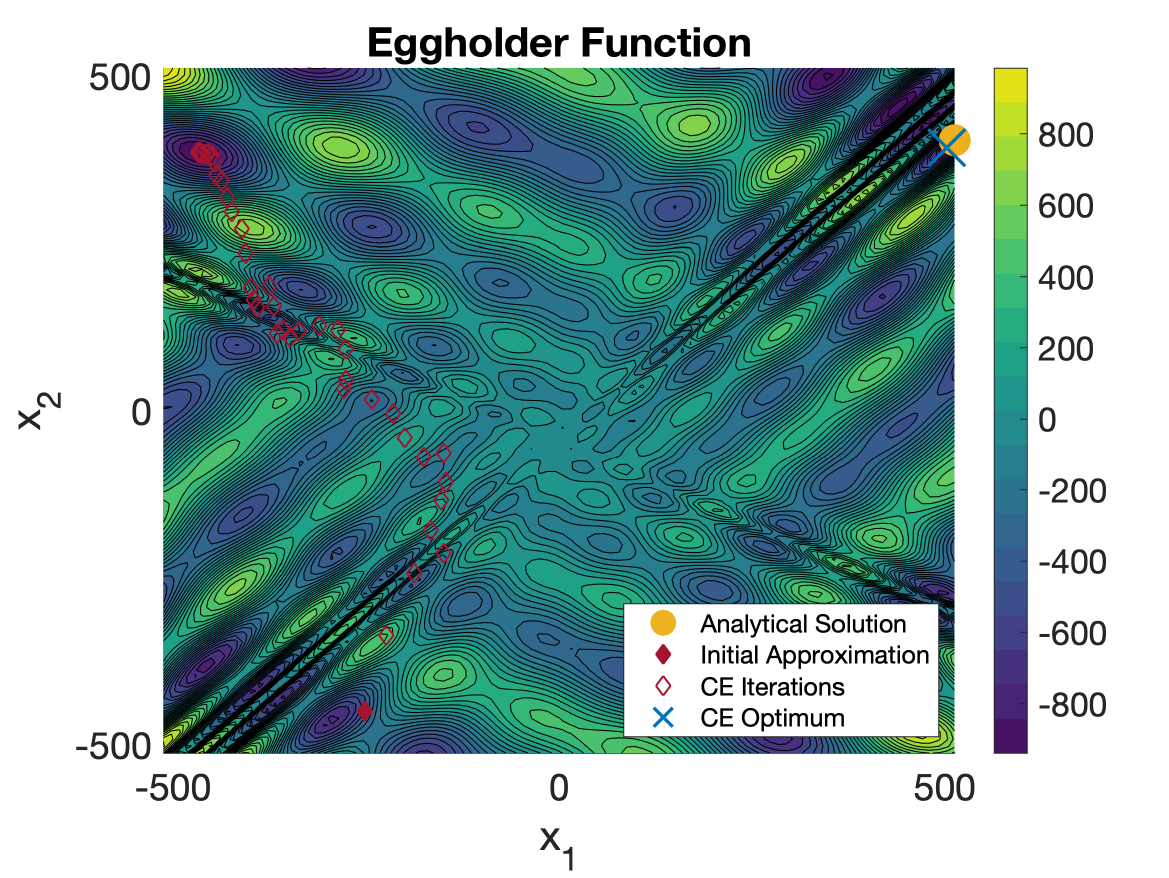}}\\
\subfloat[]{\includegraphics[width=0.5\textwidth]{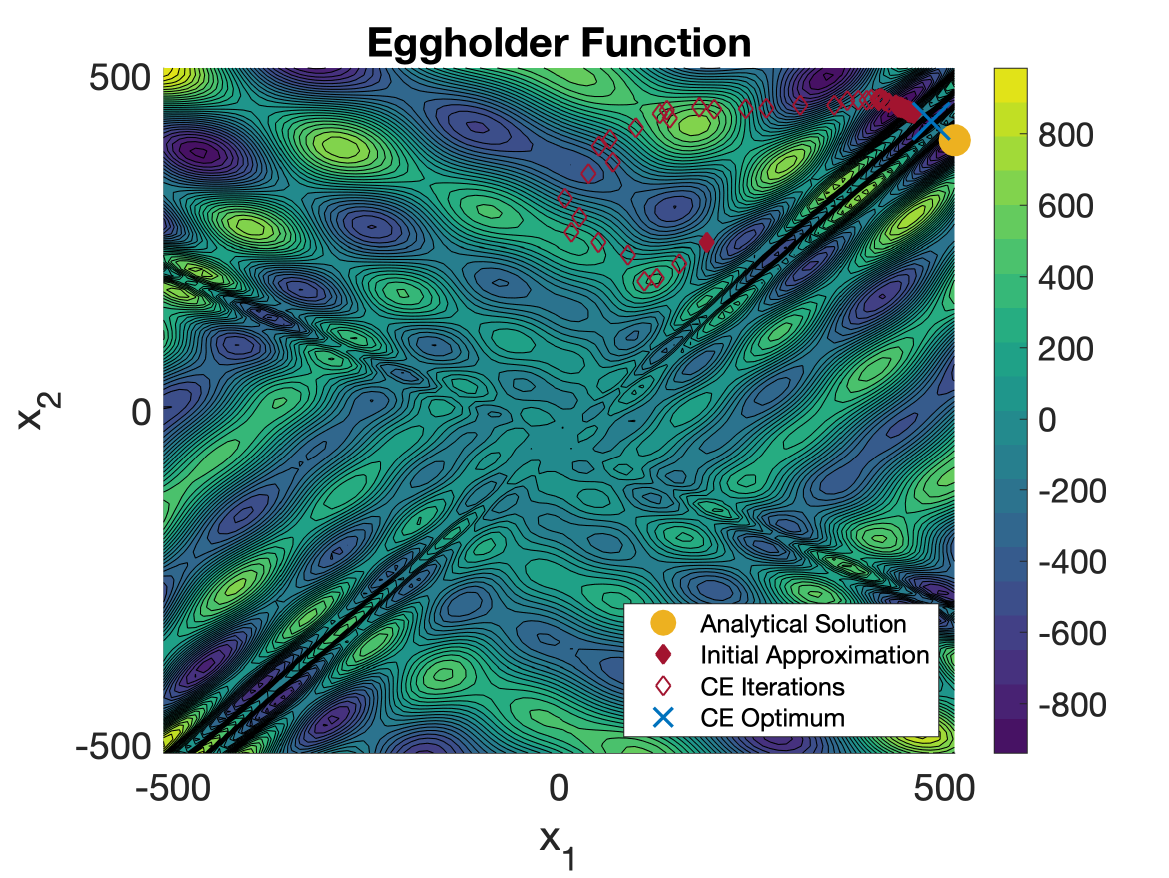}}
\subfloat[]{\includegraphics[width=0.5\textwidth]{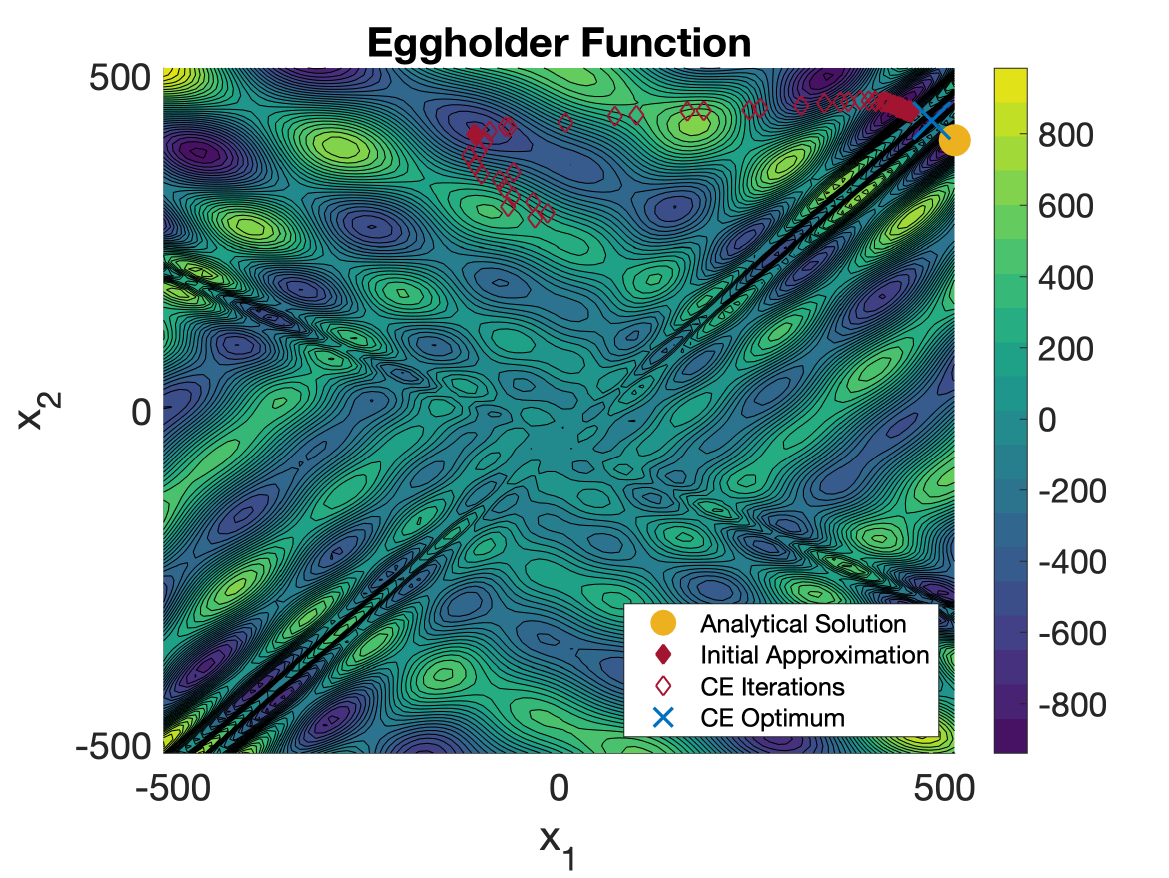}}
\caption{Illustration of multiple optimization runs on the Eggholder function using \pkg{CEopt}. Each trajectory represents a separate run, highlighting how stochastic variation can lead to different paths towards the optimum. This figure showcases the adaptive nature of \pkg{CEopt}, with each run potentially exploring new aspects of the landscape and improving the chances of locating the global optimum.}
\label{fig:Example33}
\end{figure}

In particular, the Bukin N.6 function posed a significant challenge with its steep ravines and a narrow global basin. \pkg{CEopt} consistently reached near-optimal solutions, demonstrating the importance of stochastic methods in managing functions with extreme gradients and limited areas of attraction. The performance on the Eggholder function shown in Figure~\ref{fig:Example33} further highlighted the solver's capabilities; despite the complex landscape sometimes leading the solver away from the best optimum, \pkg{CEopt} maintained a record of the best sampled solutions. This strategic retention of high-quality solutions significantly aids in achieving a superior final outcome, especially in landscapes where the global optimum is surrounded by challenging terrain.

The optimization tests conducted using \pkg{CEopt} covered a wide array of complex functions, each with its own set of peculiarities and challenges. Several key observations can be made:

\begin{enumerate}
	\item \underline{Global Search Efficiency:} For globally complex functions like Ackley and Griewank, which feature numerous local minima, \pkg{CEopt} effectively escaped local traps and converged towards global minima. This underscores the solver's ability to perform global searches efficiently, a crucial attribute for solving real-world problems where the landscape is often unpredictable and riddled with local optima.

	\item \underline{Handling Sharp Gradients:} In benchmarks like the Dixon-Price and Rosenbrock functions, which are known for their sharp, narrow valleys, \pkg{CEopt} demonstrated remarkable precision in navigating tight gradients without overshooting, a common issue in gradient-based methods. This precision is particularly advantageous in engineering applications where small deviations can lead to significantly different outcomes.

	\item \underline{Robustness in Multimodal Landscapes:} The solver's performance on multimodal landscapes such as those presented by the Rastrigin and Styblinski-Tang functions highlighted its robustness. Even in the face of severe oscillations and multiple local optima, \pkg{CEopt} managed to identify areas of interest and refine its search towards the optimum.

	\item \underline{Specialized Challenges:} Certain functions like the Shekel and Eggholder presented specialized challenges due to their peculiar landscapes. For instance, the Shekel function, with its dispersed peaks, tested the solver's ability to discern between closely spaced optima, while the Eggholder function, with its disjointed and irregular surface, required the solver to maintain a balance between exploration and exploitation.
\end{enumerate}

The diversity of these tests not only demonstrates the solver's adaptability and efficiency across various types of optimization problems but also showcases its potential for application in a wide range of disciplines, from financial modeling to bioinformatics.

Due to space constraints, detailed MATLAB implementation for this test is not included in this document. However, the code is available on GitHub, file \texttt{MainCEoptExample3Ext.m}.

The subsequent examples will feature objective functions and/or constraints defined not by algebraic formulas but through alternative representations, providing an additional layer of complexity and challenge for the \pkg{CEopt} solver to navigate.

\subsection{Example 4: Identification of a harmonic oscillator}

This example demonstrates the application of the \pkg{CEopt} solver to the system identification of a damped harmonic oscillator without external forcing. The task is to estimate parameters of the second-order ordinary differential equation (ODE) that models the physical behavior of the system.

The harmonic oscillator is described by the ODE
\begin{equation}
\ddot{y}(t) + 2 \, \zeta \, \omega_n \, \dot{y}(t) + \omega_n^2 \, y(t) = 0 \, ,
\label{eq:MSD1}
\end{equation}
where $\omega_n$ is the natural frequency, and $\zeta$ is the damping ratio. The analytic solution, assuming initial conditions $y(0) = y_0$ and $\dot{y}(0) = v_0$, is
\begin{equation}
y(t) = A  \, e^{-\zeta \, \omega_n t} \sin(\omega_d \,  t + \phi) \, ,
\label{eq:MSD2}
\end{equation}
with $\omega_d = \omega_n  \, \sqrt{1-\zeta^2}$, $A = \sqrt{y_0^2 + \left(\frac{v_0 + \zeta \, \omega_n \,  y_0}{\omega_d} \right)^2}$, and $\phi = \tan^{-1}\left(\frac{y_0  \, \omega_d}{v_0 + \zeta  \, \omega_n  \, y_0}\right)$.

Synthetic data for system identification is generated by computing the analytical solution and adding Gaussian noise, representing measurement errors. This provides realistic data, lumped into the vector $\vec{y}_{\text{data}} \in \R^{N}$, for the parameter estimation process. In this way, the goal is to minimize the objective function given by the discrepancy between the observed noisy data and the model's response, numerically computed using MATLAB's \texttt{ode45} solver
\begin{equation}
F(\vec{x}) = \frac{1}{\sqrt{N}} \left\| \vec{y}_{\text{data}} - \vec{y}_{\text{model}}(\vec{x}) \right\| \, ,
\label{eq:MSD3}
\end{equation}
where $\vec{x} = [\omega_n, \zeta, y_0, v_0]$ represents the parameters to be estimated, and $\vec{y}_{\text{model}}(\vec{x})$ is the response generated by numerically solving the ODE with these parameters for $N$ time-steps.

This example serves as a typical scenario in physics and engineering, where models are calibrated against experimental data. The test case demonstrates \pkg{CEopt}'s utility in systems where direct analytical modeling is not feasible, emphasizing its strength in handling complex systems described by differential equations. 

\begin{figure}[H]
\centering
\captionsetup[subfigure]{labelformat=empty} % Optional: removes labels from each subfigure
\subfloat[]{\includegraphics[width=0.3\textwidth]{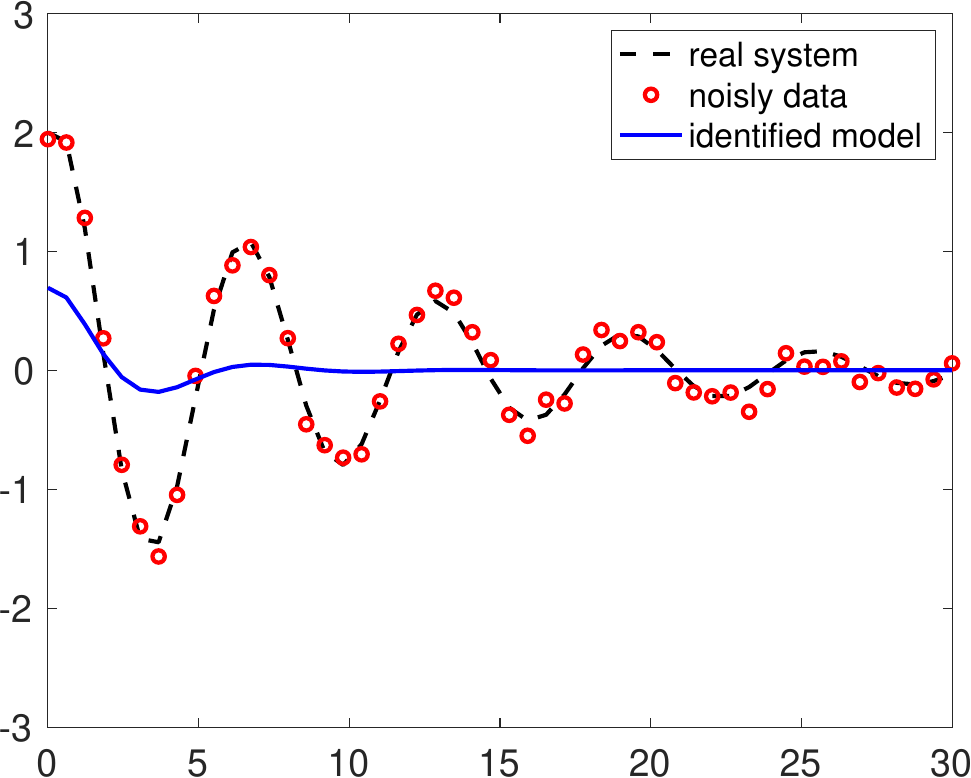}}
\subfloat[]{\includegraphics[width=0.3\textwidth]{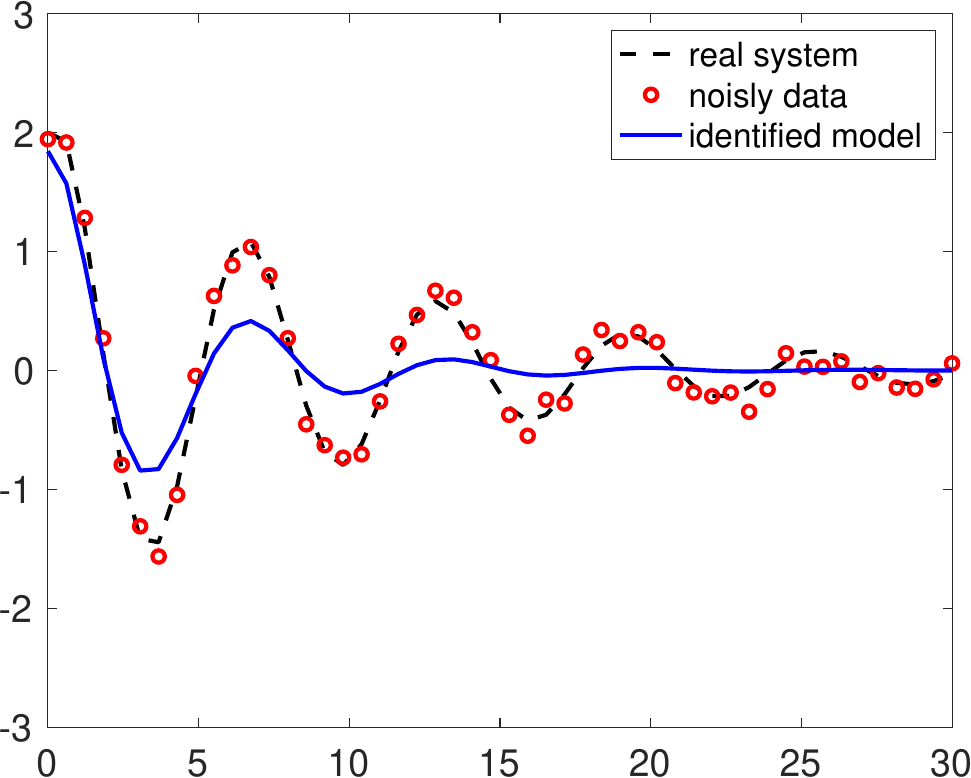}}
\subfloat[]{\includegraphics[width=0.3\textwidth]{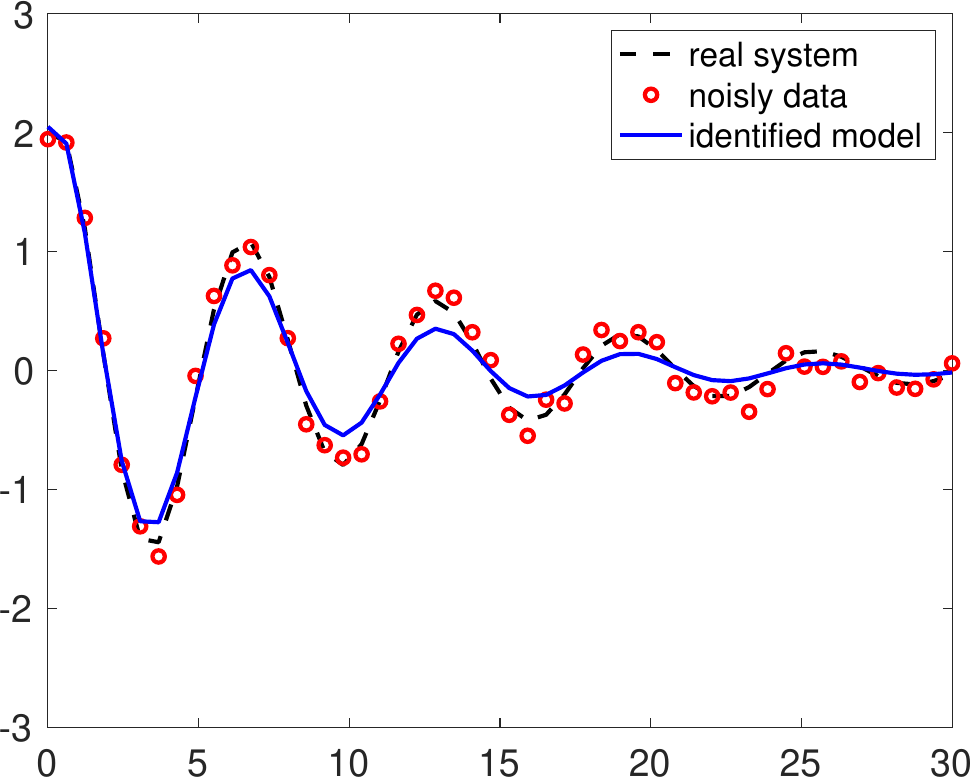}}\\
\subfloat[]{\includegraphics[width=0.3\textwidth]{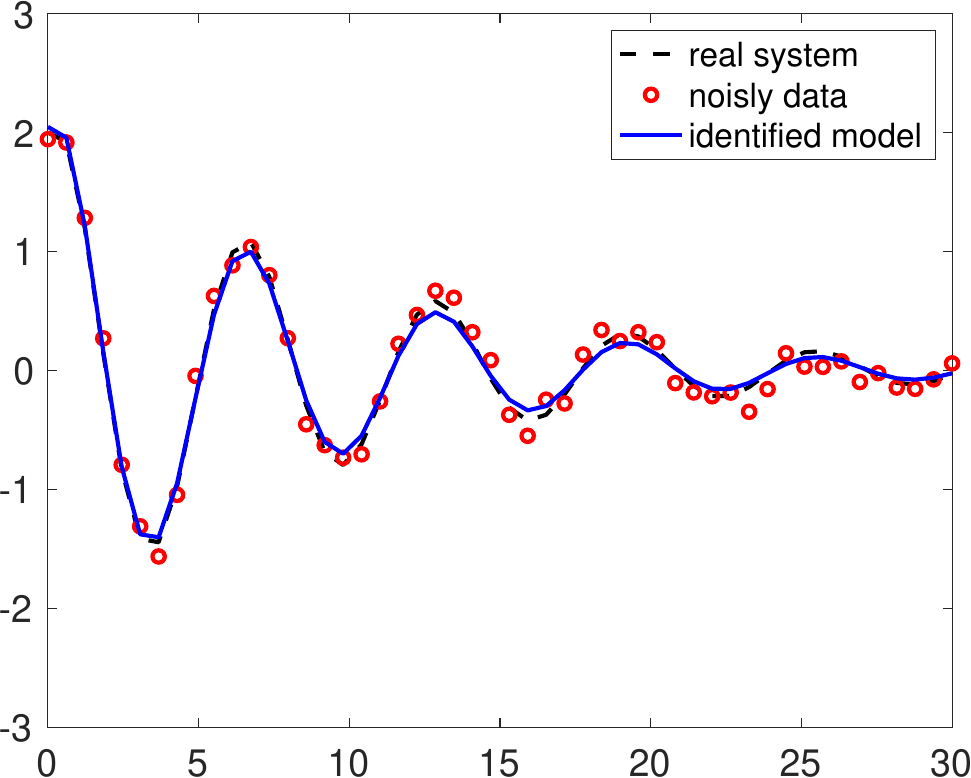}}
\subfloat[]{\includegraphics[width=0.3\textwidth]{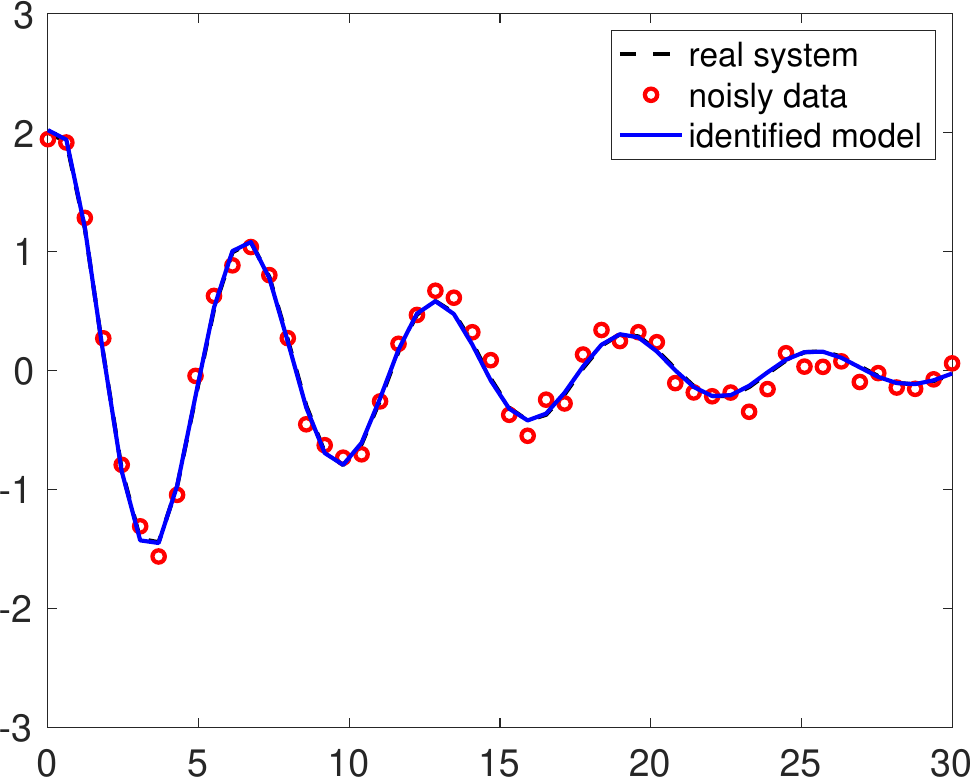}}
\subfloat[]{\includegraphics[width=0.3\textwidth]{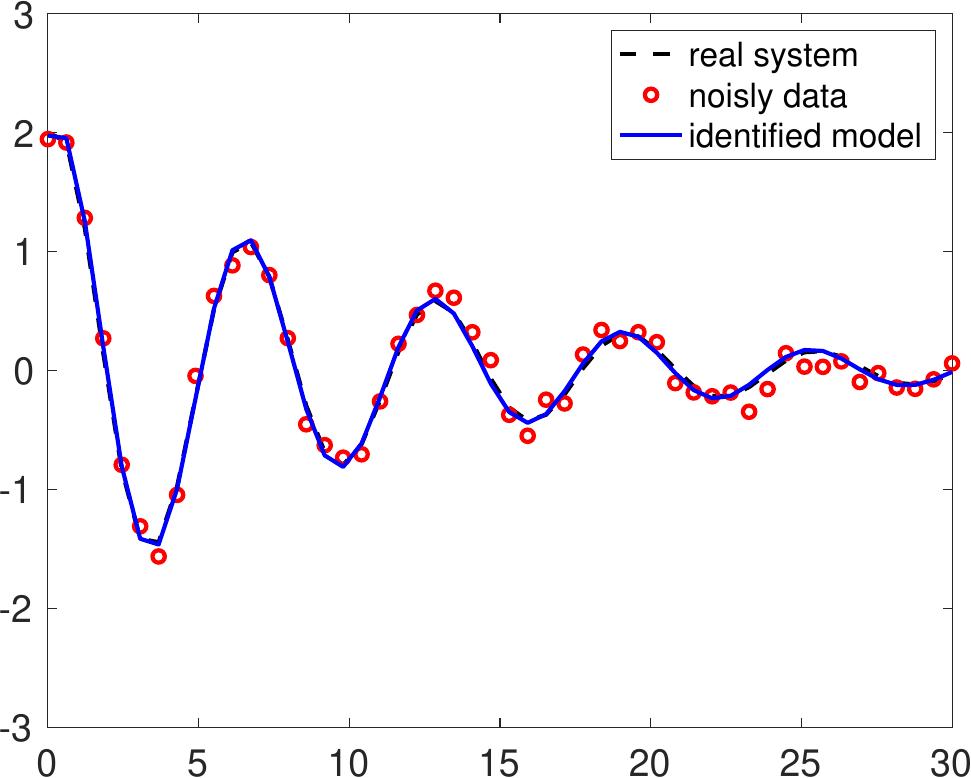}}
\caption{Progressive calibration of a damped harmonic oscillator using \pkg{CEopt}. This sequence shows the solver's iterations from initial guesses to refined solutions. Subfigures compare the groud thruth (black dashed line), the noisy data (red circles), and the identified model (blue solid line) at various stages.}
\label{fig:Example4}
\end{figure}

The Figure~\ref{fig:Example4} captures the system identification process at different iterations. These visualizations highlight the solver's ability to refine its estimates and converge toward the model that best fits the noisy data. Starting with initial guesses far from the true system dynamics, the solver iteratively adjusts the parameters, effectively reducing the discrepancy between the computed model and the observed data. This is evident in the trajectory from initial to final guesses, showing marked improvements in alignment with the real system's behavior.

\pagebreak
The MATLAB code below sets up the optimization problem by defining the objective function as the misfit between the numerical solution of the ODE, computed using MATLAB's \texttt{ode45} solver, and the noisy data. It uses the \pkg{CEopt} algorithm to guide the optimization process, aiming to minimize this misfit by adjusting the system parameters.

% Example 4
\begin{center}
\texttt{MainCEoptExample4.m}\\
\begin{minipage}{0.8\linewidth}
\begin{lstlisting}[
backgroundcolor=\color{gray!08},
%numbers=left,
numberstyle=\tiny,
%numbersep=8pt,
style=Matlab-editor,
basicstyle=\ttfamily\footnotesize,
numbersep=10pt,
frame=single]
clc; clear; close all;

disp(' ------------------- ')
disp(' MainCEoptExample4.m ')
disp(' ------------------- ')

% system response observations
wn = 1; ksi = 0.1; y0 = 2; v0 = 0.5;
t0 = 0.0; t1 = 30.0; Ndata = 50;
time = linspace(t0,t1,Ndata)';
w     = wn*sqrt(1-ksi^2);
A     = sqrt(y0^2 + ((v0+ksi*wn*y0)/w)^2);
phi   = atan(y0*w/(v0+ksi*wn*y0));
ytrue = A*exp(-ksi*wn*time).*sin(w*time+phi);
ydata = ytrue + 0.05*y0*randn(Ndata,1);

% objective function (minimize the discrepancy)
F = @(x) MyMisfitFunc(x,ydata,time);

% bound for design variables
lb = [0.0; 0.0; -3; -3];
ub = [2.0; 1.0;  3;  3];

% define parameters for the CE optimizer
CEstr.Nsamp  = 50;      % Number of samples
CEstr.TolAbs = 1.0e-3;  % Absolute tolerance
CEstr.TolRel = 1.0e-2;  % Relative tolerance

% CE optimizer
tic
[Xopt,Fopt,ExitFlag,CEobj] = CEopt(F,[],[],lb,ub,[],CEstr)
toc

% Misfit function 
function J = MyMisfitFunc(x,ydata,tspan)
    [Ns,Nvars] = size(x);     % input dimensions
             J = zeros(Ns,1); % preallocate memory for J
            wn = x(:,1);      % model parameter 1
           ksi = x(:,2);      % model parameter 2
            y0 = x(:,3);      % model parameter 3
            v0 = x(:,4);      % model parameter 4
    for n = 1:Ns
        dydt = @(t,y) [0 1; -wn(n)^2 -2*ksi(n)*wn(n)]*y;
        [time,ymodel] = ode45(dydt,tspan,[y0(n) v0(n)]);
        J(n) = norm(ydata-ymodel(:,1))/sqrt(length(ydata));
    end
end
\end{lstlisting}
	\label{Code:Example4}
\end{minipage}
\end{center}

\pagebreak
This example not only validates the effectiveness of \pkg{CEopt} in complex parameter estimation tasks but also demonstrates the solver's potential in applications requiring robust system identification. Given the stochastic nature of \pkg{CEopt}, different runs may explore various aspects of the solution space, enhancing the likelihood of approaching the global optimum—a desirable feature in real-world applications where experimental uncertainties and modeling inaccuracies must be managed effectively. Other similar applications can be seen in \citep{Dantas2019icedyn,Dantas2019cobem,Raqueti2023p4799}.

\subsection{Example 5: Design of a complex mechanism}

This example illustrates the application of the \pkg{CEopt} solver to the dimensional synthesis of a four-bar mechanism like the one shown in Figure~\ref{fig:Example6}, where one of the coupler points follows a predefined curve represented by coordinates $(x_k, y_k), ~ k=1, \cdots, N$ (shown as blue dots). This type of task is a good test for CE's capabilities in mechanical design optimization.

\begin{figure}[ht]%
\centering
\includegraphics[scale=1]{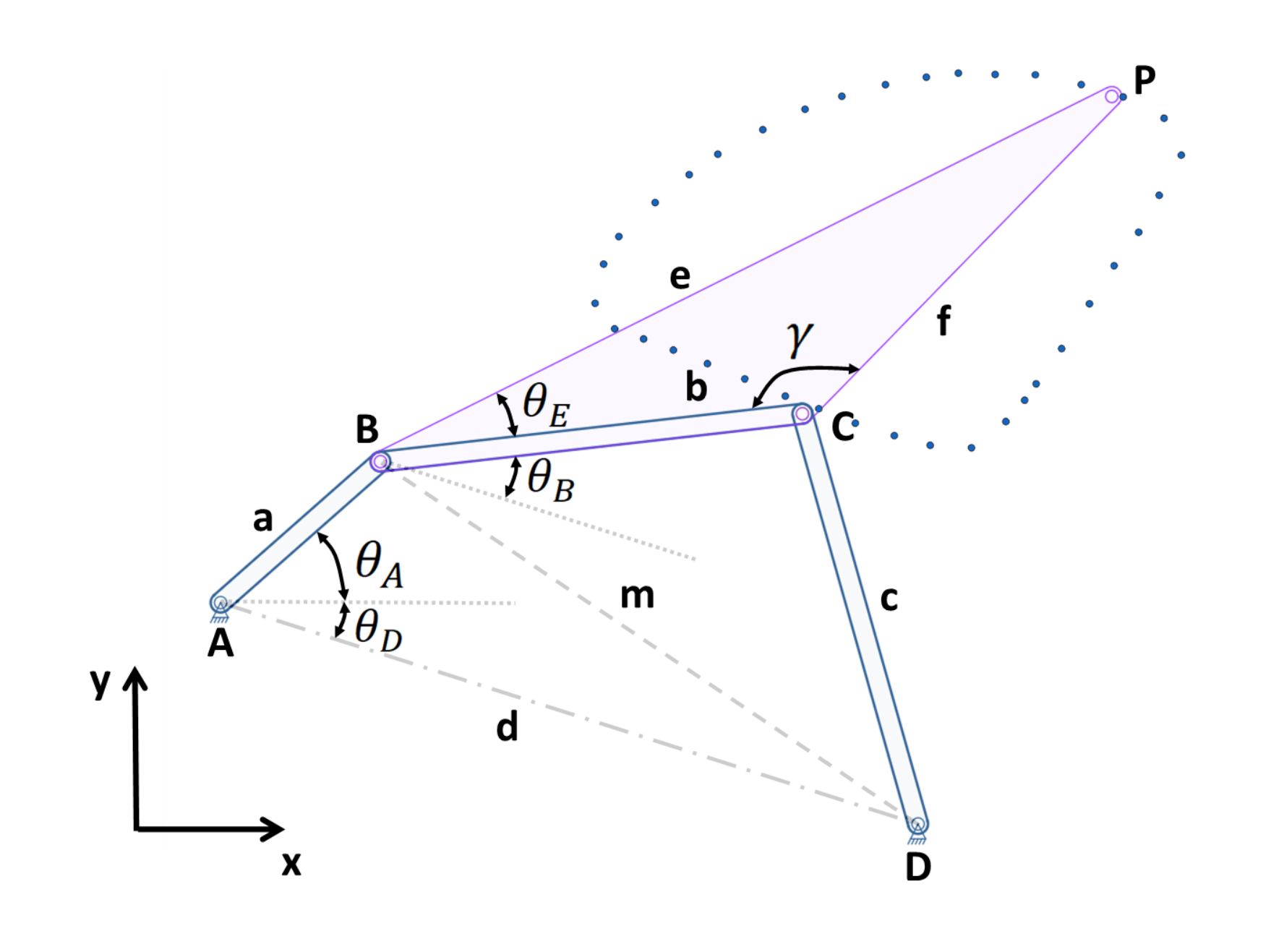}
\caption{Geometric diagram of a four-bar linkage mechanism. The figure displays the positions of the linkages and the specific trajectory (blue dots) that the coupler point $P$ is designed to follow. It shows the lengths $a$, $b$, $c$, and $d$, as well as the distances $e$ and $f$ from pivots $B$ and $C$ to point $P$, respectively. Angles $\theta_A$, $\theta_B$, $\theta_D$, $\theta_E$, and $\gamma$ are also depicted. The shaded region indicates the range of motion of point $P$.}
\label{fig:Example51}
\end{figure}

Figure~\ref{fig:Example51} depicts the geometric configuration of the four-bar mechanism, where $a$ (crank link), $b$ (coupling link), $c$ (follower link), and $d$ (ground link) are the lengths of the links. The position of the coupler point $P$ is determined by the length $e$ and angle $\theta_E$. The groud link $AD$ has its orientation relative to the x-axis is given by angle $\theta_D$. The length of the edge $BD$ is denoted by $m$. The coordinates of fixed joint A are denoted as $(x_A, y_A)$. Given the symmetry of the curve, the lengths of links $b$, $c$ and $f$ are set equal, and from the geometry we see that
\begin{equation}
e = 2 \, b \, \cos{\theta_E} \, .
\label{eq:4bar1}
\end{equation}

To simplify the parameter space, the length of link $a$ is fixed at $a=1$ without any loss of generality. The coordinates of point $P$ are derived using
\begin{equation}
x_P = x_A + a  \, \cos(\theta_A + \theta_0 + \theta_D) + e  \, \cos(\theta_B + \theta_E + \theta_D) \, ,
\label{eq:4bar2}
\end{equation}
and
\begin{equation}
y_P = y_A + a  \, \sin(\theta_A + \theta_0 + \theta_D) + e  \, \sin(\theta_B + \theta_E + \theta_D) \, ,
\label{eq:4bar3}
\end{equation}
where
\begin{equation}
\theta_E=\cfrac{\pi-\gamma}{2} \, ,
\label{eq:4bar4}
\end{equation}
\begin{equation}
\theta_B = \cos^{-1} \left(\cfrac{b^2 + m^2 - c^2}{2 \, b \, m}\right) - \beta \, ,
\label{eq:4bar5}
\end{equation}
\begin{equation}
\beta = \sin^{-1} \left(\cfrac{a  \,  \sin{\theta_A}}{m}\right) \, ,
\label{eq:4bar6}
\end{equation}
and
\begin{equation}
m = \sqrt{a^2 + d^2 - 2 \, a \, d  \, \cos{\theta_A}} \, ,
\label{eq:4bar7}
\end{equation}
being $\theta_0$ an initial angular inclination.

Normalization is applied to the reference and calculated curves to eliminate the dependence on the unknowns $x_A$ and $y_A$ so that
\begin{equation}
\hat{x}_i = \cfrac{x_k - \bar{x}}{\max(x_k) - \min(x_k)} \, ,
\label{eq:4bar8}
\end{equation}
\begin{equation}
\hat{y}_i = \cfrac{y_k - \bar{y}}{\max(y_k) - \min(y_k)} \, , 
\label{eq:4bar9}
\end{equation}
\begin{equation}
\hat{x}_{P_k} = \cfrac{x_{P_k} - \bar{x}_{P}}{\max(x_{P_k}) - \min(x_{P_k})} \, , 
\label{eq:4bar10}
\end{equation}
and
\begin{equation}
\hat{y}_{P_k} = \cfrac{y_{P_k} - \bar{y}_{P}}{\max(y_{P_k}) - \min(y_{P_k})} \, ,
\label{eq:4bar11}
\end{equation}
for $k=1, \cdots, N$,  with the upper bar denoting the ensemble average.

The design problem seeks to minimize an objective function that combines absolute values of the differences of the normalized coordinates
\begin{equation}
F(\vec{x}) = \sum_{k=1}^{N} \left( \mid \hat{x}_k - \hat{x}_{P_k} \mid + \mid \hat{y}_k - \hat{y}_{P_k} \mid  \right) \, ,
\label{eq:4bar12}
\end{equation}
where $\vec{x} = [b, d, \gamma, \theta_D, \theta_A]$ represents the design variables. This choice of objetive function tries to minimize the influence of measurement errors on the final design, once 1-norm is less sensitive to outliers than 2-norm \citep{Boyd2018}.

\begin{figure}[ht]%
\centering
\includegraphics[scale=1]{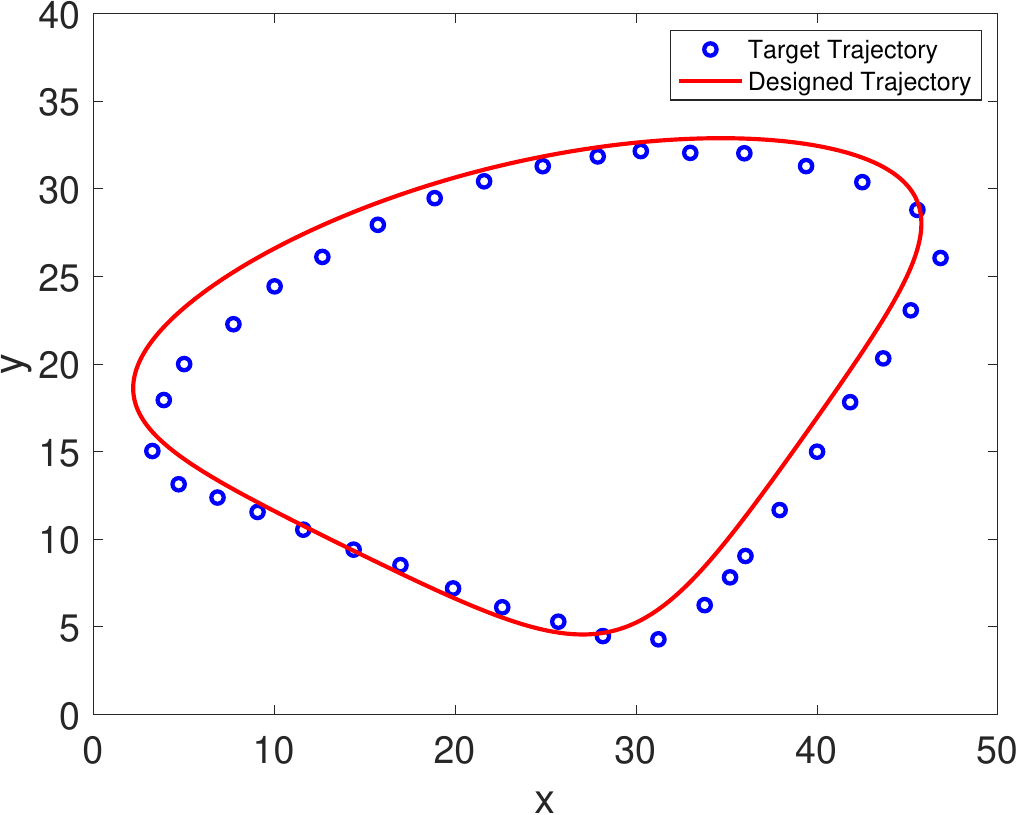}
\caption{Comparison of target and designed trajectories for coupler point $P$. This figure shows the desired trajectory (blue dots) that the coupler point $P$ is expected to follow and the trajectory achieved by the optimized four-bar mechanism (red line) using the \pkg{CEopt} solver.}
\label{fig:Example52}
\end{figure}

\pagebreak
The Figure~\ref{fig:Example52} provides a visual evaluation of the \pkg{CEopt} solver's effectiveness in the dimensional synthesis of a four-bar mechanism. It plots the target trajectory (depicted in blue) against the trajectory (shown in red) achieved through optimization. The close alignment of the two curves demonstrates the capability of \pkg{CEopt} to accurately configure the mechanism's parameters to meet the design specifications, thereby validating the optimization's success in achieving a functional and precise mechanical design.

For sake of space limitation, the MATLAB implementation for this example is not included in this document. Interested readers can access the complete source code on GitHub, available under the filename \texttt{MainCEoptExample5Ext.m}. 

This example serves as a robust demonstration of \pkg{CEopt}'s application in mechanical systems where design parameters must be optimized to achieve specific dynamic behaviors, showcasing the solver's utility in precision design tasks where analytical solutions are infeasible.

\subsection{Example 6: Nonsmooth function with conic constraints}

This example introduces the application of the \pkg{CEopt} solver to an optimization problem involving a nonsmooth function subject to conic constraints, marking the first instance in this series where constraints are explicitly considered. The objective function and constraints are formulated to test the solver's efficacy in handling lake of differentiability and complex constraint boundaries, characteristics typical of many real-world optimization problems.

% Example 6
\begin{center}
\texttt{MainCEoptExample6.m}\\
\begin{minipage}{0.85\linewidth}
\begin{lstlisting}[
backgroundcolor=\color{gray!08},
%numbers=left,
numberstyle=\tiny,
%numbersep=8pt,
style=Matlab-editor,
basicstyle=\ttfamily\footnotesize,
numbersep=10pt,
frame=single]
clc; clear; close all;

disp(' ------------------- ')
disp(' MainCEoptExample6.m ')
disp(' ------------------- ')

% objective function and constraints
F       = @PatternSearchFunc;
nonlcon = @ConicConstraints;

% bound for design variables
lb  = [-6 -4];
ub  = [ 2  4];
mu0 = [-4  2];

% cross-entropy optimizer struct
CEstr.isVectorized  = 1;       % vectorized function
CEstr.TolCon        = 1.0e-6;  % relative tolerance

tic
[Xopt,Fopt,ExitFlag,CEstr] = CEopt(F,[],[],lb,ub,nonlcon,CEstr)
toc

% objective function
function F = PatternSearchFunc(x)
    x1 = x(:,1);
    x2 = x(:,2);
    F = zeros(size(x1,1),1);
    for i = 1:size(x,1)
        if  x1(i) < -5
            F(i) = (x1(i)+5).^2 + abs(x2(i));
        elseif x1(i) < -3
            F(i) = -2*sin(x1(i)) + abs(x2(i));
        elseif x1(i) < 0
            F(i) = 0.5*x1(i) + 2 + abs(x2(i));
        elseif x1 >= 0
            F(i) = .3*sqrt(x1(i)) + 5/2 + abs(x2(i));
        end
    end
end

% equality and inequality constraints
function [G,H] = ConicConstraints(x)
    x1 = x(:,1);
    x2 = x(:,2);
    G  = 2*x1.^2 + x2.^2 - 3;
    H  = (x1+1).^2 - (x2/2).^4;
end
\end{lstlisting}
	\label{Code:Example6}
\end{minipage}
\end{center}

The objective function for this example is defined as
\begin{equation}
F(x_1, x_2) = 
\begin{cases} 
(x_1 + 5)^2 + |x_2| & \text{if } x_1 < -5 \, , \\
-2 \, \sin(x_1) + |x_2| & \text{if } -5 \leq x_1 < -3 \, , \\
0.5 \, x_1 + 2 + |x_2| & \text{if } -3 \leq x_1 < 0 \, , \\
0.3 \, \sqrt{x_1} + 2.5 + |x_2| & \text{if } x_1 \geq 0 \, .
\end{cases}
\end{equation}

The constraints include one inequality and one equality conic constraint given by
\begin{align}
G(x_1, x_2) & = 2x_1^2 + x_2^2 - 3 \leq 0 \, , \\
H(x_1, x_2) &= (x_1 + 1)^2 - \left(\frac{x_2}{2}\right)^4 = 0 \, .
\end{align}

These constraints define a feasible region that is both non-convex and challenging due to the presence of an equality constraint that introduces a higher-order nonlinearity.

The MATLAB code, provided in \texttt{MainCEoptExample6.m} and available on GitHub, implements the \pkg{CEopt} solver to minimize the objective function while satisfying the constraints.

\begin{figure}[H]
\centering
\includegraphics[width=0.95\textwidth]{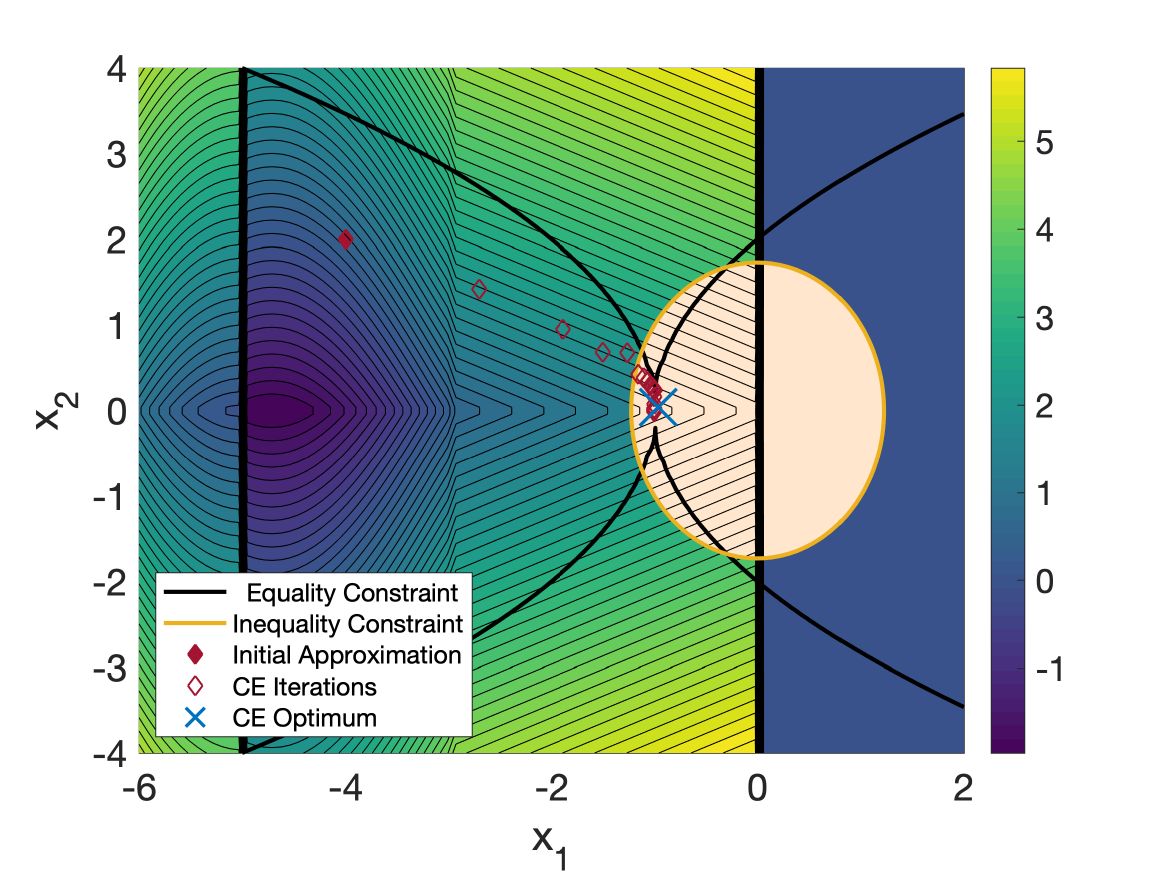}
\caption{Optimization landscape for the nonsmooth function with conic constraints. The contour map depicts the function's nonsmooth behavior, while the shaded regions represent the feasible area defined by the inequality constraint. The lines and markers indicate the levels of the equality constraint and the optimal solution found by \pkg{CEopt}, respectively. This visualization highlights the solver's capability to navigate complex constraint geometries and identify solutions near the boundary of the feasible region.}
\label{fig:Example6}
\end{figure}

Figure \ref{fig:Example6} shows the optimization landscape for this nonsmooth function with clear demarcations of the feasible regions and constraint boundaries. The contour lines provide insights into the function's complexity and its multiple discontinuities. The shaded areas depict the regions where the inequality constraints are satisfied, which are further validated by the equality constraint lines intersecting these regions. The plot visually demonstrates how \pkg{CEopt} efficiently navigates through these constraints to locate an optimal solution that lies at the intersection of these constraints, showcasing its effectiveness in managing complex constraint interactions. This is particularly notable as the optimal solution, marked on the plot, is found near the challenging boundary regions, where the constraints converge.

This example not only tests the solver's capability to handle nonsmooth landscapes and complex constraints but also serves as a benchmark for comparing the \pkg{CEopt}'s performance with MATLAB's traditional genetic algorithms (\texttt{ga}), which are frequently employed for similar types of optimization challenges. To facilitate a comprehensive evaluation, try to solve the same problem using the \texttt{ga} command in MATLAB and compare the results obtained with those from the provided code. This comparative analysis can highlight the efficiency, accuracy, and computational demands of \pkg{CEopt} relative to \texttt{ga}, offering valuable insights into their respective strengths and limitations in handling complex optimization scenarios.

\subsection{Example 7: Non-convex structural optimization}

\begin{figure}[H]
\centering
\includegraphics[width=0.95\textwidth]{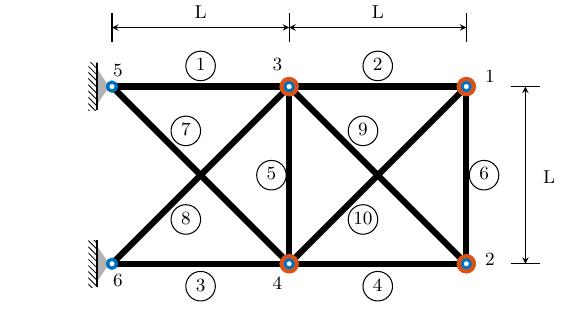}
\caption{Schematic of the ten-bar truss structure, showing the arrangement and labeling of bars, with dimensions and load conditions specified for the optimization.}
\label{fig:Example71}
\end{figure}

This example demonstrates the application of the \pkg{CEopt} solver to optimize a ten-bar truss system, Figure~\ref{fig:Example71}, aiming to minimize the structure's weight by adjusting the cross-sectional areas of the bars. The optimization incorporates constraints to ensure the first three natural frequencies exceed specified thresholds to prevent resonance and ensure dynamic stability.

%% Example 7
%\begin{center}
%\texttt{MainCEoptExample7.m}\\
%\begin{minipage}{0.8\linewidth}
%\lstinputlisting[
%backgroundcolor=\color{gray!08},
%%numbers=left,
%numberstyle=\tiny,
%%numbersep=8pt,
%style=Matlab-editor,
%basicstyle=\ttfamily\footnotesize,
%numbersep=10pt,
%frame=single]
%{MainCEoptExample7.m}
%	\label{Code:Example7}
%\end{minipage}
%\end{center}

%Mechanical equilibrium, crucial for structural integrity, is governed by the following linear system of algebraic equations
%\begin{equation}
%[\mathbf{K}] \, \mathbf{u} = \mathbf{f} \, ,
%\end{equation}
%where $[\mathbf{K}]$ represents the stiffness matrix, $\mathbf{u}$ the displacement vector, and $\mathbf{f}$ the force vector. 

The eigenvalue problem for determining natural frequencies is expressed as
\begin{equation}
[\mathbf{K}] \, \bm{\phi}_i = \omega^2_i \, [\mathbf{M}] \, \bm{\phi}_i \, ,
\end{equation}
where $[\mathbf{K}]$ represents the stiffness matrix, $[\mathbf{M}]$ is the mass matrix, $\bm{\phi}_i$ the $i$-th mode shape, and $\omega_i$ the $i$-th natural frequency. The constraints on the frequencies, to ensure they meet the dynamic stability criteria, are formulated as
\begin{equation}
\omega_i \geq \omega_i^{\dagger} 
~~ \Longleftrightarrow ~~ 
1 - \frac{\omega_i}{\omega_i^{\dagger}} \leq 0 \, ,
\end{equation}
with the given threshold values $\omega_i^{\dagger}$.

The MATLAB code implementing this optimization, \texttt{MainCEoptExample7Ext.m}, is available on GitHub. This code sets up the optimization parameters, defines the objective function and constraints, and invokes \pkg{CEopt} to solve the problem.

\begin{figure}[h]
\centering
\includegraphics[width=0.95\textwidth]{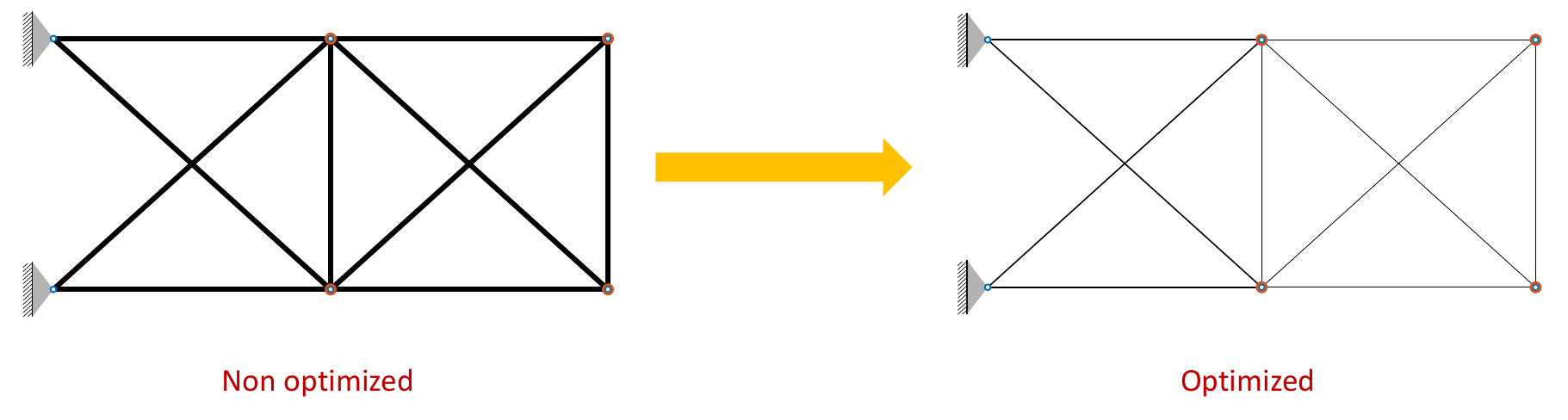}
\caption{Visualization of optimization results for the ten-bar truss system. This figure contrasts the initial and optimized configurations, emphasizing the reduction in material use while enhancing structural efficiency and meeting dynamic stability requirements.}
\label{fig:Example72}
\end{figure}

Figure~\ref{fig:Example72} provides a compelling visualization of the optimization outcomes for the ten-bar truss system, highlighting the transformation from the initial configuration to the optimized structure. This visualization demonstrates the optimization's impact, showing a significant reduction in material usage without compromising structural integrity. The optimized configuration achieves improved structural efficiency by redistributing material where it contributes most to load-bearing capacity, which is evident from the streamlined appearance of the truss. This not only ensures that the structure meets all imposed constraints, including mechanical stability and frequency requirements, but also underscores the potential of \pkg{CEopt} to enhance performance in complex engineering designs.

\subsection{Example 8: Fractional-order controller optimal tuning}

\begin{figure}[h]
\centering
\includegraphics[width=0.95\textwidth]{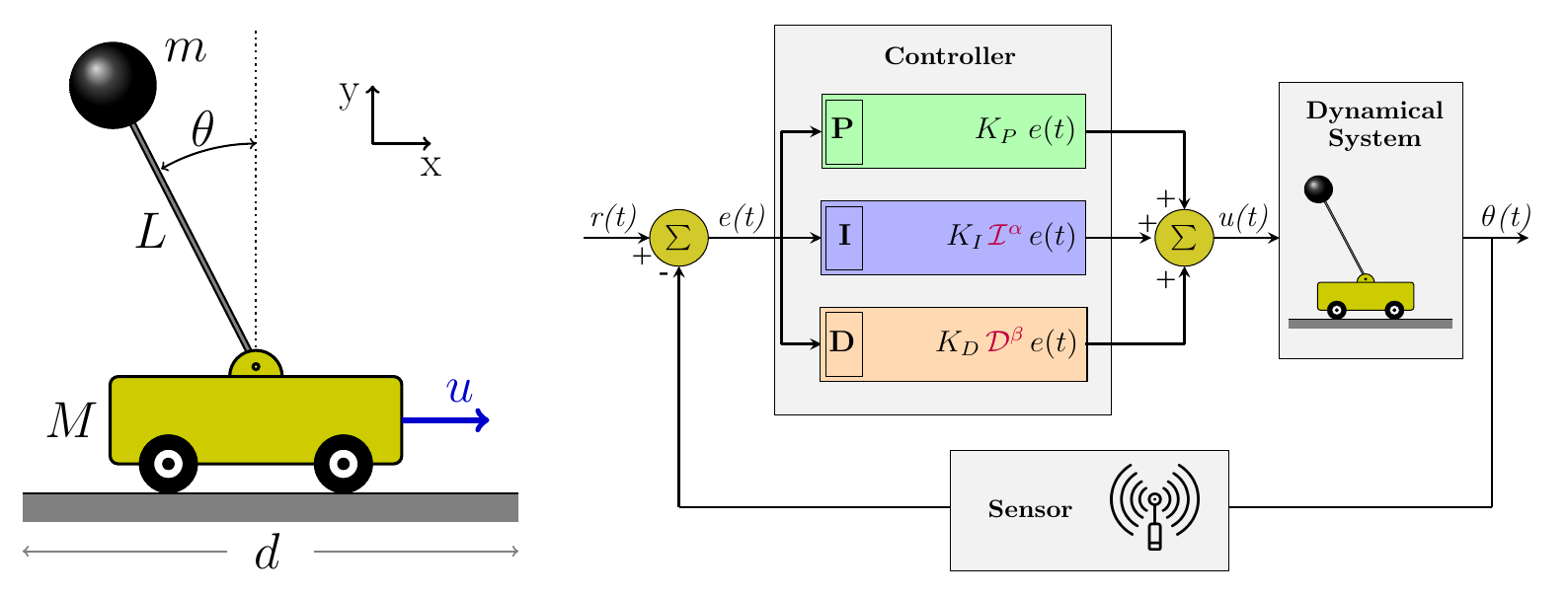}
\caption{Control system diagram showcasing the components of the fractional-order controller for the inverted pendulum over a moving cart. The diagram illustrates the feedback loop, including the fractional integrator and differentiator, demonstrating how the control action $u(t)$ is computed based on the error signal $e(t)$.}
\label{fig:Example81}
\end{figure}

This example focuses on the optimal tuning of a fractional-order controller for an inverted pendulum, a canonical control system challenge illustrated in Figure~\ref{fig:Example81} left. The inverted pendulum system, characterized by a pole balanced on a moving base, is inherently unstable and nonlinear, presenting a classic test for assessing control strategies. Fractional-order control is employed to enhance precision and adaptability, leveraging the nuanced capabilities of fractional calculus to address complexities traditional controllers may struggle with.

The dynamic behavior of the system is governed by the following differential equations, which describe the motion of the base and the angular movement of the pendulum
\begin{equation}
    m \, L \, \cos{\theta(t)} \, \ddot{x}(t) + (J + m\, L^2) \, \ddot{\theta}(t) - m \, g \, L \, \sin{\theta(t)} = 0 \, ,
\end{equation}
\begin{equation}  
	(m +M) \, \ddot{x}(t) + m\, L \, \cos{\theta(t)} \, \ddot{\theta}(t) - m \, L \, \sin{\theta(t)} \, \dot{\theta}^{\,2}(t) = u(t) \, ,
\end{equation}
where $x(t)$ represents the cart horizontal displacement; $\theta(t)$ is the inverted pendulum angular displacement; the upper dot is an abbreviation for temporal derivative; $m$ is the mass of the pendulum; $M$ is the mass of the cart; $L$ is the length of the pendulum; $J$ is the moment of inertia; $g$ is the acceleration due to gravity, and $u(t)$ is the control force applied to the cart.

The control law employs a fractional-order approach, described by
\begin{equation}
u(t) =  K_P \, e(t) + K_I \, \textcolor{purple}{\mathcal{I}^{\alpha}} \, e(t) + K_D \, \textcolor{purple}{\mathcal{D}^\beta} \, e(t) \, ,
\end{equation}
where $e(t) = r(t) - \theta(t)$ is the control error, and \textcolor{purple}{$\mathcal{I}^{\alpha}$} and \textcolor{purple}{$\mathcal{D}^\beta$} represent the fractional integral and derivative operators, respectively. The controller parameters $K_P$, $K_I$, and $K_D$, as well as the fractional-orders $\alpha$ and $\beta$, are tuned to optimize system response.

Fractional calculus  introduces flexibility to the controller by accounting for past states of the system \citep{Ortigueira2011,ortigueira2015p4,DElia2021p1301}, a feature that is vital for handling dynamic complexities like those of an inverted pendulum. The fractional operators are implemented using the FOMCON toolbox \citep{Tepljakov2011p51,Tepljakov2011}, using the Riemann-Liouville definition
\begin{equation}
\textcolor{purple}{\mathcal{I}^{\alpha}} e(t) = \frac{1}{\Gamma(\alpha)} \int_{a}^{t} (t - \tau)^{\alpha-1} \,  e(\tau) \, d\tau \, ,
\end{equation}
\begin{equation}
\textcolor{purple}{\mathcal{D}^{\beta}} e(t) = \frac{d^n}{dt^n} \left( \frac{1}{\Gamma(n-\beta)} \int_{a}^{t} (t - \tau)^{n-\beta-1} \, e(\tau) \, d\tau \right) \, ,
\end{equation}
where $\Gamma$ denotes the Gamma function.

The control cost function to be minimized for $\vec{x} = [K_P, K_I, K_D, \alpha, \beta]$, which emphasizes both performance and energy efficiency, integrates the square of the error and the square of the control signal over time
\begin{equation}
\displaystyle{F}(\vec{x}) = \int_0^t e(\tau)^{2} \, d\tau + \int_0^t \left[ \frac{u(\tau)}{100}  \right]^{2} \, d\tau \, ,
\end{equation}
subject to the constraints that cart displacement must remain within specified bounds
\begin{equation}
-d/2 \leq x(t) \leq d/2
~~ \Longleftrightarrow ~~ 
\mid x(t) \mid - d/2 \leq 0 \, .
\end{equation}

\begin{figure}[h]
\centering
\includegraphics[width=0.5\textwidth]{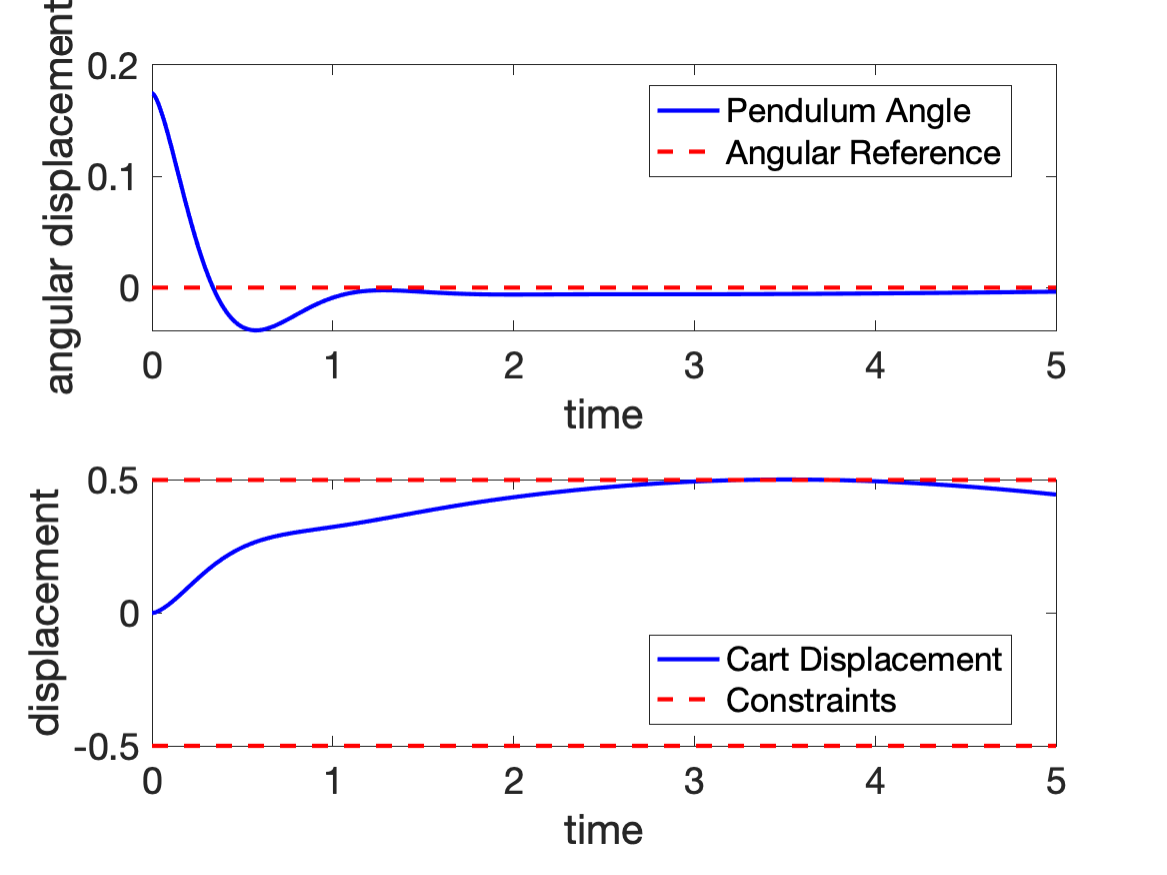}
\caption{Controlled Dynamics of the Cart-Pendulum System. This figure illustrates the controlled trajectories of the cart position $x(t)$ and the pendulum angle $\theta(t)$ over time, showcasing the effectiveness of the optimized fractional-order controller in stabilizing the inverted pendulum over a moving cart.}
\label{fig:Example82}
\end{figure}

Figure~\ref{fig:Example82} displays the dynamic behavior of both the cart position $x(t)$ and the pendulum angle $\theta(t)$ under the influence of an optimized fractional-order controller. The trajectories demonstrate how the controller effectively stabilizes the pendulum in an upright position while managing the horizontal motion of the cart. These results validate the controller's performance in real-time, emphasizing its ability to maintain stability and respond appropriately to the system's inherent instabilities and nonlinear characteristics. The successful regulation of both position and angle highlights the practical applicability and efficiency of \pkg{CEopt} to tune the optimized control strategy in a challenging dynamic environment.

The MATLAB code \texttt{MainCEoptExample8Ext.m} provided on GitHub facilitates the parameter tuning process using the \pkg{CEopt} solver, demonstrating the effectiveness of fractional-order control in complex scenarios. For further details on the applications of fractional-order control, see \cite{cunhajr_vetomac2019_1, cunhajr_nodycon2021_2, cunhajr_vss2022}.

\pagebreak
\subsection{Example 9: Sparse Identification of Nonlinear Dynamics -- SINDy}

The final example discusses the application of the Sparse Identification of Nonlinear Dynamics (SINDy) algorithm \citep{Brunton2016p3932,Brunton2022} using the \pkg{CEopt} solver. SINDy is an innovative machine learning method for discovering governing differential equations from time-series data of a system's state variables. This approach is particularly potent in scenarios where the underlying dynamics are complex but sparsely driven by a few active terms in a potentially large system of possible dynamics.

\begin{figure}[h]
\centering
\includegraphics[width=0.9\textwidth]{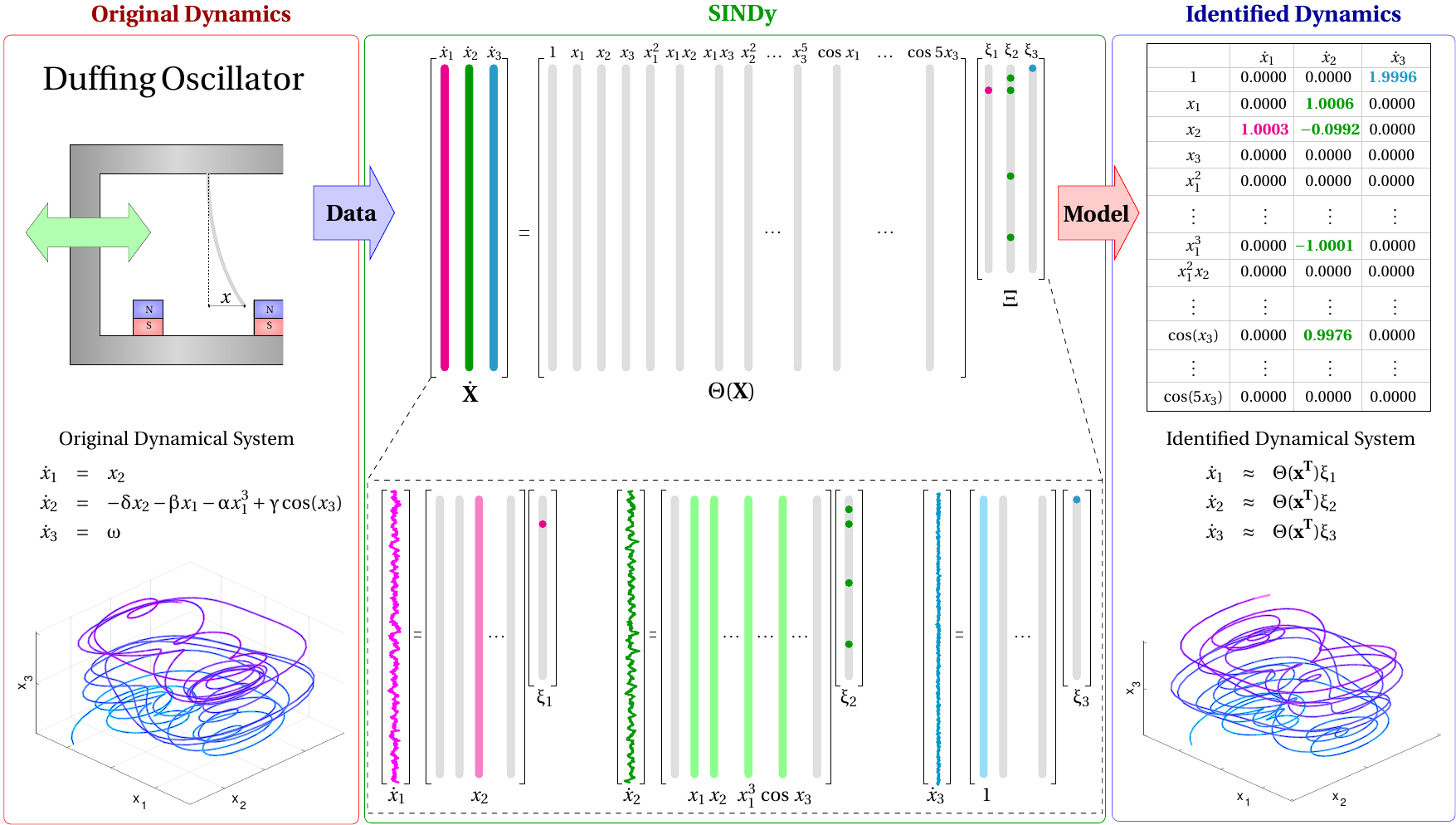}
\caption{Application of SINDy algorithm to discover the underlying differential equations associated with a given time-series. In the left part one has the original dynamical system equations and a plot of the system's trajectory in the phase space. The middle section shows the library of functions $\Theta(\vec{X})$, alongside the coefficients matrix $\Xi$ representing the selected terms that best describe the system dynamics. The right section displays the identified dynamical equations resulting from the SINDy process.}
\label{fig:Example9}
\end{figure}

The central challenge in SINDy is to construct an evolutionary equation that accurately captures the dynamics of the system from noisy data. The original method, which is schematically illustrated in Figure~\ref{fig:Example9}, involves:
\begin{itemize}
	\item \underline{Data Collection:} Gather time-series measurements of the system’s state variables. In this example, synthetic data is generated using the forced Duffing oscillator, a chaotic system known for its intricate nonlinear dynamics.
	\item \underline{Library Construction:} From the time-series data, a comprehensive library of potential functions (polynomials, trigonometric functions, logarithms, etc.) is constructed. These functions are candidate terms that may appear in the true governing equations.
	\item \underline{Regression Problem:} The next step is to solve a regression problem where the coefficients of these functions are determined such that they best fit the observed data. The fit is quantified by how well the derivatives (computed from the data) can be approximated by linear combinations of the library functions.
\end{itemize}

In this sense, SINDy employs misfit function that balances fidelity to the data with sparsity in the model representation. It combines the mean squared error of the model fitting with a penalty proportional to the number of non-zero coefficients, being formulated as
\begin{equation}
    \mathcal{J}(\Xi) = \| \dot{\vec{X}} - \Theta(\vec{X}) \, \Xi \|_2 + \lambda \, \| \Xi \|_0,
\end{equation}
where $\vec{X}$ and $\dot{\vec{X}}$ are matrices which contain the state variables and their derivatives, respectively, among the columns; $\Theta(\vec{X})$ is a dictionary of functions used to reconstruct the vector field\footnote{Here we assume the dynamics is defined by an autonomous system of differential equations $\dot{\textbf{x}}(t) = \mathcal{F}(\textbf{x}(t))$, where $\textbf{x}(t)$ is a state vector that lumps the state coordinates and the function $\mathcal{F}$, known as the system vector field, gives the evolution law of the dynamical system.} of the underlying dynamical system, with each column representing an independent function; $\Xi$ is the matrix of coefficients for $\Theta(\vec{X})$; $\lambda$ is a regularization parameter that controls the trade-off between model complexity (sparsity) and fit to the data; $\| \cdot \|_0$ denotes the ``zero-norm'', which counts the number of non-zero entries, promoting sparsity.

The primary challenge here arises from the use of the ``zero-norm'' for regularization, promoting sparsity in the coefficients during the regression process. The $L^0$ count, indicating the number of non-zero coefficients, transforms the optimization problem into a non-convex NP-hard one, notoriously difficult to solve due to the potential for multiple local minima.

Obviously, the CE method may be employed to tackle this non-convex optimization problem, despite not be the standard choice. The usual procedure to handle this problem in machine learning literature is to use convex relaxation techniques, like Sequential Threshold Least-Squares (STLS) \citep{Brunton2016p3932} or Sparse Relaxed Regularized Regression (SR3) \citep{Zheng2019p1404}. Here we decided to use CE method to test its capacity in such a difficult problem. Therefore, \pkg{CEopt} solver evolves a population of solutions towards the optimum by iteratively learning from the best performers, making it suitable for finding global optima in complex landscapes posed by $L^0$ regularization.

\begin{figure}[H]
\centering
\captionsetup[subfigure]{labelformat=empty} % Optional: removes labels from each subfigure
\subfloat[]{\includegraphics[width=0.5\textwidth]{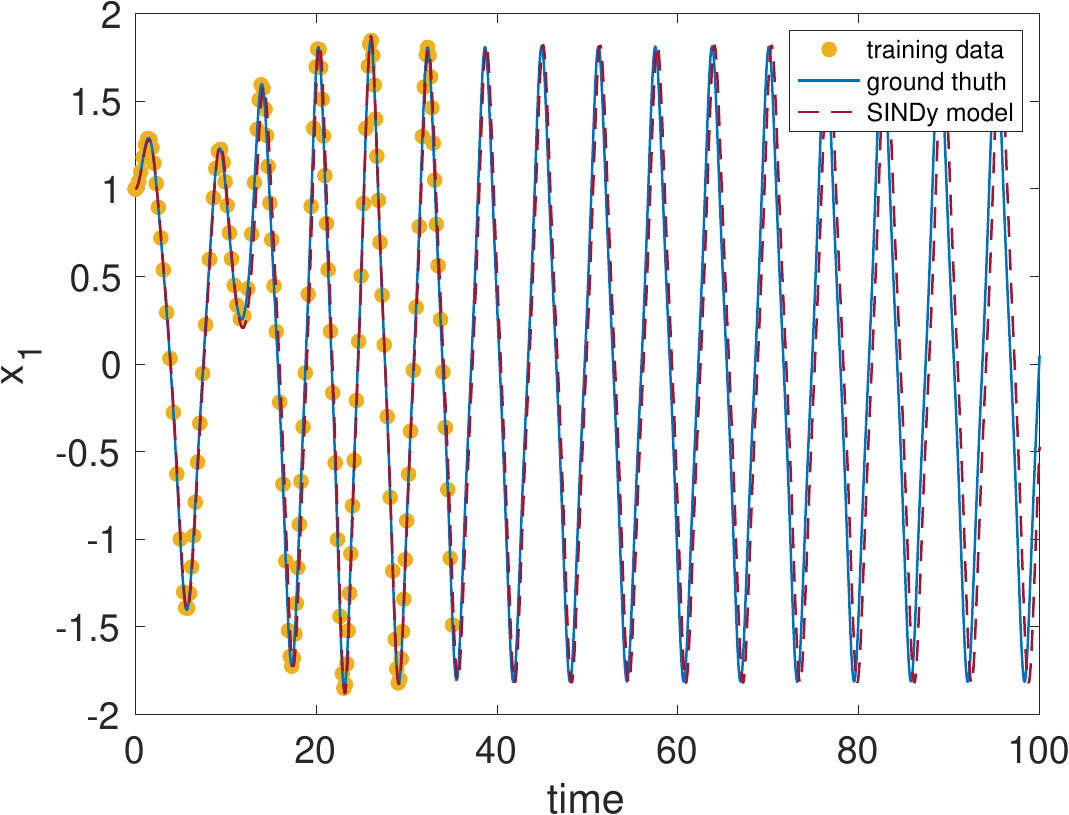}}
\subfloat[]{\includegraphics[width=0.5\textwidth]{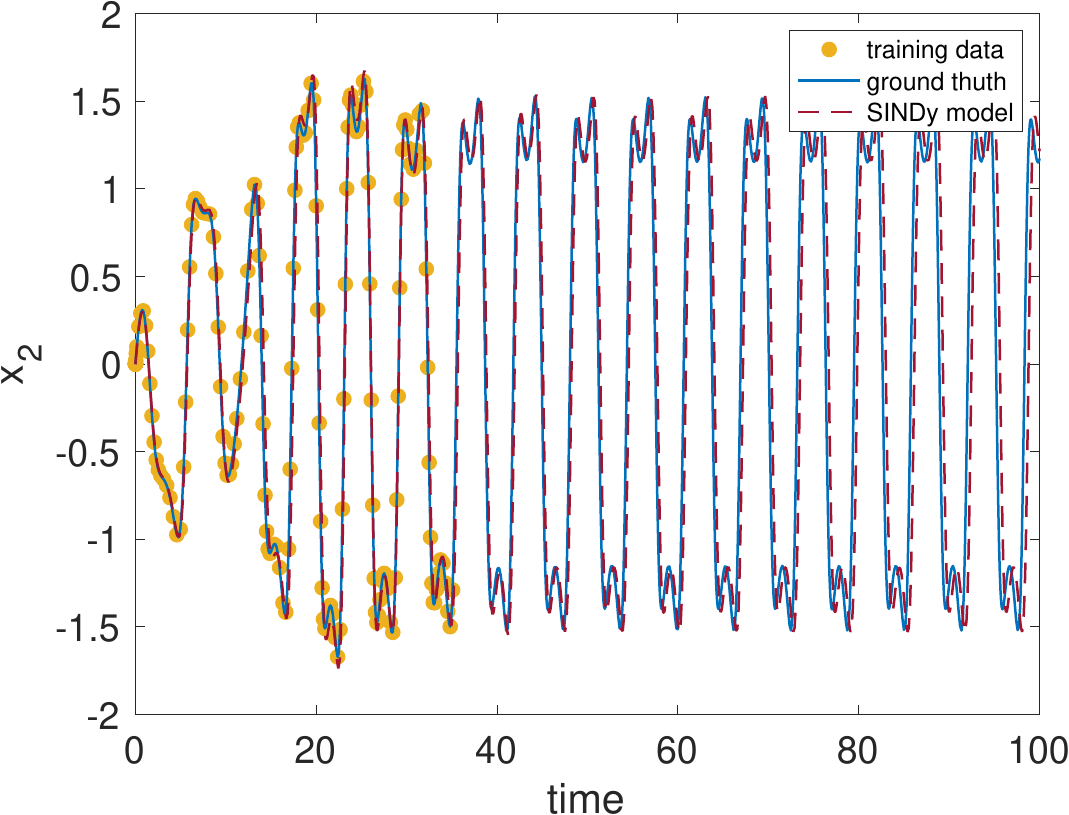}}
\caption{Results of applying the CE-SINDy algorithm to the forced Duffing oscillator. From left to right: state variable $x_1$, and state variable $x_2$. The plots compare the ground truth, training data, and the SINDy data-driven model predictions.}
\label{fig:Example92}
\end{figure}

In this example, synthetic data was generated using the forced Duffing oscillator, which is known for its chaotic and nonlinear dynamics. The parameters used were $\alpha = 1.0$, $\beta = -1.0$, $\delta = 0.3$, $\gamma = 0.65$, and $\omega = 1$. Initial conditions were set to $[x_1(0), x_2(0), x_3(0) ] = [1, 0, 0]$, and the temporal interval of analysis was from $t_0 = 0$ to $t_1 = 100$. The ground truth trajectories of the system were obtained using MATLAB's \texttt{ode45} function, and noisy training data was generated by sampling the ground truth data and adding Gaussian noise.

The MATLAB code \texttt{MainCEoptExample9.m} implements SINDy using \pkg{CEopt} as optimization solver. It constructs the library of functions from the noisy data, sets up the optimization problem with $L^0$ regularization, and uses \pkg{CEopt} to find the sparsest set of coefficients that can model the data accurately. The code is available on GitHub, for replication and exploration.

\begin{table}[h]
\centering
\caption{Identified coefficients of Duffing oscillator obtained via CE-SINDy, employing a dictionary of functions with linear and cubic polynomials, and cosine, as well as $\lambda = 0.25$.}
\begin{tabular}{cccc}
\toprule
Dictionary & \multicolumn{3}{c}{Coefficients} \\
functions  & $x_1(t)$ & $x_2(t)$ & $x_3(t)$ \\
\midrule
$1$              & -0.0160 & 0.1226 & 0.9947 \\
$x_1$          & -0.0097 & 0.9841 & -0.0122 \\
$x_2$          & 0.9910 & -0.2687 & -0.0385 \\
$x_3$          & -0.0063 & -0.0025 & -0.0039 \\
$x_1^3$      & -0.0090 & -0.9912 & -0.0048 \\
$x_2^3$      & -0.0376 & -0.0138 & -0.0089 \\
$x_3^3$      & -0.0000 & -0.0000 & -0.0000 \\
$\cos{x_1}$ & -0.0064 & -0.1018 & -0.0196 \\
$\cos{x_2}$ & -0.0197 & -0.0868 & -0.0105 \\
$\cos{x_3}$ & -0.0393 & 0.6229 & -0.0075 \\
\bottomrule
\end{tabular}
\label{tab:Example91}
\end{table}

\begin{table}[h]
\centering
\caption{Identified coefficients of Duffing oscillator obtained via CE-SINDy after thresholding, employing a dictionary of functions with linear and cubic polynomials, and cosine, as well as $\lambda = 0.25$.}
\begin{tabular}{cccc}
\toprule
Dictionary & \multicolumn{3}{c}{Coefficients} \\
functions  & $x_1(t)$ & $x_2(t)$ & $x_3(t)$ \\
\midrule
$1$              & 0 & 0 & 0.9947 \\
$x_1$          & 0 & 0.9841 & 0 \\
$x_2$          & 0.9910 & -0.2687 & 0 \\
$x_3$          & 0 & 0 & 0 \\
$x_1^3$      & 0 & -0.9912 & 0 \\
$x_2^3$      & 0 & 0 & 0 \\
$x_3^3$      & 0 & 0 & 0 \\
$\cos{x_1}$ & 0 & 0 & 0 \\
$\cos{x_2}$ & 0 & 0 & 0 \\
$\cos{x_3}$ & 0 & 0.6229 & 0 \\
\bottomrule
\end{tabular}
\label{tab:Example92}
\end{table}

\pagebreak
The identified coefficients before and after applying a threshold $\lambda = 0.25$ are presented in Tables~\ref{tab:Example91} and \ref{tab:Example92}. The coefficients obtained before thresholding indicate the presence of several non-zero terms, reflecting the complexity of the system's dynamics. However, after applying a threshold, many coefficients are set to zero, retaining only the most significant terms that contribute to the system's behavior. This sparsity is crucial for interpretability and reducing overfitting, demonstrating SINDy can be used combined with the \pkg{CEopt} solver to handle identification problems for complex nonlinear systems.

The results presented in Figure~\ref{fig:Example92} show the comparison between the training data, ground truth, and the predictions made by the thresholded SINDy model for the state variables $x_1$ and $x_2$. The SINDy model accurately captures the dynamics of the system, even with the presence of noise in the training data.

Although the CE method is not the standard solver for SINDy, this example highlights its robustness and capability to achieve good results in a challenging optimization problem. We do not recommend using CE in place of STLS or SR3 for SINDy. However, this example shows that CE can handle the problem when good bounds for the parameters and a good guess for the mean are prescribed.
% --------------------------------------------------------------

% --------------------------------------------------------------
\section{Final remarks}
\label{sec:FinalRemarks}

This paper introduces \pkg{CEopt}, a MATLAB-based implementation of the Cross-Entropy (CE) method for solving non-convex optimization problems. Through detailed theoretical foundations, historical development, and practical examples, we have demonstrated the method's robustness and versatility in addressing complex optimization challenges.

The CE method stands out for its ability to transform deterministic optimization problems into stochastic ones, leveraging adaptive importance sampling to iteratively hone in on the global optimum. This probabilistic framework simplifies navigating complex, multidimensional landscapes and ensures a focused and efficient search process. The method's simplicity, minimal set of control parameters, and robust convergence properties make it accessible and effective across various applications.

Our implementation, \pkg{CEopt}, extends the CE method's capabilities by incorporating advanced features such as robust handling of both equality and inequality constraints through an augmented Lagrangian approach. This enhancement broadens the method's applicability, particularly in engineering and control systems where such constraints are critical. Additionally, \pkg{CEopt} offers user-friendly interfaces and detailed diagnostic outputs, facilitating ease of use and comprehensive analysis of the optimization process.

The numerical experiments conducted illustrate \pkg{CEopt}'s effectiveness across various domains, including mechanical equilibrium, fractional-order control, and machine learning. These case studies underscore the method's capacity to provide high-quality solutions in scenarios characterized by non-convex, multidimensional, and noisy data landscapes.

%While \pkg{CEopt} provides a robust and user-friendly implementation of the CE method, it is crucial for users to carefully specify the bounds for the design variables using the \texttt{lb} (lower bound) and \texttt{ub} (upper bound) parameters. Without well-defined bounds, the solver may struggle to find high-quality solutions, especially in high-dimensional spaces where the search space can become overwhelmingly large and complex.

While \pkg{CEopt} provides a robust and user-friendly implementation of the CE method, it is crucial for users to carefully specify the bounds for the design variables using the \texttt{lb} (lower bound) and \texttt{ub} (upper bound) parameters. Without well-defined bounds, the solver may struggle to find high-quality solutions, especially in high-dimensional spaces where the search space can become overwhelmingly large and complex. For more challenging problems, fine-tuning the tolerances and smoothing parameters within the \texttt{CEstr} structure is often necessary. This fine-tuning process can be iterative and requires a degree of trial and error to balance the convergence speed and solution accuracy. Users must be prepared to adjust these settings based on the specific characteristics of their optimization problems to fully leverage the potential of \pkg{CEopt}.

In conclusion, \pkg{CEopt} represents a significant addition to the optimization toolkit, bridging the gap between theory and practical application. Its development enhances researchers' and practitioners' ability to engage with and solve complex non-convex optimization problems, providing a reliable, transparent, and user-friendly tool. Future work will focus on further enhancing the algorithm's efficiency, expanding its applicability, and integrating it with other optimization frameworks to address the evolving needs of the optimization community.

We invite the numerical optimization community to explore \pkg{CEopt} and contribute to its ongoing development, ensuring it remains a cutting-edge tool for addressing the increasingly complex optimization challenges of our time. For more information, updates, and educational content, please visit \url{https://ceopt.org}. This website and the solver will be continuously updated to include improvements, bug corrections, and new educational content as the use of the solver expands. We encourage the community to explore, use, and contribute to \pkg{CEopt}, enhancing its capability to tackle optimization problems across various domains.

% --------------------------------------------------------------

% --------------------------------------------------------------
\section*{Acknowledgments}

The first author acknowledge the support given by by the Brazilian agencies Coordena\c{c}\~{a}o de Aperfei\c{c}oamento de Pessoal de N\'{\i}vel Superior (CAPES) under Finance Code 001, Conselho Nacional de Desenvolvimento Cient\'{i}fico e Tecnol\'{o}gico under the grant 305476 /2022-0, and the Carlos Chagas Filho Research Foundation of Rio de Janeiro State (FAPERJ) under grants 210.167/2019, 211.037/2019, and 201.294/2021. The authors thank Mr. Diego Matos (UERJ) for the elaboration of Figure~\ref{fig:Example9} and Prof. Welington Oliveira (MINES Paris-Tech) for an insightful discussion about augmented Lagrangian methods.
% --------------------------------------------------------------

% --------------------------------------------------------------
%\bibliographystyle{jss}
%\bibliography{references}

% --------------------------------------------------------------

\end{document}